\documentclass[12pt]{iopart}

\usepackage{graphicx}
\usepackage{epstopdf}
\usepackage{iopams}

\usepackage{footmisc}

\usepackage{natbib}

\usepackage{color}

\usepackage{xspace}
\usepackage{subfig}
\usepackage{accents}

\def \kbar {\bar{k}}
\def \qbar {\bar{q}}
\def \e { \mbox{$\mathrm{e}$} }
\def \vth {v_{\mathrm{th}}}
\def \vcut {v_{\mathrm{cut}}}

\def \W {\mathcal{W}}
\def \rhovec { \mbox{\boldmath $\rho$} }
\def \ZNS {Z_{\mbox{\scriptsize{NS}}}}
\def \ENS {E_{\mbox{\scriptsize{NS}}}}
\def \EGHM {E_{\mbox{\scriptsize{GHM}}}}
\def \ZGHM {Z_{\mbox{\scriptsize{GHM}}}}
\def \qGHM {q_{\mbox{\scriptsize{GHM}}}}
\def \Egf {E_{\mbox{\scriptsize{gf}}}}

\def \Tau {\mbox{$\mathcal T$}}

\newcommand{\poiss}[2]{\{#1,\;#2\}}
\newcommand{\zon}[1]{\overline{#1}}
\newcommand{\nzo}[1]{\tilde{#1}}

\newcommand{\Aref}[1]{\ref{#1}}

\newcommand{\cfrac}[3][c]{{\displaystyle\frac{%
  \strut\ifx r#1\hfill\fi#2\ifx l#1\hfill\fi}{#3}}%
  \kern-\nulldelimiterspace}

\newcommand\etc{\textit{etc.}\xspace}
\newcommand\eg{\textit{e.g.}\xspace}
\newcommand\ie{\textit{i.e.}\xspace}
\newcommand\cf{\textit{cf.}\xspace}

\newcommand{\gyroavg}[1]{\left\langle#1\right\rangle_{\bf R}}
\newcommand{\angleavg}[1]{\left\langle#1\right\rangle_{\bf r}}
\newcommand{\CollisionOp}[1]{\gyroavg{C[#1]}}

\newlength{\dhatheight}
\newcommand{\dblhat}[1]{%
    \settoheight{\dhatheight}{\ensuremath{\hat{#1}}}%
    \addtolength{\dhatheight}{-0.35ex}%
    \hat{\vphantom{\rule{1pt}{\dhatheight}}%
    \smash{\hat{#1}}}}

\bibpunct{[}{]}{,}{n}{}{;}

\begin{document}

\title{Considering Fluctuation Energy as a Measure of Gyrokinetic Turbulence}



\author{G G Plunk$^{1, 2}$, T Tatsuno$^{2, 3}$ and W Dorland$^2$} 
\address{$ö1$ Max-Planck-Institut f\"{u}r Plasmaphysik, EURATOM-Assoziation, Wendelsteinstr. 1, 17491 Greifswald, Germany} 
\address{$ö2$ Department of Physics and IREAP, University of Maryland, College Park, Maryland 20742, USA} 
\address{$ö3$ Department of Communication Engineering and Informatics, The University of Electro-Communications, 1-5-1 Chofugaoka, Chofu, Tokyo 182-8585, Japan} 
\ead{gplunk@ipp.mpg.de} 

\begin{abstract}
In gyrokinetic theory there are two quadratic measures of fluctuation energy, left invariant under nonlinear interactions, that constrain the turbulence.  The recent work of \citet{plunk-prl} reported on the novel consequences that this constraint has on the direction and locality of spectral energy transfer.  This paper builds on that work.  We provide detailed analysis in support of the results of \citet{plunk-prl} but also significantly broaden the scope and use additional methods to address the problem of energy transfer.  The perspective taken here is that the fluctuation energies are not merely formal invariants of an idealized model (two-dimensional gyrokinetics \citep{plunk-jfm}) but are general measures of gyrokinetic turbulence, \ie quantities that can be used to predict the behavior of the turbulence.  Though many open questions remain, this paper collects evidence in favor of this perspective by demonstrating in several contexts that constrained spectral energy transfer governs the dynamics.
\end{abstract}

\maketitle

\section{Introduction}

The free energy (also referred to as incremental or perturbed entropy) has been identified by many authors as an important measure of fluctuations in gyrokinetic turbulence.  Collisional dissipation of this quantity is known to be a necessary feature of the true steady state for gyrokinetic turbulence \citep{krommes-hu}.  It enters standard expressions for entropy balance and is directly connected to transport \citep{sugama}.  Its usefulness is identified by \citet{hallatschek}, who calls it the ``generalized grand canonical potential.''  It is the central quantity in the free energy (or entropy) cascade \citep{schek-ppcf, schekochihin, howes, tatsuno-prl, plunk-jfm, tatsuno-jpf, banon-prl} and is also used in a non-cascade theory \citep{terry-hatch} that describes gyrokinetic turbulence as a process of both injection and decay on the same scale.

Studies of two-dimensional gyrokinetics \citep{schek-ppcf, plunk-jfm} have focused on another quantity that is distinct from free energy but is also a fundamental measure of fluctuations.  It measures the intensity of fluctuations in the electromagnetic fields and is conserved under nonlinear interactions (\citet{candy-waltz-entropy} call it ``field energy'').  In the electrostatic approximation, we refer to it as electrostatic energy.

The nonlinear conservation of free energy and electrostatic energy, both being quadratic measures of fluctuation intensity, constitutes a constraint on nonlinear interactions in electrostatic gyrokinetics.  It is the goal of this paper to study the implications of this constraint.  We build on the recent publication of \citet{plunk-prl}, although the scope of the present work is quite a bit broader.  As with \citet{plunk-prl}, we make repeated reference to the theory of \citet{fjortoft} and its generalization to gyrokinetic plasmas.  This theory is simple, being essentially just the careful accounting of the conservation laws imposed on nonlinear energy transfer between different modes of the system.  We wish to answer the basic question ``how will the energy of fluctuations redistribute spectrally from a given initial state?"  We believe that the answer to this question can lead to insight into the broader question:  ``How can one predict the features of fully developed gyrokinetic turbulence?''

Predicting the features of turbulent states (\ie statistical properties such as saturation amplitudes, frequency and wavenumber spectra, transport fluxes, \etc) is a basic goal in studies of gyrokinetic turbulence.  Ideally, one would like to identify generic measures that are useful for various instability drives, magnetic geometries and basic plasma parameters.  The present work identifies a predictor, namely the spectral distribution of free energy.  We will give evidence that a single measure of this distribution, the ratio of free energy to electrostatic energy, seems by itself to be a fairly good predictor of the direction of the nonlinear energy transfer.

We focus on a minimal form of gyrokinetics that retains nonlinear interactions but neglects other effects such as linear instability, collisionless damping, and wave phenomena.  This is the two-dimensional electrostatic gyrokinetic system in a homogeneous background.  It has been argued before that two-dimensional gyrokinetics is a paradigm for kinetic magnetized plasma turbulence \citep{plunk-jfm} and that the tendency toward nonlinear transfer exhibited by this system should be retained in the full three-dimensional system \citep{plunk-prl}.  This second assertion is a conjecture.  However, there is a precise feature of gyrokinetics that motivates it.  The peculiar spectral transfer that occurs in two-dimensional ideal fluids, leading \eg to self-organization into large vortices and inverse cascade, can be traced to the existence of global integrals that are conserved under nonlinear interactions in two dimensions, but not in three dimensions.  The nonlinearity of gyrokinetics takes the same form in both two and three dimensions, and for this reason the nonlinear invariants of two-dimensional gyrokinetics are retained in three dimensions -- that is, they continue to be conserved under nonlinear interactions.  In particular, these quantities are preserved separately {\em for each drift plane}, (the plane perpendicular to the magnetic field).  

\subsection{Overview and results}

The structure of the paper is as follows.  \Sref{prelim-sec} discusses the basic assumptions and then provides the definitions and equations that will be needed throughout the paper.  In \Sref{gk-energy-sec} we discuss energy generally in gyrokinetics.  This material only serves as background for the remainder of the paper but should be interesting to a broad audience as the subject of energy and conservation laws in gyrokinetic theory is a matter of longstanding debate.  We note that although the parallel nonlinearity introduces a non-trivial energy invariant (closely related to the electrostatic energy on which we focus) this quantity is not suitable as a measure of fluctuation energy as it is not quadratic in fluctuations and so does not provide a useful constraint on spectral energy transfer.   We then demonstrate how the quantity we call electrostatic energy enters in ``physical'' energy balance, showing that although the energy of the electric field is negligible in a non-relativistic plasma, changes in the electrostatic energy do track the flow of physical kinetic energy between the parallel and perpendicular directions.

\Sref{spectral-sec} is concerned with the spectral representation for the velocity-space dependence of the distribution function.  Because we put considerable focus on spectral energy transfer, the spectral theory is a very central element to the work and we provide as much detail as possible, without sparing technical aspects.  We find that the basic requirement that finite free energy be finite significantly constrains the space of allowable distribution functions and enables us to derive an appropriate set of basis functions.  Establishing the space of functions also leads to limits on the ratio of free energy to electrostatic energy.  We call this ratio the ``$\kappa$ factor.''  It is a quantity that enters frequently in the following analysis and so the rigorous bound proves useful.

In \Sref{spectral-redist-sec} we present Fj{\o}rtoft's theory, generalized to gyrokinetics.  This is in line with the analysis presented in \citet{plunk-prl} but the scope is broadened to include scales above and below the Larmor radius and also the case of a modified adiabatic response that leads to preferential generation of zonal flows.  In \Sref{three-scale-sec} we first review the elementary three-scale transitions considered by \citet{plunk-prl}.  In \Sref{gen-transfer-sec} we generalize the arguments to energy transfer involving an arbitrary number of scales, which is an important extension for application to turbulence.  The point of \Sref{three-scale-sec} and \Sref{gen-transfer-sec} is to establish precise limits on the spectral redistribution of electrostatic energy subject to a fixed amount of free energy.  A quantity that naturally arises in the derivation is the $\kappa$ factor, which we identify as a predictor of nonlinear-transfer direction.  The strength of these derivations lies in their simplicity and freedom from assumptions.  The weakness is that the result only establishes constraints and limits on how the energy transfer may proceed.  In \Sref{interp-sec} we extend these results with additional arguments to make a prediction of cascade direction.   We review the arguments of \citet{plunk-prl} and also discuss our expectations for scales above the Larmor radius.

In \Sref{super-larmor-sec} we turn our attention to scales larger than the Larmor radius, giving special attention now to a modified Boltzmann or adiabatic response, which captures the enhanced role of zonal flows in closed flux surface geometry.  (Note that we do not introduce any non-uniformity in the background but simply represent the ``flux surface'' average as an average in the $y$-direction.)  We make some general comments about the long-wavelength limit and then proceed  to the case of the modified response.  This presents an opportunity to compare the so-called generalized Hasegawa-Mima (GHM) equation with gyrokinetics and in particular contrast the role of inverse cascade as a mechanism for zonal flow regulation.  We show that the GHM equation generates zonal flows by inverse cascade but that the appearance of finite Larmor radius (FLR) effects in gyrokinetics can drastically modify the nonlinear behavior allowing for both quiescent states (in the absence of collisional dissipation) composed only of stationary zonal flows and states of suppressed zonal flows characterized by large fluctuations.  

In \Sref{gf-sec} we investigate a simple two-field gyrofluid model, derived from the two-dimensional gyrokinetic equation but with the addition of {\it ad-hoc} linear drive terms.  We discuss the derivation of this model in detail in \Aref{lw-app} and conclude that the long wavelength limit $k^2\rho^2 \ll 1$ (with the modified Boltzmann electron response) is a singular limit, casting doubt on the (quantitative) validity of any gyrofluid model that relies on a small argument expansion of the Bessel function.  Nevertheless, we argue that such simple models can be used to study qualitative behavior if finite Larmor radius (FLR) terms are retained at sufficient order.  We perform direct numerical simulations of these gyrofluid equations, and find a nonlinear critical transition associated with the presence of the relative amplitude of the temperature fluctuations.  We interpret the results in terms of our generalized Fj{\o}rtoft theory:  What might otherwise be identified as a reversal in the sign of the turbulent viscosity on the zonal flows, we interpret as a cascade reversal induced by the increase of the $\kappa$ factor of the unstable eigenmodes.

In \Sref{gf-sec} we investigate spectral transfer by linearizing the gyrokinetic equation about a monochromatic initial condition.  We modify the theory of \citet{plunk-pop} to allow for arbitrary $\kappa$ factor of the initial condition, $\kappa_p$.  We consider three cases: (1) the sub Larmor scales with zero response (for comparison with \citet{plunk-prl}) (2) super Larmor scales assuming a standard Boltzmann response and (3) super Larmor scales assuming a modified response.  For this final case we assume a zonal mode initial condition (this problem was previously investigated by \citet{rogers-prl} and the instability termed ``tertiary instability'').  In each case we find a transition associated with the parameter $\kappa_p$, which we interpret in terms of the Fj{\o}rtoft theory developed in \Sref{spectral-redist-sec}.  Previously reported properties of this tertiary instability are given a new explanation in terms of basic considerations of mode coupling and energy conservation.

In \Sref{numeric-sec} we present previously unpublished results of nonlinear gyrokinetic simulations corresponding to the numerical runs of \citet{plunk-prl}.  We compute the shell-filtered spectral transfer function, which is based on a standard summation of the elementary triad interaction terms used in computations of neutral fluid turbulence.  The results demonstrate that the spectral evolution of free energy observed by \citet{plunk-prl} was due to nonlinear interaction and clearly show the signatures of the three modes of spectral transfer: (1) nonlocal inverse transfer, (2) local inverse transfer and (3) forward transfer.

In \Sref{discussion-sec} we discuss the results of this paper and point out possible applications and future work.  We argue that two-dimensional gyrokinetics is relevant to driven systems with nontrivial magnetic geometry and point to recent evidence in support of this view.

\section{Preliminary considerations}\label{prelim-sec}

\subsection{The Larmor scale}

For a given particle species, we assume there is a characteristic velocity associated with fluctuations in the distribution function.  We take the characteristic velocity to be the thermal velocity of the background distribution, $\vth = \sqrt{T/m}$.  We make this assumption by conjecture, but argue as follows that it is reasonable:  First, we require on physical grounds that the free energy of the fluctuations (henceforth referred to simply as the free energy) be a finite quantity.  Therefore the perturbed distribution function must fall off in velocity space reasonably fast as compared to the background distribution function, which itself falls off on the scale of $\vth$.  (At minimum, finite free energy requires $|\delta f|^2 \exp(v^2/(2\vth^2)) \rightarrow 0$ as $v \rightarrow \infty$.)

\begin{center}
\begin{figure*}
\subfloat[Particle motion.]{\label{vel-scale-illustration-fig-a}\includegraphics[width=0.5\textwidth]{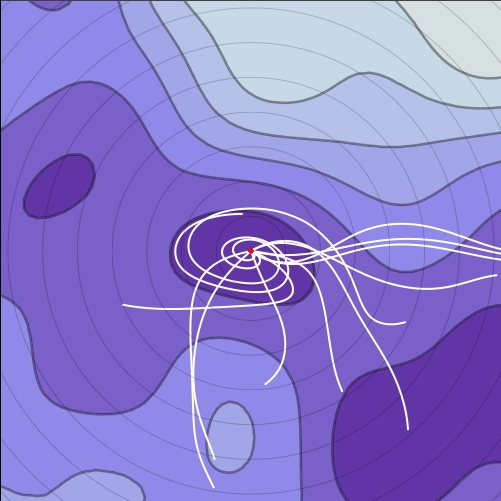}}
\subfloat[Evolution of local distribution function.]{\label{vel-scale-illustration-fig-b}\includegraphics[width=0.5\textwidth]{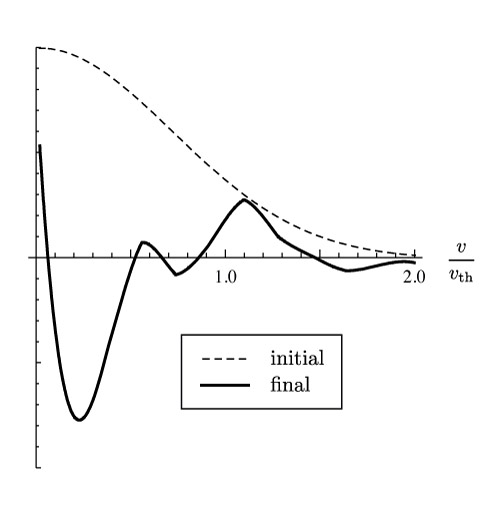}}
\caption{Advection of passive particles by gyro-averaged $E\times B$ motion due to a random static electrostatic potential:  The process is depicted with (a) the motion of a sample of tracer particles and (b) initial and final distribution of tracer particles as measured at the central (red) point.  The local gyrocenter distribution of tracers is initialized to be Maxwellian, with density proportional to local value of the electrostatic potential, which is illustrated by contour plot and colored level sets.  Tracer particles which arrive simultaneously at the central red point are selectively plotted, with paths shown in white.  The tracer distribution function at the central point is shown for initial and final times.  Phase mixing evolves distribution of traces away from the initial Maxwellian, but a Maxwellian envelope is retained.}
\label{vel-scale-illustration-fig}
\end{figure*}
\end{center}

Also note that the drive terms in the gyrokinetic equation are proportional to the background distribution function (\ie they are ``Maxwellian-like'') and thus so are the unstable linear eigenmodes modes, which stimulate the turbulence.  Although the distribution function will evolve (by nonlinear interaction) away from the form of linear modes, the distribution function should fall off in velocity space in a manner that is consistent with Maxwellian-like source terms.  Indeed, modification of the distribution function by the nonlinear term is done by the incompressible flow of gyrocenters around the domain, which leaves the density of gyrocenters unchanged along the motion of the flow.  This implies that a distribution function evolving only under the influence of the nonlinearity, must retain the amplitude set by an initial condition, locally in velocity space.  For example, consider the motion of a collection of tracer particles, representing positive and negative gyrocenter density, distributed initially in velocity space as a local Maxwellian and having magnitudes proportional to the value of a random electrostatic potential.  As depicted in \Fref{vel-scale-illustration-fig-b}, the tracer distribution function will develop structure in velocity space as particles are advected by $E\times B$ motion, but an envelope is retained that reflects the initially local Maxwellian shape.  Thus, we expect that for turbulence driven by ``Maxwellian-like'' source terms, the distribution function will attenuate in velocity space on a scale comparable to the Maxwellian.

The assumption of the characteristic velocity scale $\vth$ implies a corresponding spatial scale, the thermal Larmor radius $\rho$.  This scale marks the transition between two asymptotic regimes, the sub-Larmor and super-Larmor ranges.  We will give some results that apply in both regimes, and also study them separately.

\subsection{Normalized Equations and Definitions}\label{eqns-sec}

Let us consider a two-dimensional slab geometry, where fields vary only in the direction perpendicular to a mean magnetic field, which is itself uniform and points in the $\hat{z}$-direction.  Wherever integration over velocity space is present, integration over the velocity component parallel to the guide field $v_{\parallel}$ is implied; the same goes for integration over position space where integration over $z$ is implied.  We consider a single species to be kinetic, with the second species satisfying a simple adiabatic response model.  Thus the regime of applicability is limited to scale ranges where an adiabatic response is valid to reduce the dynamics to a single kinetic species, that is $k \sim \rho_i^{-1} \ll \rho_e^{-1}$ or $k \sim \rho_e^{-1} \gg \rho_i^{-1}$.

Henceforth, we use the normalized variables and notation of \citet{plunk-jfm}.  Thus $v_{\perp}/\vth \rightarrow v$ (with $\vth = \sqrt{T/m}$) is the normalized perpendicular velocity and the normalized wavenumber is $k_{\perp}\rho \rightarrow k$ where thermal Larmor radius is $\rho \equiv \vth/\Omega_c$ and $\Omega_c = qB/m$.  The two-dimensional gyrokinetic equation in nondimensional form is written as follows in terms of the gyrocenter distribution function $g({\bf R}, v, t)$, where ${\bf R} = \hat{\bf x} X + \hat{\bf y} Y$ is the gyrocenter position:

\begin{equation}
\frac{\partial g}{\partial t} +  \poiss{\gyroavg{\varphi}}{g} = \CollisionOp{h}.
\label{gyro-g}
\end{equation}

\noindent where the Poisson bracket is $\poiss{a}{b} = \hat{\bf z}\times\bnabla a \cdot \bnabla b$ and the gyro-average is defined $\gyroavg{A({\bf r})} = \frac{1}{2\pi}\int_0^{2\pi} d\vartheta A({\bf R} + \rhovec(\vartheta))$, where the Larmor radius vector is $\rhovec(\vartheta) = {\bf {\hat z}}\times{\bf v} = v_{\perp}({\bf {\hat y}}\cos{\vartheta} - {\bf {\hat x}}\sin{\vartheta})$ and $\vartheta$ is the gyro-angle.  (Note also that the quantity inside of the collision operator is $h = g + \gyroavg{\varphi}F_0$.)  We leave the collision operator unspecified, other than noting that it is in general an integrodifferential operator (with appropriate conservation properties) that acts to smooth phase-space structure of the distribution function, evolving the plasma to thermodynamic equilibrium.  Quasi-neutrality, under the assumption of an adiabatic response, yields the electrostatic potential $\varphi({\bf r}, t)$, where ${\bf r} = \hat{\bf x} x + \hat{\bf y} y$ is the position-space coordinate:

\begin{equation}
2\pi\int_0^{\infty} v dv \angleavg{g} = (1 + \Tau)\varphi - \Gamma_0\varphi,
\label{qn-g}
\end{equation}

\noindent where the angle average is defined $\angleavg{A({\bf R})} = \frac{1}{2\pi}\int_0^{2\pi} d\vartheta A({\bf r} - \rhovec(\vartheta))$, and the term $\Tau\varphi$ is the adiabatic density response.  The constant $\Tau$ is typically the temperature ratio of the background plasma, but can also be taken as an operator to capture modifications to the conventional Boltzmann response.  The operator $\Gamma_0\phi = 2\pi\int_0^{\infty} v dv \;F_0(v)\angleavg{\gyroavg{\phi}}$ is easily evaluated in Fourier space assuming a Maxwellian background $F_0 = Exp[-v^2/2]/(2\pi)$:

\begin{equation}
\hat{\Gamma}_0(k) = \int_0^{\infty} v dv \;\e^{-v^2/2}J_0^2(k) = I_0(k^2)e^{-k^2},\label{gamma0-def}
\end{equation}

\noindent where $I_0$ is the zeroth-order modified Bessel function.  Note that this is the only explicit place (other than the collision operator) where the background distribution function enters the theory.  Indeed, the $k$-dependence of $\hat{\Gamma}_0$ contributes to the transition in the physics of the turbulence across the Larmor scale.  However, to satisfy finite (perturbed) entropy, as discussed later, the distribution function $g$ also must have dependence on the characteristic scale of the background $\vth$, which imposes a characteristic spatial scale, the thermal Larmor radius $\rho = \vth/\Omega_c$. Using \Eref{gamma0-def}, we can write quasi-neutrality in Fourier space:

\begin{equation}
\hat{\varphi} = \beta({\bf k})\int_0^{\infty} v dv J_0(k v)\hat{g}({\bf k}, v),
\label{qn-g-k}
\end{equation}

\noindent where

\begin{equation}
\beta = \frac{2\pi}{1+\Tau-\hat{\Gamma}_0(k)}.\label{beta-def}
\end{equation}


We use the term ``nonlinear invariant'' to mean a quantity that is conserved under the sole action of the nonlinearity, \ie in the absence of collisions, linear instability or linear collisions damping.  These quantities are exact invariants of the collisionless two-dimensional gyrokinetic system with a homogeneous equilibrium (uniform background Maxwellian and uniform guide magnetic field).  The kinetic free energy is a nonlinear invariant, 

\begin{equation}
G = \int \frac{d^2{\bf R}}{V} \frac{g^2}{2} = \displaystyle{\sum_{\bf k}} \frac{|\hat{g}({\bf k}, v)|^2}{2},
\end{equation}

\noindent as is any suitably weighted velocity integral $\int v dv G(v) w(v)$, where $w(v)$ is a weighting function; indeed, the volume average of any power of $g$ is a nonlinear invariant but we focus here on quadratic invariants.  The electrostatic energy is also invariant under nonlinear interactions, 

\begin{equation}
E = \frac{1}{2}\int \frac{d^2{\bf r}}{V}[(1+\Tau)\varphi^2 - \varphi \Gamma_0 \varphi] = \displaystyle{\sum_{\bf k}} \frac{2\pi}{\beta}\frac{|\hat{\varphi}({\bf k})|^2}{2}.\label{E-def}
\end{equation}

\noindent The presence of these two quadratic invariants establishes the analogy with two-dimensional fluid turbulence that has inspired work of this and other papers.  Note that our arguments for the tendencies of spectral transfer, based on those of Fj{\o}rtoft, depend only on the existence of these invariants (and the spectral representations of the following section) but do not depend on details of the actual dynamics; indeed the gyrokinetic equation, \Eref{gyro-g}, is not needed for these arguments and did not even appear in \citet{plunk-prl}.

\subsection{Energy in gyrokinetics}\label{gk-energy-sec}

In this section we discuss the role of energy in gyrokinetic turbulence and explain why our choice is appropriate for the study of turbulence.  By comparing different meanings of ``energy'' in gyrokinetics we also place our work in a broader context of gyrokinetic theory.  The remaining parts of the paper do not depend on this discussion, but it is recommended for those readers who are generally interested in gyrokinetic conservation laws.  We also recommend general discussions of energy by other authors such as those found in Refs.~\citep{schekochihin, hallatschek, parra-lagrangian, abel-plunk}

We have identified two quantities, $G$ and $E$, each which represent a kind of fluctuation energy, being quadratic measures of fluctuations.  The quantity $G$ is related to the free energy $W$, which is the traditional focus of what is called the turbulent ``free energy cascade'' (or entropy cascade) in the electrostatic limit (for purely electrostatic fluctuations, this quantity is proportional the perturbed or incremental entropy):

\begin{equation}
W = W_{g0} + E,\label{W-def}
\end{equation}

\noindent where

\begin{equation}
W_{g0} = 2\pi\int_0^{\infty} du\; G(u)/F_0.\label{Wg0-def}
\end{equation}

\noindent The electrostatic energy $E$ is a nonlinear invariant of collisionless 2D gyrokinetics, but has more general meaning in gyrokinetics.

There are two general camps in gyrokinetic theory.  One we will call Hamiltonian gyrokinetics (some use the terms Lagrangian or ``modern'' gyrokinetics) \citep{sugama-gyro-field-theory, brizard-2000, brizard-hahm}.  The other is sometimes referred to as ``iterative gyrokinetics'' \citep{frieman-chen} (since the asymptotic derivation is done by iterative procedure, order-by-order) and its extension to transport is called ``multiscale'' gyrokinetics \citep{sugama-transport, sugama-transport-2, parra-catto, abel-plunk} because it assumes scale separation additional to that of traditional gyrokinetic ordering.  These approaches differ in the treatment of dissipation and conservation laws.  

Multiscale gyrokinetics gives energy balance in the form of transport equations by extending the gyrokinetic derivation to an order higher than that needed to solve for fluctuations.  Collisional dissipation is included at each order.  Energy balance is derived in a straightforward fashion by taking the kinetic energy moment of the Vlasov-Maxwell equations in the phase-space variables of the particles, \ie $\sum_s \int d^3{\bf v}\frac{mv^2}{2} (Df_s/dt = \sum_r C[f_s, f_r])$, after which Poynting's theorem can be used to evaluate the exchange terms that describe flow between particle kinetic energy and electromagnetic field energy.  (In the limit we consider, \ie the non-relativistic electrostatic limit in the presence of a static guide field, this energy is only due to particle kinetic energy as the energy of the electric field is negligible.)  This is not a procedure for deriving dynamical invariants of the system.  It is a way to establish the balance of what, for lack of a better term, we will call ``physical'' energy (a quantity that is already known to be conserved by the collisional Vlasov-Maxwell system) under the application of the various ordering assumptions of gyrokinetics.

As is standard with Hamiltonian theories, Hamiltonian gyrokinetics includes collisional dissipation after the equations of motion have already been derived.  In this approach, invariants (analagous to those of the Vlasov-Maxwell system) are derived in the absence of collisional effects and written in terms of gyrokinetic variables, \ie the phase-space variables of quasi-particles.  This leads to a pleasantly self-contained theory, but also one where familiar quantities such as energy and momentum take on a more abstract meaning.  One might ask whether these quantities are the same as the ``physical'' conserved quantities of the Vlasov-Maxwell system, just simply written in different variables.  This is not the case for the very simple example below.

Extending the homogeneous 2D gyrokinetic equation, \Eref{gyro-g}, into 3D (where $g = g({\bf R}, z, v_{\perp}, v_{\parallel})$) we may write

\begin{equation}
\frac{\partial g}{\partial t} + v_{\parallel}\frac{\partial g}{\partial z} + \poiss{\gyroavg{\varphi}}{g} + v_{\parallel}\frac{\partial \gyroavg{\varphi}}{\partial z}F_0 - \epsilon \frac{\partial \gyroavg{\varphi}}{\partial z}\frac{\partial g}{\partial v_{\parallel}} = \gyroavg{{\mathcal C}[h]}\label{gyro-3D-PNL}
\end{equation}

\noindent where we have retained the so called parallel-nonlinearity, which has an explicit factor of $\epsilon = \rho/L$ due to the normalization we have taken.  In accordance with the asymptotic limit $\epsilon \rightarrow 0$, iterative gyrokinetics doesn't include this term at this order but includes it at higher order to correctly calculate the energy balance.  Indeed, numerical studies confirm that this formally small quantity does not have a significant effect on the turbulence \citep{candy-PNL}.  However, by including the term here, a non-trivial energy is conserved \footnote{In absence of the parallel nonlinearity, the first order physical energy is trivially conserved $K_1 = \int d^3{\bf r} \int d^3{\bf v}\frac{mv^2}{2}\delta f_1$.  Note also that we consider only a very simple case of gyrokinetics and the subject of exact conservation laws for general $\delta f$ gyrokinetics -- that is, gyrokinetic theory that treats fluctuations separately -- has been covered extensively in the recent work of \citet{brizard-energy}.}

\begin{equation}
H = \frac{1}{2V}\int d^3{\bf r} \left((1+\Tau)\varphi^2 - \varphi\Gamma_0\varphi +  \frac{1}{\epsilon}\int d^3{\bf v}\; v_{\parallel}^2 \angleavg{g} \right)\label{gyro-ham}
\end{equation}

\noindent To demonstrate that $H$ is conserved, combine the $v_{\parallel}^2/2$ moment of \Eref{gyro-3D-PNL} with \Eref{gk-es-energy-bal-eqn}.  Note that despite the factor of $\epsilon^{-1} = L/\rho$, these terms are of the same order if we take the spatial integral of $g$ to vanish at dominant order (which we can as the evolution of this quantity is order $\epsilon$).  Thus $g$ must absorb part of $\delta f_2$ to enforce the conservation of $H$.  Although a conserved quantity like $H$ may be useful for numerical application, we note that it contains a term that is linear in fluctuations.  It thus does not constitute a useful measure for constraining spectral energy transfer in the way quantities like the free energy and electrostatic energy do; see also \citet{hallatschek} who discusses the value of measures that are quadratic in fluctuations.  Also, it is worth noting that $H$ is not the physical energy, which in the quasi-neutral approximation is just the kinetic energy of the particles.



We can identify the terms involving $\varphi$ in \Eref{gyro-ham} as the electrostatic energy $E$ defined in \Eref{E-def} for our 2D system.  What is this quantity $E$?  It may seem paradoxical that there is an energy associated with the electric potential, since the physical energy of the electric field ($\propto |{\bf E}|^2$) of a non-relativistic plasma is negligible compared to the other contributions to energy (kinetic energy or energy of magnetic fluctuations, if included).  But in gyrokinetics the electrostatic field does carry some kind of ``energy.''  In what we've just presented, and generally in Hamiltonian gyrokinetics (\cf \citep{dubin-krommes}), $E$ is part of the total conserved energy  and in that context it can be interpreted as the interaction energy of gyrocenters (or simply the gyrocenter potential energy), from which the $E\times B$ nonlinear term (and also higher order nonlinear terms) in the gyrokinetic equation originates.

Multiscale gyrokinetics also has something to say about energy conservation.  It provides a framework for tracking irreversible and reversible energy flows on both turbulent and heating timescales.  The electrostatic energy $E$ can be interpreted in this context as well.  We assume periodicity in position space (thus fluxes give no contribution in the global energy balance).  A key observation is that, as a simple consequence of charge neutrality and continuity, the electrostatic field ${\bf E} = -\bnabla\varphi$ can do no work on the particles:

\begin{equation}
\int d^3{\bf r} {\bf J}\cdot{\bf E} = \int d^3{\bf r} \varphi \bnabla\cdot{\bf J} = 0,\label{no-work-eqn}
\end{equation}

\noindent  For this reason, the electrostatic work does not enter the global energy balance equation (nor does the electrostatic energy).  However, separating out the part of energy balance parallel to the guide field (\ie, the $mv_{\parallel}^2/2$-moment of the Vlasov-Maxwell equation), one finds

\begin{equation}
\frac{dK_{\parallel}}{dt} = \int d^3{\bf r}E_{\parallel}J_{\parallel},\label{parallel-work-eqn}
\end{equation}

\noindent where $E_{\parallel} = - \partial_z\varphi$ is the parallel electrostatic field and $K_{\parallel}$ is the parallel kinetic energy of the particles (summed over species) \footnote{The quantity $K_{\parallel}$ formally contains contributions from the bulk plasma (slow evolution of background) and second order fluctuations on the turbulent timescale, \ie the fluctuations in $\delta f_2$ for which one traditionally does not solve.  As the volume average of $\delta f_1$ is invariant, the total particle kinetic energy contributed at first order is invariant and generally taken to be zero.}.  The work done by the parallel electric field also appears in the balance equation for $E$ derived by multiplying the gyrokinetic equation by $\gyroavg{\varphi}$, integrating over the velocity and spatial domains and summing over species:

\begin{equation}
\frac{dE}{dt} + \frac{1}{V}\int d^3{\bf r} E_{\parallel} J_{\parallel} = \frac{1}{V}\displaystyle\sum_s \int d^3{\bf r} \int d^3{\bf v}q_s\varphi\angleavg{\gyroavg{{\mathcal C}[h_s]}},\label{gk-es-energy-bal-eqn}
\end{equation}

\noindent Note that evidence from driven \citep{banon-pop} and decaying \citep{tatsuno-pop} simulations indicates that the collisional dissipation of $E$, written on the right-hand side of this equation, seems to be exceedingly weak in comparison to that of $W$ \footnote{The weak collisional damping of $E$ is expected from the cascade theory of $W$ because the amount of $E$ that reaches the dissipation scale is asymptotically less than the amount of $W$, as the collision rate is sent to zero.}.  On the other hand, the parallel work term is quite substantial in the overall energy balance as it must balance input (not included here) of electrostatic energy in stationary state.  We expect this term to be positive, on average, when Landau damping of the electrostatic field is operating, but can also be negative in the presence of instability.  Balancing the change in gyrokinetic electrostatic energy against the parallel work, we see by Equations \eref{parallel-work-eqn} and \eref{no-work-eqn} that ``collisionless'' damping of $E$ must be accompanied by a simultaneous flow of (physical) kinetic energy between the parallel and perpendicular directions (acceleration of particles in the direction parallel to the magnetic field and deceleration of particles in the perpendicular direction).  Since the electrostatic energy is quadratic in fluctuations, the kinetic energy that it balances with must appear at higher order in the distribution function, \ie $\delta f_2$ fluctuations, or on timescales much longer than those of the turbulence.

In other words, gyrokinetics shows a tendency toward anisotropization of the kinetic energy of the particles.  And it is this anisotropization that provides a physical interpretation of both the gyrokinetic electrostatic energy $E$ and the parallel nonlinearity.  In terms of physical energy, the role of $E$ is to track the higher order exchange of particle kinetic energy between the parallel and perpendicular components, as we've just described.  Inclusion of the parallel nonlinearity forces this exchange to be accounted for in small fluctuations $\delta g \sim \epsilon g$, though it does not affect the conservation of physical energy, which is satisfied automatically in iterative gyrokinetics at each order in $\epsilon$.\footnote{Now that we have examined the electrostatic case, the electromagnetic version of \Eref{gk-es-energy-bal-eqn} can be interpreted in the same manner by identifying work terms due to magnetic fluctuations in the parallel energy balance equation.  Because the magnetic field does no work on particles, these work terms again function as anisotropic energy exchange.}

Having put the energetics into context, let us now return to the two-dimensional system introduced in \Sref{eqns-sec}.  We are interested in how this system nonlinearly redistributes energy among modes and so we will first need to establish some details of a spectral theory.

\section{Discrete Spectral Representations}\label{spectral-sec}

In \citet{plunk-jfm}, a spectral formalism for velocity space was introduced based on the Hankel transform, which has a Bessel function as its kernel.  This approach is mathematically convenient due to the frequent appearance of Bessel functions in gyrokinetics.  However, physically realizable systems are finite, so the continuous-variable Fourier and Hankel transforms are, in some sense, more than is needed (but are useful sometimes for simplifying mathematical arguments).  In this section we will present discrete-variable spectral representations for velocity space.  In addition to aiding in theoretical arguments, these representations may prove useful for numerical solutions of the gyrokinetic equation.

In position space we assume finite volume, $V = L^2$, where $L$ is the size of the two-dimensional domain.  The velocity space domain, however, is formally semi-infinite $[0, \infty)$ but there are physical constraints, such as finite density and energy moments, which put definite limits on the distribution function.  The constraint we are interested in is finite free energy, which implies $W_{g0} \equiv \int v dv G(v)/F_0(v) < \infty$.  This defines the space of functions in which $g(v)$ must be contained.  In \Sref{bessel-sec} we will use this condition to construct a simple Bessel series representation of $g(v)$.  In \Sref{orth-poly-sec}, we will present another spectral representation based on orthogonal polynomials that span the space of functions $g(v)$ having finite $W_{g0}$.  We will arrange the notation so that either representation can be used with the subsequent results of the paper.

\subsection{Bessel Series}\label{bessel-sec}

In \citet{plunk-prl} we introduced a discrete spectral theory for velocity space using a Bessel series.  It proved useful for studying sub-Larmor fluctuations, and gives an intuitive definition of scale due to its close relationship to the Fourier series.  The Bessel-series representation relies on the approximation that the distribution function is zero above a cutoff in velocity space $\vcut$.  We justify this approximation, as noted above, by reasoning that if $W_{g0}$ is finite then $g(v)$ must fall off at least as fast as a Maxwellian at large velocity.  Thus we take $\vcut \gg 1$, and argue that the error introduced should be negligible, falling off exponentially in $\vcut$.  In \Sref{orth-poly-sec} we will introduce an exact spectral representation that does not assume a velocity cutoff.

The Bessel series is built on the orthogonality relation

\begin{equation}
\int_0^{\vcut} J_0(\lambda_n v/\vcut)J_0(\lambda_m v/\vcut) v dv = \frac{1}{2}\vcut^2 J_1^2(\lambda_n)\delta(m - n),\label{bessel-orthog}
\end{equation}

\noindent where $\lambda_n$ is the $n$th zero of $J_0(x)$.  We use the unit step function $\Theta$ in the Bessel series expansion of $g$ to provide the velocity cutoff explicitly:

\begin{equation}
\hat{g}({\bf k}, v) = \displaystyle \sum_{j} \frac{2J_0(p_j v)}{\vcut^2J_1^2(\vcut p_j)} \Theta(\vcut-v)\; \dblhat{g}({\bf k}, p_j),\label{g-bessel-series}
\end{equation}

\noindent where $p_j = \lambda_j/\vcut$.  Now, taking the limit $p_j\vcut \gg 1$ \footnote{This may be justified by confining attention to sub-Larmor scales $k \gg 1$ and ordering $p \sim k$.  Alternately, we may assume $\vcut$ to be large enough to ensure $p_j\vcut \gg 1$ for all $p_j$ of interest.}, we may simplify \Eref{g-bessel-series} by evaluating $J_1$ in the coefficient of the expansion using the large argument expression $J_n(x) \approx \sqrt{2/(\pi x)}\cos(x - n\pi/2 -\pi/4)$ and correspondingly $\lambda_j \approx \pi(j+ 3/4)$.  Thus we obtain the form of the Bessel series used in \citet{plunk-prl},

\begin{equation}
\hat{g}({\bf k}, v) = \displaystyle\sum_{j} \frac{\pi p_j}{\vcut} \Theta(\vcut-v)J_0(p_j v)\dblhat{g}({\bf k}, p_j).\label{approx-bessel}
\end{equation}

\noindent Using \Eref{approx-bessel}, the free energy $W_{g1} = 2\pi \int_0^{\infty} v dv \;G(v)$ can be written compactly as a summation over the spectral density in ${\bf k}$-$p$ space:

\begin{eqnarray}
W_{g1} &= 2\pi \int_0^{\infty} v dv \;G(v)\nonumber\\
&= \displaystyle{\sum_{{\bf k}, j}} \frac{\pi^2}{\vcut}p_j|\dblhat{g}({\bf k}, p_j)|^2\label{wg1-def}\\
&= \displaystyle{\sum_{{\bf k}, j}} W_{g1}({\bf k}, p_j).\nonumber
\end{eqnarray}

\noindent A subtlety now arises.  We require $J_0(k \vcut) = 0$, so that quasi-neutrality, \Eref{qn-g-k}, can be written in terms of a single component of the Bessel series.  (The reason for this will become apparent in \Sref{fjortoft-constraint-sec}.)  This, however, cannot be satisfied uniformly for all $k$ with a single value of $\vcut$.  Thus, we actually need $k$-dependence in the cutoff, $\vcut = \vcut(k)$.  However, we can ensure this dependence is weak by writing $\vcut(k) = \vcut^{(0)} + \delta \vcut$, where $|\delta \vcut| \ll \vcut^{(0)}$ and $\delta \vcut$ is the smallest quantity that satisfies $p_j = k$ for some $j$ \footnote{Since the period of oscillation of $J_0$ is roughly $2\pi$, the quantity $k\vcut$ can be made a zero of $J_0$ with a magnitude of $\delta \vcut$ that is at most $\pi/(2k)$.  Thus, since $\vcut \gg 1$, we have $\delta \vcut \ll \vcut$ as long as $k$ is not too small.  If $k \gg 1$, as assumed in \citet{plunk-prl}, then $\delta \vcut/\vcut^{(0)} \sim 1/(\vcut^{(0)}k)$ is indeed quite small.}.  We argue that if the system we are studying is well approximated with a finite velocity cutoff, then it will remain well-approximated under small $k$-dependent adjustments to this cutoff velocity.  Combining Equations \eref{bessel-orthog} and \eref{g-bessel-series} we find that quasi-neutrality becomes simply

\begin{equation}
\hat{\varphi}({\bf k}) = \beta \;\dblhat{g}({\bf k}, k),\label{qn-g-k-p}
\end{equation}

\noindent Note that to obtain \Eref{qn-g-k-p} we need not modify the bounds of integration in \Eref{qn-g} because the step function in \Eref{g-bessel-series} provides explicit velocity cutoff.  From \Eref{qn-g-k-p} we can express the spectral density of electrostatic energy as follows

\begin{equation}
E({\bf k}) = \frac{1}{2}\frac{2\pi}{\beta}|\hat{\varphi}({\bf k})|^2 = \pi \beta |\dblhat{g}({\bf k}, k)|^2.\label{E-k-def}
\end{equation}

\noindent Now, taking the limit $k \gg 1$ and $\vcut^{(0)}/\vcut \approx 1$, and combining Equations \eref{wg1-def} and \eref{E-k-def} we can write

\begin{equation}
\frac{2\vcut^{(0)}}{(1 + \Tau)}\frac{W_{g1}({\bf k}, k)}{E({\bf k})} \approx k,\label{bessel-fjo-constraint}
\end{equation}

\noindent which is equivalent to Equation (6) of \citet{plunk-prl}.

\subsection{Representation Based on Orthogonal Polynomials}\label{orth-poly-sec}

As noted above, physical realizations of the distribution $g$ correspond to finite free energy, \ie finite $W_{g0}$.  Equivalently, the normalized distribution function $g^{\prime} = g/F_0$ must be a member of the weighted Hilbert space with inner product defined $(g_1, g_2) = \int_0^{\infty} \e^{-u}g_1(u) g_2(u) du$, where $u = v^2/2$.  That is, $g^{\prime}$ has finite norm $||g^{\prime}|| = \sqrt{(g^{\prime},g^{\prime})} < \infty$.  It follows that we can decompose $g^{\prime}$ in a series of orthogonal functions that are a basis for this space.  One choice is the Laguerre polynomials.  However, we would also like a simple relationship between the distribution function and the electrostatic potential as provided by the Bessel series; see \Eref{qn-g-k-p}.  We can achieve this by constructing a set of orthogonal functions using the Bessel function $J_0 (k v)$.  The set is denoted

\begin{equation}
\mathcal{G} = \{P^{(k)}_0; P^{(k)}_1, P^{(k)}_2, P^{(k)}_3, ... \}
\end{equation}

\noindent where $P^{(k)}_0 = \hat{\Gamma}_0(k)^{-1/2} J_0(k v)$, (the factor $\hat{\Gamma}_0(k)^{-1/2}$ comes from normalizing $||P^{(k)}_0|| = 1$) and $P^{(k)}_n$ is a normalized polynomial of degree $n$.  We construct $P^{(k)}_1$, $P^{(k)}_2$, $P^{(k)}_3$, \etc, in the Gram-Schmidt fashion as follows.  The polynomial $P^{(k)}_1$ is determined by requiring it has unit norm and is orthogonal to $P_0^{(k)}$; the higher order polynomials $P^{(k)}_n$ are then determined, in increasing order, by requiring orthogonality with lower order $P^{(k)}_0, P^{(k)}_1, ... , P^{(k)}_{n-1}$.  In \Aref{G-completeness-sec} we prove that the set $\mathcal{G}$ is complete.  We may now express $\hat{g}$ as a series in these functions.  Defining $u = v^2/2$ we have

\begin{equation}
\hat{g}({\bf k}, u) = \displaystyle{\sum_{j=0}^{\infty}} \hat{g}_j({\bf k}) \e^{-u} P_j^{(k)}(u),\label{poly-series}
\end{equation}

\noindent and, recalling $F_0 = \e^{-u}/2\pi$, we have

\begin{equation}
W_{g0} = 2\pi\int_0^{\infty} du\; G(u)/F_0 = 4\pi^2\displaystyle{\sum_{{\bf k}}\sum_{j=0}^{\infty}}|\hat{g}_j({\bf k})|^2/2.\label{Wg0-def2}
\end{equation}

\noindent Plugging \Eref{poly-series} into \Eref{qn-g}, we have another compact expression of quasi-neutrality,

\begin{equation}
\hat{\varphi}({\bf k}) = \beta({\bf k}) \hat{\Gamma}^{1/2}_0(k) \;\hat{g}_0({\bf k}),\label{qn-g-k-n}
\end{equation}

\noindent which implies (see \Eref{E-def})

\begin{equation}
E({\bf k}) = \pi\hat{\Gamma}_0\beta\;|\hat{g}_0({\bf k})|^2 = \hat{\Gamma}_0\frac{\beta}{2\pi}W_{g0}({\bf k}, 0)
\end{equation}

In addition to $W_{g1}$ and $W_{g0}$, a third ``free energy'' quantity will be useful, which we denote $W_z$ because its role appears analogous to enstrophy (conventionally denoted $Z$) at both sub-Larmor and super-Larmor scales.  For constant $\Tau$, the following quantity is an invariant:

\begin{equation}
W_z = W_{g0} - \Tau E.\label{wz-def}
\end{equation}

\noindent Let us accordingly define $W_z({\bf k}, j) = W_{g0}({\bf k}, j) - \Tau \delta(j) E({\bf k})$ so that $\sum W_z({\bf k}, j) = W_z$.  Then, because of the asymptotic form of $\hat{\Gamma}_0$, \Eref{gamma-large}, it is easy to see that $W_z({\bf k}, j) \approx W_{g0}({\bf k}, j)$ for $k \gg 1$.  However, the two quantities diverge at $k^2 \ll 1$.  

\subsection{The Fj{\o}rtoft Constraint}\label{fjortoft-constraint-sec}

Many approaches are available for explaining energy flows in 2D fluids.  Examples include absolute statistical equilibria of the ideal (inviscid) fluid \citep{onsager49d, kraichnan1967, montgomery-joyce-1974, edwards-taylor, kraichnan-montgomery-1980}, the dual cascade (including non-zero viscosity) \citep{kraichnan1967, leith, batchelor1969} and variational methods that minimize enstrophy while conserving energy and other constants of motion \citep{leith1984}.  Such approaches generally seek equilibrium solutions subject to some non-trivial assumptions.  Impressively, precise predictions have been validated in experimental systems, though none universally due to the non-universal applicability of the fundamental assumptions, \eg ergodicity or conservation laws.  We argue that Fj{\o}rtoft's theory remains an attractive way to view tendencies of energy transfer because it lays bare the basic mechanism of constrained nonlinear energy transfer without additional assumptions.  Supplemented with simple hypotheses, this constraint yields a quick and (mathematically) intuitive prediction of transfer direction.

We are interested in the particular relationship between the spectral densities of free energy and electrostatic energy $E$.  We call this relationship the ``Fj{\o}rtoft constraint.''  This constraint affects the tendencies of spectral energy transfer.  However, a spectral representation is non-unique, in the sense that we can choose (or invent) any one we like.  Furthermore, each representation, along with each weighted integral of $G(v)$, or free energy, yields a different relationship or constraint.  Thus, to keep the discussion general let us define a generic ``free energy'' quantity $\W$, which can represent any of those free energies defined, \ie $W_{g1}$, $W_{g0}$ or $W_z$, each having been assigned an appropriate spectral representation.  In \Sref{spectral-redist-sec} we will then be able to consider the constrained spectral transfer of $E$ and this generic free energy $\W$. \footnote{The question may be raised, "why limit consideration to only one free energy quantity when there are an infinite number of choices?"  There is no rigorous justification.  However, we note that the full hierarchy of invariants is not retained under spectral truncation \citep{zhu-priv}.  For instance the invariance of $W_{g0}$ is retained under the formal assumption of a finite extent of the spectral domain $[j_{min}, j_{max}]$, but other measures of free energy are not.  Thus $W_{g0}$ can be considered ``rugged'' \citep{kraichnan-montgomery-1980} and perhaps in some sense more important.}

We denote the spectral density of free energy as $\W({\bf k}, j)$, where $j$ is an index for the velocity-space spectrum and $j = 0$ represents the density component.  In the case of the Bessel representation this is the $p = k$ component.  We will also call this component the ``density component,'' departing from the convention of \citet{plunk-prl}.  The previous terminology (``diagonal'' component) was justified because this component is due to integration over velocity space with a factor of $J_0(k v)$ in the integrand (see \Eref{qn-g-k}), and so represents an effective velocity-space wavenumber $p = k$.  However, our use of the generic spectral index $j$ in this paper makes the new terminology more appropriate.  With this general notation we may write the Fj{\o}rtoft constraint as


\begin{equation}
\frac{\W({\bf k}, 0)}{E({\bf k})} = q({\bf k}).\label{fjortoft-constraint}
\end{equation}

\noindent This establishes a constraint on how nonlinear interactions can spectrally redistribute $E$ and $W$ simultaneously.  We would like to understand precisely how it does this; this is a central goal of this paper.

The asymptotic form of $q$ at small and large scales affects the physics of the turbulence at those scales.  We calculate this for the various spectral densities and collect the results here for reference.  For the Bessel series, we are only interested in the $k \gg 1$ limit.  Re-arranging \Eref{bessel-fjo-constraint} we have

\begin{equation}
\frac{W_{g1}({\bf k}, k)}{E({\bf k})} \approx k \left(\frac{1 + \Tau}{2\vcut^{(0)}}\right),\label{bessel-fjo-constraint-2}
\end{equation}

\noindent Evaluating $W_{g0}({\bf k}, j)$ at $j = 0$ and taking the ratio with $E({\bf k})$, using \Eref{beta-def}, we obtain the expression

\begin{equation}
\frac{W_{g0}({\bf k}, 0)}{E({\bf k})} = \frac{1 + \Tau - \hat{\Gamma}_0}{\hat{\Gamma}_0}.\label{invariant-ratio-plunk-basis}
\end{equation}

\noindent Note that unlike \Eref{bessel-fjo-constraint-2}, \Eref{invariant-ratio-plunk-basis} is valid at all scales.  Using the asymptotic form of $\hat{\Gamma}$ for small and large argument (see \Eref{gamma-large}), we can obtain the following asymptotic forms of this ratio

\begin{equation}
\frac{(1 + \Tau - \hat{\Gamma}_0)}{\hat{\Gamma}_0} \approx \cases{(1 + \Tau)\sqrt{2\pi}k & for $k \gg 1$,\\
\Tau + (1+ \Tau) k^2 & for $k^2 \ll 1$.\\}
\label{q-asymptotic}
\end{equation}

\noindent Note that at large scales, $k^2 \ll 1$, the ratio tends to a constant.  By subtracting the part of $W_{g0}$ that is proportional to $E$ at small $k$, we can obtain a free energy that has non-trivial scaling $q(k)$ at all $k$.  This quantity is $W_z$, defined already in \Eref{wz-def}.  We have

\begin{equation}
\frac{W_z({\bf k}, 0)}{E({\bf k})} = (1 + \Tau)[\hat{\Gamma}_0^{-1} -1],\label{wz-constraint}
\end{equation}

\noindent which for $k^2 \ll 1$ implies $W_z({\bf k}, 0)/E({\bf k}) \approx  (1+\Tau)k^2$, and for $k \gg 1$ the ratio becomes $(1 + \Tau)\sqrt{2\pi}k$ as expected from \Eref{q-asymptotic}, since $W_z({\bf k}, 0) \approx W_{g0}({\bf k}, 0)$ in that limit.

It is comforting to note that at sub-Larmor scales, all of these definitions of $q$ have linear scaling in $k$ \footnote{For this and other reasons, the sub-Larmor limit is more straightforward, which is the reason it was the focus of the earlier work \citet{plunk-prl}.}, giving consistency across arbitrary definitions.



\section{On the spectral redistribution of energy}\label{spectral-redist-sec}

The main object of study in this work is energy transfer by the nonlinear interaction via the quadratic nonlinearity of the gyrokinetic equation.  Spectrally, this occurs by particular three-wave interactions.  We will present numerical work in \Sref{numeric-sec} that examines nonlinear transfer directly in terms of a sum of three wave interaction terms.  We approach the problem another way in \Sref{secondary-sec} by solving a linearization of the gyrokinetic equation directly to find unstable modes; these solutions are then examined to draw conclusions about spectral energy transfer.  The question of energy transfer can also be approached in an ``equation-free'' sense by simply considering how energy transfer is constrained by the existence of the two invariants, $\W$ and $E$.  This approach, initiated by \citet{fjortoft}, is the subject of \Sref{spectral-redist-sec}.  We are not concerned here with how precisely the transfer occurs (in terms of three-wave interactions), but only with the quantities of energy transfered between different scales.  We begin by revisiting the three-scale theory of \citet{plunk-prl} and then in \Sref{gen-transfer-sec} we will generalize it to an arbitrary number of scales.

\subsection{Three-scale transitions}\label{three-scale-sec}

In \citet{plunk-prl}, we considered energy transfer among three scales.  We repeat this analysis here, but with some simple modifications.  As mentioned we would like to keep this discussion general, so let us refer to the general quantity $\W$, along with the general constraint given by \Eref{fjortoft-constraint}.

Let us define a scale, denoted by the pair $(q_i, j_i)$, to be the collection of Fourier components having wavenumbers ${\bf k}$ satisfying $q({\bf k}) = q_i$ and velocity-space index $j_i$.  We keep the form of $q({\bf k})$ general but note that for the isotropic $\beta({\bf k})$ defined in \Eref{beta-def}, there is a mapping $q \rightarrow k$ since $q$ is a monotonic function of the modulus $k$.  However, in order to handle anisotropic cascades (\ie those which generate zonal flows; see \Sref{zonal-sec}), it is convenient to allow for anisotropic $q$.

We assume that, by some nonlinear interactions, some amount of $\W$ (and the corresponding amount of $E$) is transfered between three different scales, $(q_1, j_1)$, $(q_2, j_2)$ and $(q_3, j_3)$.  We define the free energy at $(q_i, j_i)$ to be $\W_i = \sum_{{\bf k}} \delta(q({\bf k}) - q_i) \W({\bf k}, j_i)$ and the corresponding electrostatic energy to be $E_i = \sum_{{\bf k}}  \delta(j_i)\delta(q({\bf k}) - q_i) E({\bf k})$, where $\delta$ is the discrete delta function.  Define $\Delta \W_i = \W_i(t_1) - \W_i(t_0)$ and $\Delta E_i = E_i(t_1) - E_i(t_0)$.  Conservation of $\W$ and $E$ implies that the changes sum to zero:

\begin{eqnarray}
\sum_i \Delta \W_i = 0,\label{W-conserve}\\
\sum_i \Delta E_i = 0.\label{E-conserve}
\end{eqnarray}

\begin{center}
\begin{figure*}
\subfloat[I: Fj{\o}rtoft-type]{\label{transitions-fig-a}\includegraphics[width=0.3\textwidth]{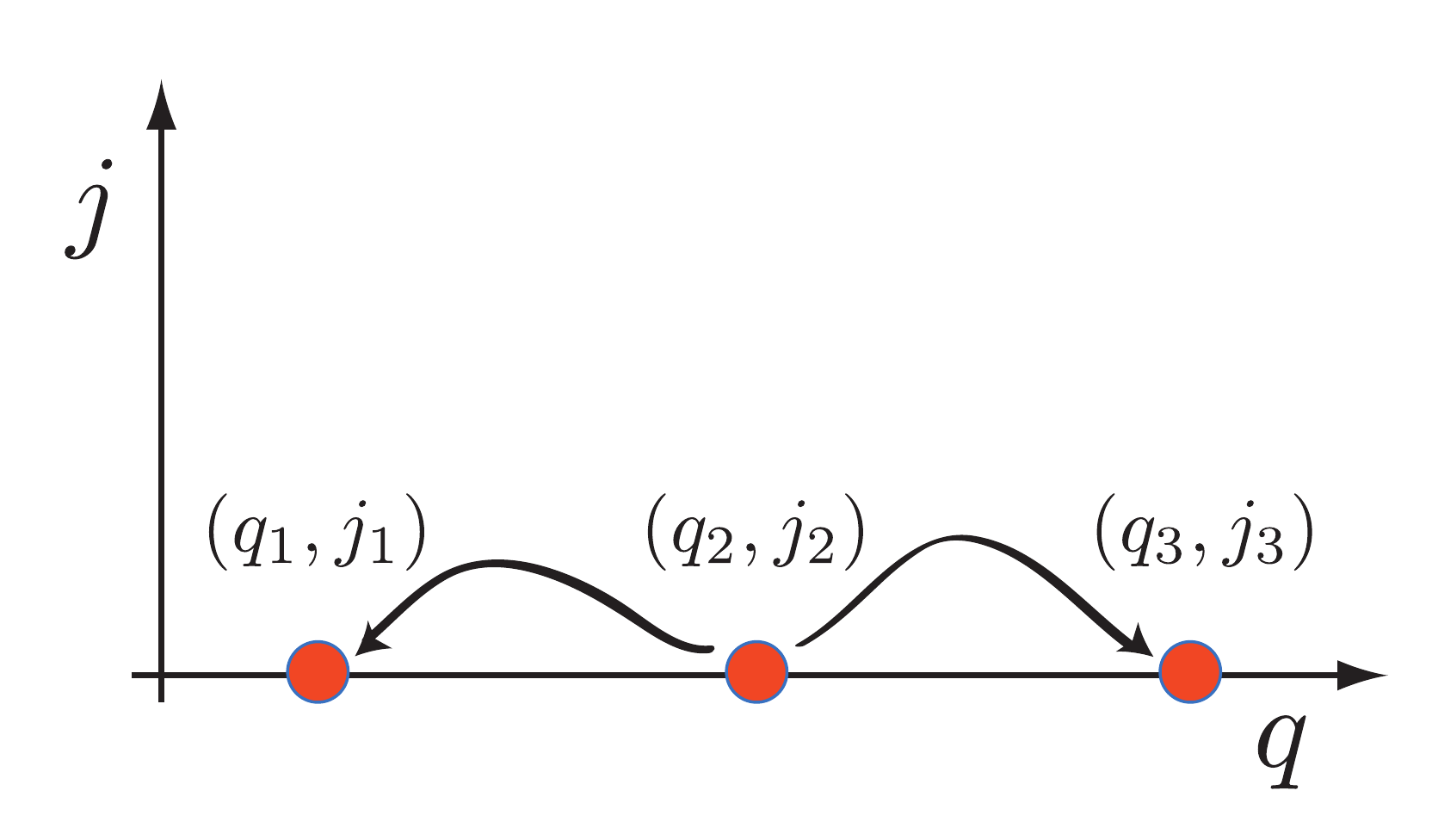}}
\subfloat[II: Kinetic-type]{\label{transitions-fig-b}\includegraphics[width=0.3 \textwidth]{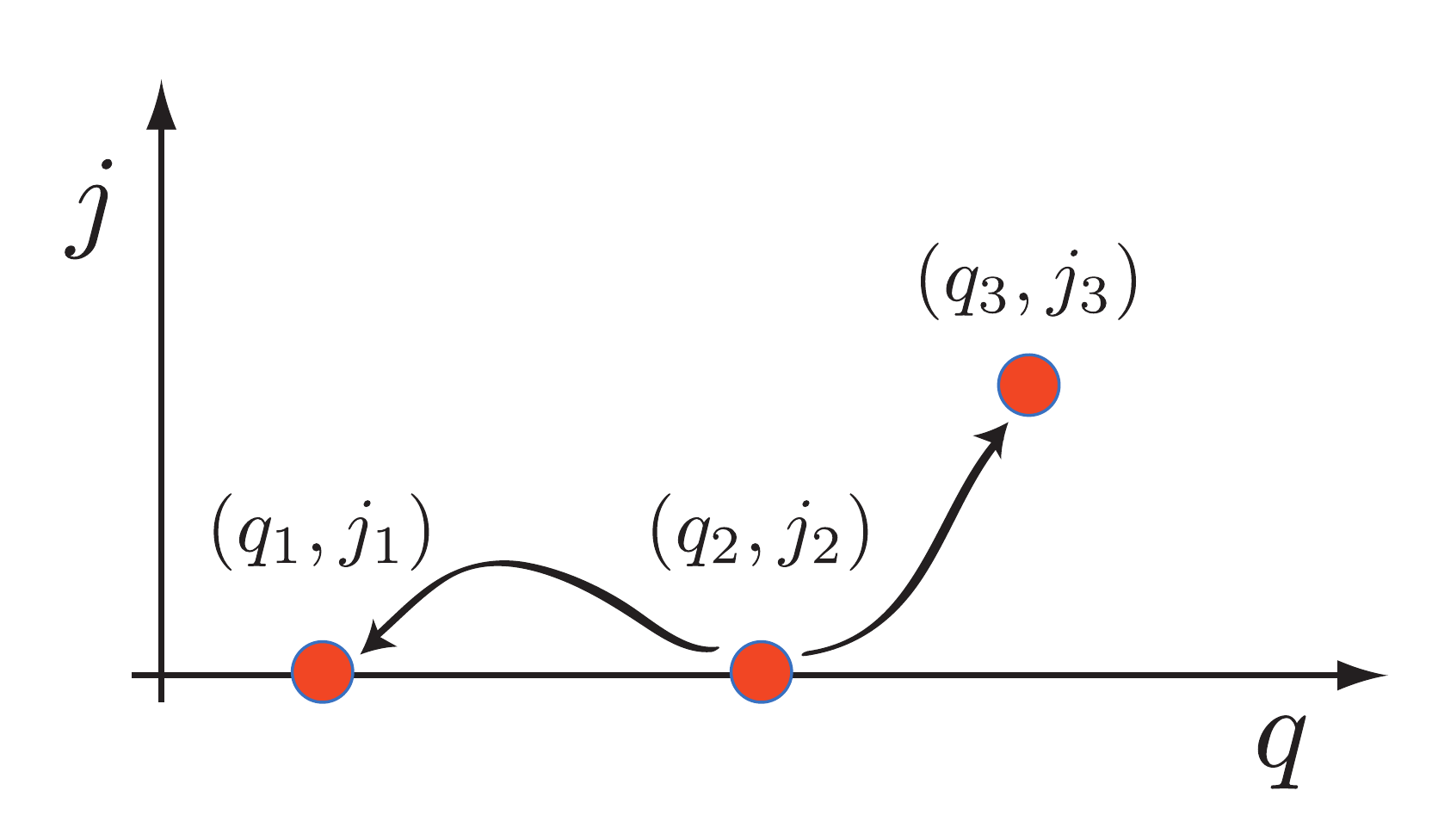}}
\subfloat[III: Unconstrained]{\label{transitions-fig-c}\includegraphics[width=0.3 \textwidth]{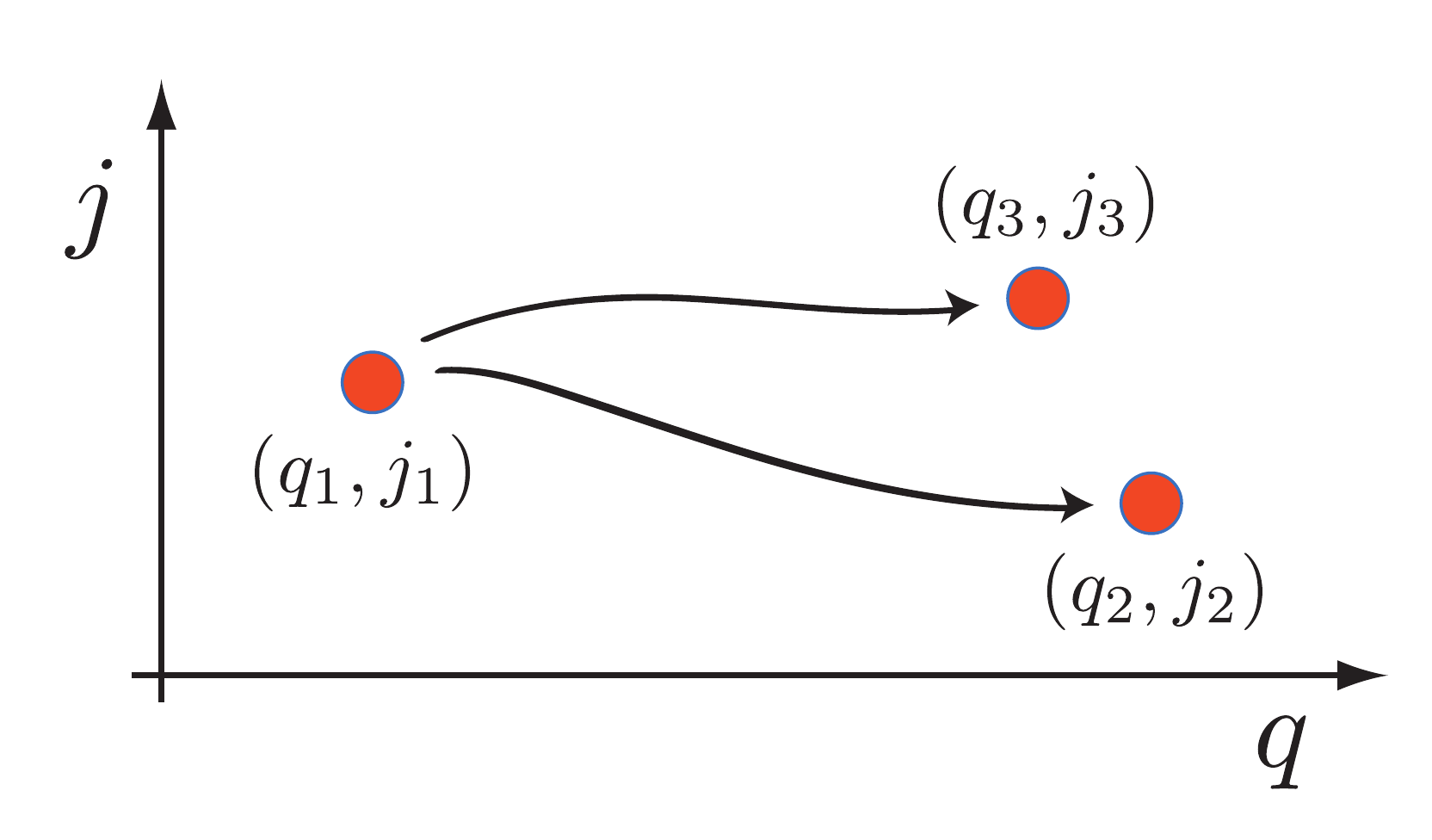}}
\caption{Energetic transitions involving three scales.  The horizontal axis is the density component $j = 0$.}
\label{transitions-fig}
\end{figure*}
\end{center}

\noindent As depicted in \Fref{transitions-fig} there are three types of transitions possible .  Without loss of generality we take $q_1 < q_2 < q_3$; for isotropic cases (that is, for $\beta = \beta(|{\bf k}|$), these $q$ correspond to $k_1 < k_2 < k_3$, so $q$ serves as proxy for $k = |{\bf k}|$.  In type I transitions (see \Fref{transitions-fig-a}) all three scales correspond to density components, $j_1 = j_2 = j_3 = 0$.  In this case, transfers are constrained in precisely the same way as considered by Fj{\o}rtoft, with the substitution $k_i^2 \rightarrow q_i = q(k_i)$.  We call these Fj{\o}rtoft-type transitions.  From Equations \eref{fjortoft-constraint}, \eref{W-conserve} and \eref{E-conserve} we can obtain the expressions equivalent to those obtained in \citep{fjortoft}:

\begin{eqnarray}
\Delta E_1 = -\left(\frac{q_3 - q_2}{q_3 - q_1}\right)\Delta E_2\label{fjortoft-type-eqn-1}\\
\Delta E_3 = -\left(\frac{q_2 - q_1}{q_3 - q_1}\right)\Delta E_2\label{fjortoft-type-eqn-2}
\end{eqnarray}

\noindent The quantities in parentheses are positive (because $q_1 < q_2 < q_3$).  Thus, as noted by Fj{\o}rtoft, it is only the intermediate component, $q_2$, which can be a source for both the two remaining components.

Type II transitions (\Fref{transitions-fig-b}) involve two density components ($j_1 = j_2 = 0$ but $j_3 \neq 0$) and also lead to inverse transfer of $E$ but only if the quantity of free energy transfered to the non-density component is positive.  To see this, first note that $\Delta E_3 = 0$ because $j_3 \neq 0$.  Solving Equations \eref{W-conserve} and \eref{E-conserve}, we find $\Delta E_1 = -\Delta E_2$ and

\begin{eqnarray}\label{kinetic-type-eqn}
\Delta E_1 = \Delta \W_3/(q_2 - q_1),\\
\Delta E_2 = -\Delta \W_3/(q_2 - q_1).
\end{eqnarray}

That is, any transfer of free energy from density to non-density components must be accompanied by a simultaneous {\em inverse} transfer of $E$ (\ie, free energy, {\em along $j = 0$}, to scales of smaller $q$); note this is unlike the case of a transition involving three density components, where some non-zero amount of energy travels to both smaller and larger $k$

Clearly, transitions involving only one density component (having non-zero energy change) are forbidden as they do not conserve $E$.  Type III transitions (\Fref{transitions-fig-c}) involve no density components, so are unconstrained.  Of course, actual turbulence involves a large number of scales simultaneously and in the next section, we will generalize these arguments to an arbitrary number of scales.  

\subsection{Transitions involving an arbitrary number of scales}\label{gen-transfer-sec}

Using \Eref{fjortoft-constraint}, we can derive a bound on the ratio of the invariants as follows.  For fixed ${\bf k}$ we clearly have $\W({\bf k}) = \sum_j \W({\bf k}, j) \geq q({\bf k}) E({\bf k})$, with equality only when all of the free energy is in the density component.  Thus the ratio of the invariants $\kappa$ is bounded below by the extreme case where all of the free energy $\W$ is in the density moment $j = 0$ (that is, $\W({\bf k}, j) = 0$ for $j \neq 0$):

\begin{equation}
\kappa \equiv \frac{\W}{E} \geq \frac{\displaystyle{\sum_{{\bf k}}} q({\bf k}) E({\bf k})}{\displaystyle{\sum_{{\bf k}}}E({\bf k})} \equiv \qbar. \label{kappa-min}
\end{equation}

\noindent The quantity on the right-hand side, $\qbar$, is an energy-weighted average of the function $q(k)$.  Note that from the large-$k$ asymptotic form of $q(k)$ given in \Eref{q-asymptotic}, $\qbar$ is proportional to the energy centroid for fluctuations at sub-Larmor scales, \ie for $W_{g0}$ we have $\qbar \approx \kbar\sqrt{2\pi} (1 + \Tau)$ where $\kbar = \sum k E({\bf k}) /\sum E({\bf k})$.

We now consider redistribution of $E$ and $\W$ involving an arbitrary number of scales.  Let us assume an ordering $q_1 < ... < q_n < q_{n+1} < ... < q_M$, where $q_M$ is the maximum $q$.  We have defined $\kappa = \W/E$, which, being the ratio of two invariants, is also an invariant itself.  In hydrodynamics, the analogue to this quantity is the enstrophy divided by the energy, which is equal to the energy-averaged squared wavenumber.  The fact that $q \propto k$ at sub-Larmor scales suggests that $\kappa$ may, analogously, be interpreted as something like a wavenumber there.



Let us define $q_N$, which we will use to divide the system into small- and large-wavenumber components.  Our goal is to establish an upper bound on the fraction of energy that can be found above $q_N$, as a function $\kappa$.  We write $E$ as a sum of large- and small-scale components

\begin{equation}
E = {\displaystyle\sum_{i = 1}^{N}} E_i + {\displaystyle\sum_{i = N + 1}^{M}} E_i\label{E-sum}
\end{equation}

\noindent We rewrite the free energy, separating the sum into the low-$q$ $j = 0$ terms (domain $\mathbb{G} = \{(q_i, 0)\}, q_i \leq q_N$), and the remainder of wavenumber pairs (domain $\mathbb{P}$):


\begin{equation}
\W = {\displaystyle\sum_{\mathbb{G}}} \W(q_i, 0) + {\displaystyle\sum_{\mathbb{P}}} \W(q_i, j_i).\label{W-sum}
\end{equation}

\noindent \Eref{W-sum} can be rewritten as

\begin{equation}
\kappa E = q^{\star}\displaystyle\sum_{i = 1}^{N}E_i + q^{\star\star}\displaystyle\sum_{i = N + 1}^{M}E_i,\label{W-sum-2}
\end{equation}

\noindent where $q^{\star}$ is defined to be the ratio of the first sum in \Eref{W-sum} to the first sum in \Eref{E-sum}, and $q^{\star\star}$ is defined to be the ratio of the second sum in \Eref{W-sum} to the second sum in \Eref{E-sum}.  Combining Equations \eref{E-sum} and \eref{W-sum-2}, we find

\begin{equation}
F_N = \frac{\kappa - q^{\star}}{q^{\star\star} - q^{\star}}\label{n-scale-result}
\end{equation}

\noindent where $F_N = \sum_{i = N + 1}^{M}E(k_i)/\sum_{i = 1}^{M}E(k_i)$ is the fraction of $E$ contained at wavenumbers larger than $q_N$.

Let's consider the consequences of \Eref{n-scale-result}.  It is easy to see that $q^{\star} < \qbar <  q^{\star\star}$ (and so $q^{\star} < \kappa$) and $q^{\star} \leq q_N <  q^{\star\star}$.  Thus we may infer that the maximum of $F_N$ is reached by taking $q^{\star} \rightarrow q_1$ and $q^{\star\star} \rightarrow q_N$.  Thus we obtain the bound $F_N < (\kappa - q_1)/(q_N - q_1)$.  If we consider $q_N \gg q_1$, this bound becomes

\begin{equation}
F_N <  \frac{\kappa}{q_N}.\label{n-scale-ineq}
\end{equation}

\noindent From this result we conclude that the fraction of energy that can be found at large (effective) wavenumbers $q > q_N$ is limited by the parameter $\kappa$ and becomes negligible with increasing $q_N$.  Note that although \Eref{n-scale-ineq} (and also \Eref{n-scale-result}) is valid for any value of $\kappa$, it is trivial if $\kappa > q_M$ since, by definition $F_N \leq 1$.  On the other hand, this bound is strong for $\kappa \ll q_M$ and indicates that the electrostatic energy content at large (effective) scales is preserved during dynamical evolution; it can be inferred also that if a turbulent cascade is driven with a sufficiently small $\kappa$, then $E$ will cascade inversely (to smaller $q$).

\subsection{Interpretation}\label{interp-sec}

Let us now discuss the implications that the expressions derived in the previous section have for the direction of spectral energy transfer.  The original work of Fj{\o}rtoft, concerning two-dimensional fluid turbulence, is simple to interpret:  Energy at a wavenumber $k$, if it is to be redistributed spectrally, must be transfered simultaneously to both smaller and larger wavenumbers.  If a turbulent cascade develops, such as that proposed by Kraichnan \citep{kraichnan1967} (with a driven stationary state characterized by a constant fluxes of energy/enstrophy through inertial scales larger and smaller than the injection scale) nearly all of the energy flux will be carried to large scales and nearly all of the enstrophy flux to small scales.  This is a basic consequence of the constraint $\ZNS(k) = k^2 \ENS(k)$ where $\ENS(k)$ and $\ZNS(k)$ are the spectral densities of energy and enstrophy respectively, for two-dimensional Navier-Stokes turbulence.  Viscosity acting at large wavenumbers dissipates asymptotically more enstrophy than energy, due to the factor of $k^2$, and so asymptotically more enstrophy must flow to large $k$ in the steady state.

In fact a similar conclusion for gyrokinetics can be drawn from \Eref{fjortoft-constraint}.  That is, the amount of free energy cascading to fine (collisional) scales must exceed that of $E$ by at least a factor of $q \propto k$ so that we can expect $E$ to be effectively conserved in the weakly collisional limit, even if the collisional dissipation of $\W$ tends to a non-zero constant in this limit.

In the case of gyrokinetics, the Fj{\o}rtoft constraint only governs transfer among components that contribute to particle density, \ie density components.  Thus, we can infer that Fj{\o}rtoft-type transitions (see Equations \eref{fjortoft-type-eqn-1}-\eref{fjortoft-type-eqn-2}) will promote inverse transfer of $E$, but it is not {\it a priori} obvious how the kinetic-type transitions, involving non-density components, will affect this tendency.  From \Eref{kinetic-type-eqn}, we can see that free energy transfer from density to non-density components will (further) cause inverse transfer of $E$, \ie to components of smaller $q$.  But it is also possible for free energy to flow in the opposite sense, from non-density to density components, inducing the reverse effect in the transfer of $E$.  

In \Sref{gen-transfer-sec}, we identified the ratio of free energy to electrostatic energy $\kappa$ as a control parameter that limits spectral redistribution of $E$.  The maximal fraction of free energy $F_N$ identified by \Eref{n-scale-ineq} corresponds to an extreme configuration where all the free energy is in the density components, all of the energy at large scales (small $q$) resides at the absolute largest scale available to the system $q_1$ and all of the energy at small scales (large $q$) is found just above the cutoff $q_N$.  This extreme distribution of energy is not likely to be spontaneously generated, and so we generally expect the fraction of $E$ that will be transfered to small scales to be significantly smaller than the maximum of \Eref{n-scale-ineq}.  Generally, we will see that the parameter $\kappa$ seems to be a good predictor of transfer direction.  We argue that this is because it measures the relative distribution of free energy between density and non-density components and so it determines whether the kinetic type transitions (see \Eref{kinetic-type-eqn}) will occur in the positive or negative direction (\ie positive or negative $\Delta \W_3$).

The limits set by the Fj{\o}rtoft constraint are clearly not sufficient to alone predict transfer direction.  To complete the argument for transfer direction of $E$, a simple conjecture was made in \citet{plunk-prl} that free energy that is initially concentrated in the density component about a single wavenumber $k_0$ will spontaneously ``spread out'' in $k$-$p$ space.  Because of the volume factor of $k$ in the two-dimensional $(k, p)$-plane, we estimated that this would lead to a distribution of free energy $W(k) \sim k q(k) E(k) \sim k^2 E(k)$ for $k \sim k_0$ (this is also the expected scaling from entropy cascade theory \citep{schek-ppcf, plunk-jfm}).  Thus, for initial conditions composed of energy around $k_0$, we identified the parameter $\kappa/k_0^2$ for the sub-Larmor range to delineate three regimes; see \Sref{secondary-sec} and \Sref{numeric-sec} for more details.  When this quantity is much less than 1, we found that there is strong nonlocal inverse transfer of $E$.  As $\kappa/k_0^2$ approached 1 the behavior changed to a more conventional local inverse cascade.  For $\kappa/k_0^2 > 1$, a transition was found where the transfer completely changed directions.

In the present work we are also interested in what happens at $k < 1$.  This is an important range for fusion plasmas, where instability at scales somewhat larger than the ion Larmor radius drive turbulence.  Generally speaking, we expect that low-$\kappa$ forcing should lead to inverse transfer of $E$ while high-$\kappa$ forcing should cause a cascade reversal.  It is, however, not obvious what constitutes ``high'' and ``low'' $\kappa$ at $k < 1$.  We investigate this question in \Sref{super-larmor-sec} and \Sref{secondary-sec} and find evidence that the transition occurs at $\kappa \sim {\cal O}(1)$.



\section{Super-Larmor scales}\label{super-larmor-sec}

At scales larger than the Larmor radius, nonlinear phase mixing loses potency.  This is because the Larmor orbit of a typical particle does not allow it to sample much variation in the electrostatic potential, and nearly all particles move with approximately the same ${\bf E}\times {\bf B}$ velocity.  Thus, nonlinear interactions take on a fluid character.  However, generally speaking, the long-wavelength limit does not admit rigorous closures in the fluid moment hierarchy, and the phase-space cascade is preserved in some form.  To what extent this cascade remains relevant is a subtle question that we will explore in this section and in \Aref{lw-app}.

In a particular limit, 2D gyrokinetics reduces to a single-field fluid equation (the HM equation), which possesses a nonlinear invariant (enstrophy) additional to those exhibited by 2D gyrokinetics \citep{plunk-jfm}.  This fact warns us generally against applying gyrokinetic results directly to the long-wavelength regime without considering how the system is reduced in that limit.  In the special HM limit, the non-density components of the distribution function (\ie those which give zero particle density) are dynamically decoupled from the density components \citep{plunk-jfm}.  That is, the mechanism for transfer of free energy between density and non-density components is removed entirely.  Thus $W_z$ (see \Eref{wz-def}) can be written as a sum of two quantities, one being the enstrophy, that are individually conserved.

For finite $\Tau$, the components do not decouple and the non-density components can become dynamically important, controlling the spectral transfer of $E$ (and, crucially, breaking the conservation of enstrophy that occurs in the HM limit).  In taking moments of the gyrokinetic equation, one encounters an infinite hierarchy of equations (moment hierarchy) where the different fields are coupled by finite-Larmor radius effects (phase mixing terms).  This fluid moment hierarchy is, of course, equivalent to the full gyrokinetic equation and no reduction is necessarily achieved this way.  (However, viewing the system in this manner shows that the coupling between velocity components becomes a formally local process, in that the effect of nonlinear phase mixing is to couple neighboring moments in the moment hierarchy.)

We will consider this moment hierarchy in detail in \Aref{lw-app} in a special limit (modified Boltzmann/adiabatic electron response) where zonal flows are preferentially generated.  But first, let us consider the form of the Fj{\o}rtoft constraint in the limit $k \ll 1$.  We have from \Eref{wz-constraint} that $W_z({\bf k}) \propto k^2 E({\bf k})$.  It could be argued that the fact that $k^2$ appears in this relationship (instead of $k$ in the sub-Larmor case) should strengthen the tendency toward inverse cascade of $E$ at super-Larmor scales.  Indeed, consider three-scale transitions involving a fixed set of wavenumbers $k_1 < k_2 < k_3$.  For Fj{\o}rtoft-type transitions, Equations \eref{fjortoft-type-eqn-1}-\eref{fjortoft-type-eqn-2} imply that the ratio of energy transfered inversely to that transfered forward is $R(q_1, q_2, q_3) = E_1/E_3 = (q_3 - q_2)/(q_2 - q_1)$, which is greater if $q \propto k^2$ than if $q \propto k$ (\ie $R(k_1^2,k_2^2,k_3^2)/R(k_1, k_2, k_3) = (k_3 + k_2)/(k_2 + k_1) > 1$).  However, though the stronger scaling of $q$ at $k^2 \ll 1$ may promote inverse cascade of $E$, the weakening of phase mixing at $k < 1$ may act counter to this by suppressing transfer of free energy from density to non-density components.  Thus, it is not {\it a priori} apparent what to expect at $k < 1$.

Let us now proceed to the case of modified Boltzmann electrons, where zonal flows play a special role in the turbulence.  This is especially relevant for fluctuations at scales somewhat larger than the ion Larmor radius, which strongly affect the confinement properties of tokamaks and other devices with closed flux-surface geometry.  

\subsection{Zonal Flows}\label{zonal-sec}

The so-called adiabatic response (contained within the constant $\beta({\bf k})$) strongly affects the asymptotic analysis of the long-wavelength limit.  The adiabatic response of electrons (\ie the response to ion-scale gyrokinetic fluctuations) can be corrected to take into account the special role of zonal components \citep{dorland-hammett-93, cohen} (see also section 8.1 of \citep{dorland-thesis}).  The correction dramatically changes the nonlinear dynamics, rendering the turbulence strongly anisotropic.  We can include this correction by giving $k$-dependence to $\Tau$,

\begin{equation}
\Tau = \tilde{\tau} =  \tau(1 - \delta(k_y)),\label{Tau-zonal-eqn}
\end{equation}

\noindent where $\delta$ is the discrete delta function and $\tau$ is a constant.  For small $\tau$ a simple fluid model may be derived called the generalized Hasegawa Mima (GHM) equation \citep{smolyakov-prl, manfredi}.  This system exhibits zonal flow generation by an anisotropic inverse cascade, which can be understood as a simple extension of the inverse cascade that occurs in fluid turbulence.  We review this mechanism in \Aref{ghm-theory-app}, noting that it lacks an essential mechanism to bring about zonal flow saturation, and so we now turn to gyrokinetics for a more nuanced description of how cascade dynamics can induce and regulate zonal flows.  The subject of zonal flow generation and regulation in magnetized plasma turbulence is a subject that has been studied intensely.  We note, however, that a clear description is still lacking of the precise role of inverse cascade (or more generally, constrained spectral transfer of energy) for generation and regulation of zonal flows in a fully gyrokinetic system.  This description requires that we consider, once again, the constraint placed on spectral energy transfer by the gyrokinetic conservation laws.


Let us introduce a free energy quantity suited for the long-wavelength regime with the modified Boltzmann response.  We define $\bar{W}_z = W_{g0} - \tau E$ and find 

\begin{equation}
\frac{\bar{W}_z({\bf k}, 0)}{E({\bf k})} = \frac{1 + \tilde{\tau}}{\hat{\Gamma}_0} - (\tau + 1).\label{q-zonal-z}
\end{equation}

\noindent Although the effect of the modified Boltzmann response is modest at $k \gg 1$, it is very important at $k < 1$.  Indeed, for $k^2 \ll 1$ we have

\begin{equation}\label{q-zonal-z-asymp}
\frac{1 + \tilde{\tau}}{\hat{\Gamma}_0} - (\tau + 1) \approx \cases{-\tau & for $k_y = 0$,\\
k^2(1 + \tau) & for $k_y \neq 0$.\\}
\end{equation}

\noindent which can be compared with \Eref{wz-constraint}.\footnote{Note that by comparing Equations \eref{wz-constraint} and \eref{q-zonal-z-asymp}, one can identify the well-known difference between the nonlinear physics of electron-temperature-gradient (ETG) and ion-temperature-gradient (ITG) driven turbulence that originates from the modified Boltzmann response, favoring generation of strong zonal flows in the ITG case.}  Thus the long-wavelength zonal component corresponds to the smallest possible value of $q$ and thus represents the largest effective scale of the system.  So, we conclude that if an inverse cascade occurs then it is anisotropic, leading to the accumulation of $E$ into components with $k_x^2 \ll 1$ and $k_y = 0$.  However, just as we have seen with the isotropic sub-Larmor cascade \citep{plunk-prl}, the inverse cascade is not guaranteed, but in fact could be tempered or reversed by an appropriate choice of forcing.  We will demonstrate this in \Sref{secondary-mod-response-sec} using linear instability theory.  Below we demonstrate this with simulations employing a simple fluid model.

\subsection{Gyrofluid simulations of stationary driven turbulence}\label{gf-sec}

Unlike the small-$\tau$ limit of the HM equation, the finite-$\tau$ long-wavelength limit has a closure problem in the fluid moment hierarchy.  Still, low order truncations or other closure schemes can yield a simple set of equations, which may provide physical insight into plasma behavior.

In the long-wavelength limit ($k\rho_i \ll 1$), so-called finite Larmor radius (FLR) terms account for nonlinear phase mixing and the cascade in velocity-space formally becomes local in the sense that the evolution equation for each moment involves only neighboring moments in the hierarchy.  We write down this moment hierarchy and describe some of its properties in \Aref{lw-app}.  The important conclusion is that there are critical problems with performing an FLR expansion because it will fail to model nonlinear phase mixing in a quantitatively correct way.  Nevertheless, we derive a gyrofluid system, which truncates the fluid moment hierarchy in an ad-hoc manner, as a simple toy model to address basic questions.

In this section we use this gyrofluid model to demonstrate the ``cascade reversal'' phenomenon in a driven and anisotropic context.  The model satisfies exact nonlinear conservation of an approximate version of $E$ and approximate nonlinear conservation of a ``truncated'' version of $W_{g0}$; see \Aref{lw-app}.  The gyrofluid model is

\begin{eqnarray}
&\frac{\partial B\varphi}{\partial t} + \poiss{\varphi}{B\varphi}  + N_2[\varphi, T_{\perp}] = B \left (A_{11} \nzo{\varphi} + A_{12} \nzo{T_{\perp}} \right) + D\tilde{\varphi},\label{gf-eqn-1}\\
&\frac{\partial T_{\perp}}{\partial t} + \poiss{\varphi}{T_{\perp}} + \poiss{\nabla^2\varphi}{2T_{\perp}- \tau \nzo{\varphi}/2} = A_{22} \nzo{T_{\perp}} + A_{21}\nzo{\varphi} + DT_{\perp},\label{gf-eqn-2}
\end{eqnarray}

\noindent where the zonal and non-zonal parts of $\varphi$ are denoted $\zon{\varphi}$ and $\nzo{\varphi}$, respectively (see Equation \eref{zon-op-def-eqn}), and we define

\begin{equation}
B = \tilde{\tau} - \nabla^2 \\
\end{equation}

\noindent and

\begin{equation}
N_2[\varphi, T_{\perp}] = \nabla^2\poiss{\varphi}{T_{\perp}} + \poiss{\nabla^2\varphi}{T_{\perp}} - \poiss{\varphi}{\nabla^2T_{\perp}}.\label{N2-def}
\end{equation}

\noindent Note that the so-called finite Larmor radius (FLR) terms are contained in \Eref{N2-def}; these nonlinearly couple the potential and the ion perpendicular temperature perturbations.  The operator $D = \nu_D \nabla^2 L_D $ (where $\nu_D$ is a constant) provides high-$k$ dissipation to the system by filtering out $|{\bf k}| < k_d$, this filtering being performed by $L_D$.  Note that we implement dissipation so that it does not act on $\zon{\varphi}$; this allows us to focus on nonlinear damping of zonal flows as a saturation mechanism.  The linear terms involving operators $A_{11}$, $A_{12}$, $A_{21}$ and $A_{22}$ on the right-hand side of Equations \eref{gf-eqn-1} and \eref{gf-eqn-2} are constructed to give (for $\nu_D = 0$) linear modes having complex frequencies

\begin{equation}
\omega = \frac{k_y}{2}\left(v_{*} \pm G \sqrt{(k/k_w)^2 - 1}\right).
\end{equation}

\noindent and growing eigenmodes with components that have the ratio

\begin{equation}
\frac{\hat{T}_{\perp}}{\hat{\varphi}} = R_0[\sin(\phi_0)\sqrt{(k/k_w)^2 - 1} + \cos(\phi_0)].
\end{equation}

\noindent The left-hand-sides of Equations \eref{gf-eqn-1} and \eref{gf-eqn-2} are derived by truncating the moment hierarchy above the perpendicular temperature in such a way as to approximately conserve a truncated version of the free energy $W_{g0}$, \ie the first two terms in the summation of \Eref{laguerre-Wg0}.  The model is related closely to the gyrofluid model of \citep{rogers-prl} (derived from gyro-fluid equations of \citep{dorland-hammett-93}) but deviates (somewhat arbitrarily) in this closure (and also in the {\it ad-hoc} linear drive terms).

\begin{figure}
\begin{center}
\subfloat[Zonal Energy at saturation.]{\label{R0-scan-fig-a}\includegraphics[width=0.504\columnwidth]{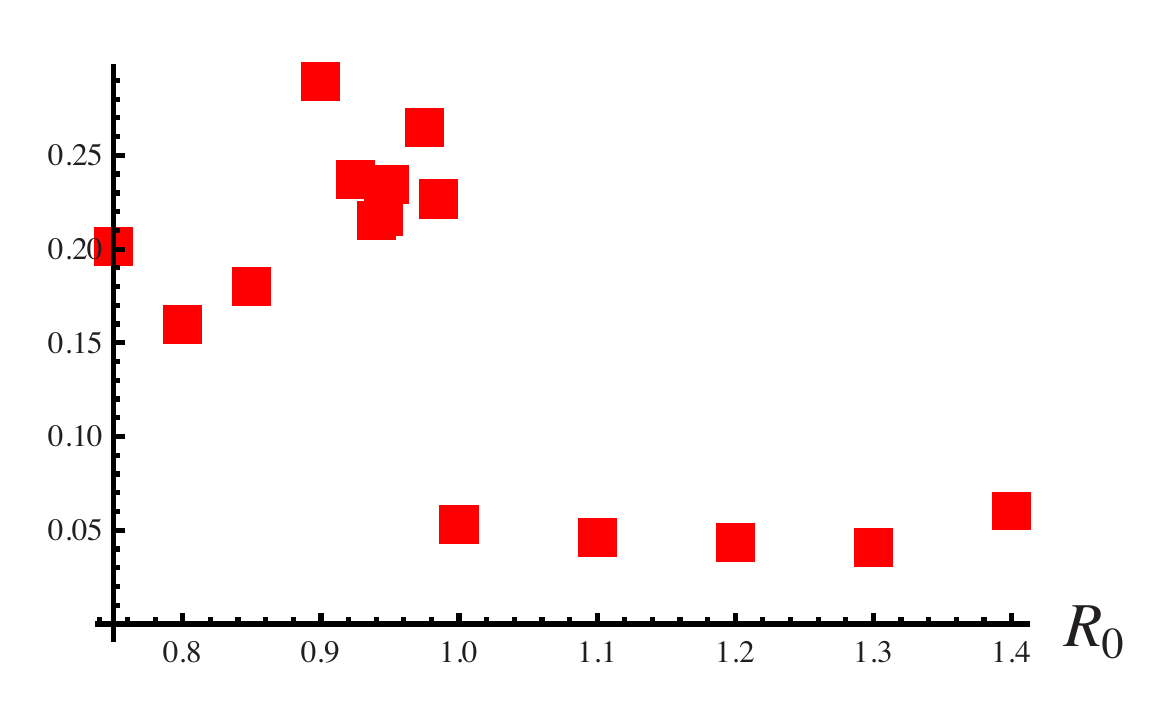}}
\subfloat[Non-zonal Energy at saturation.]{\label{R0-scan-fig-b}\includegraphics[width=0.504\columnwidth]{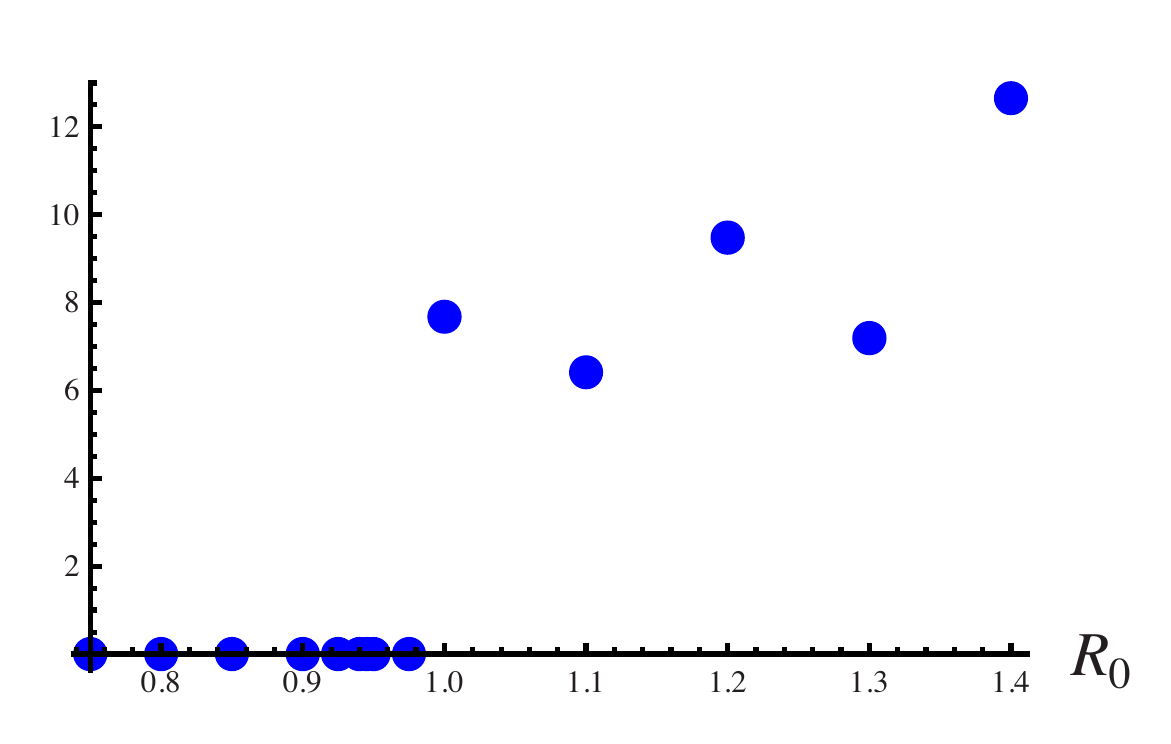}}
\end{center}
\caption{Nonlinear critical transition: enhanced turbulent viscosity acting on zonal flows with increase of parameter $R_0$.}\label{gf-transition-fig}
\end{figure}

We solve Equations \eref{gf-eqn-1} and \eref{gf-eqn-2} by a fully spectral method.  For the runs of \Fref{gf-transition-fig} we use 220 Fourier modes (with a spectral domain corresponding to the upper-half plane in $k_x$-$k_y$ space with grid space of $0.1$ and maximum wavenumber $1.0$) and linear drive parameters $v_{*} = 1$, $G = 0.3$, $\phi_0 = -\pi/2$, $k_w = 0.25$, $k_d = 0.3$ and $\nu_D = 0.05$.  Viscous dissipation is provided at large wavenumbers but set to zero for the zonal component of the potential.

\begin{figure}
\begin{center}
\subfloat[Subcritical saturated state ($R_0 = 0.85$).]{\label{GF-phi-fig-a}\includegraphics[width=0.504\columnwidth]{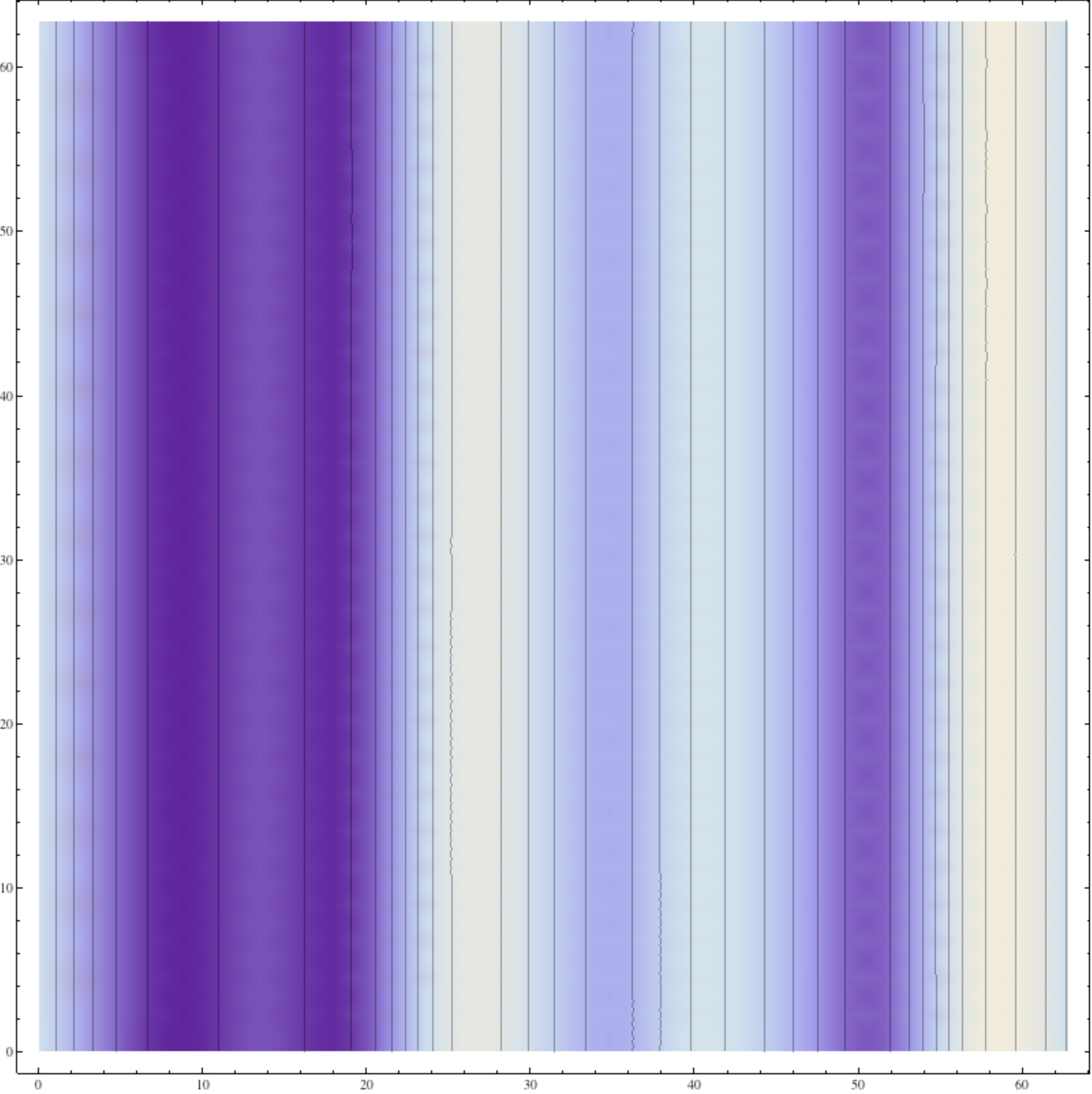}}
\subfloat[Supercritical saturated state ($R_0 = 1.2$).]{\label{GF-phi-fig-b}\includegraphics[width=0.504\columnwidth]{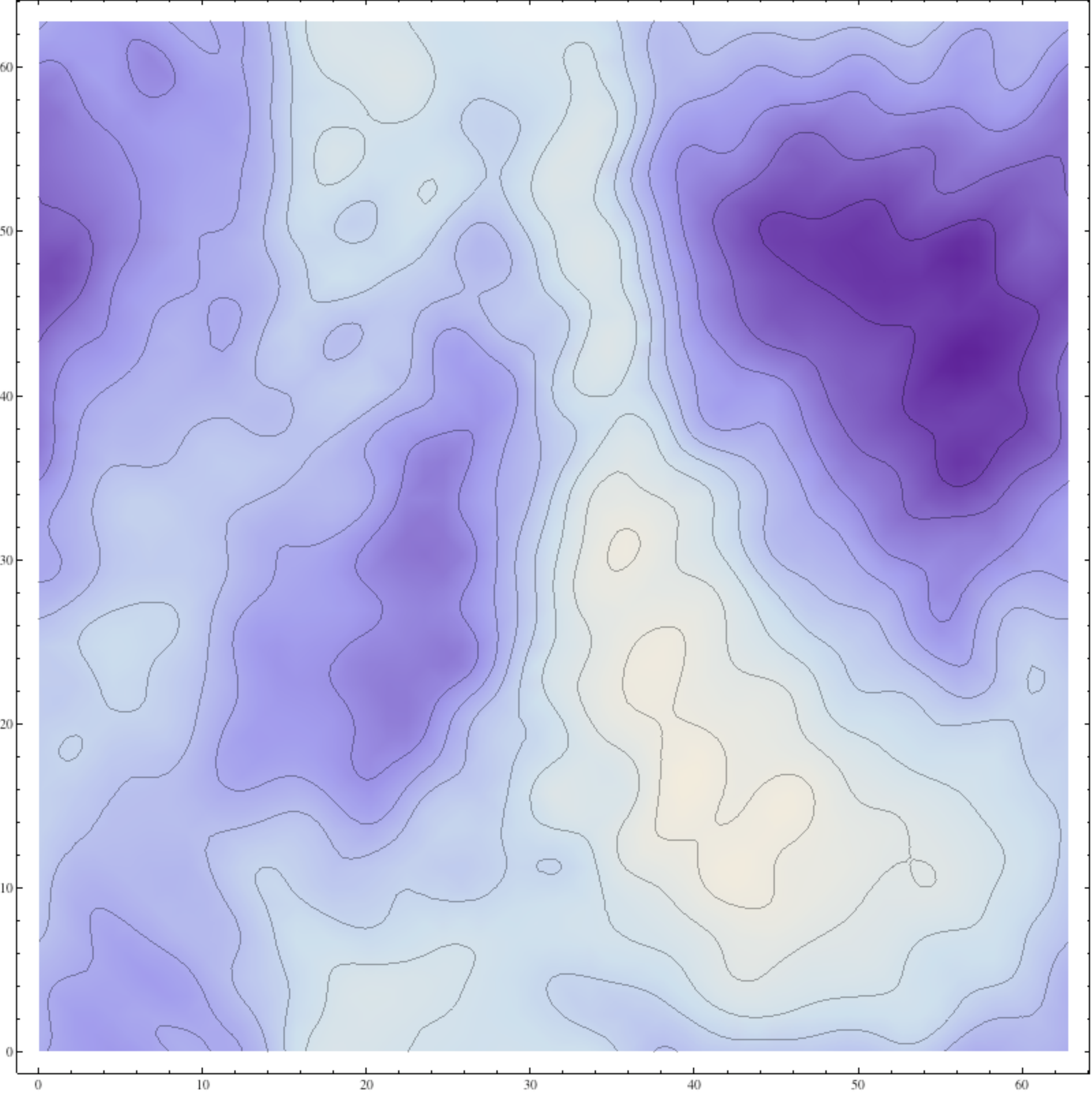}}
\end{center}
\caption{Electrostatic potential of saturated states for supercritical and subcritical cases: density plot with overlaid contours of constant potential.}\label{gf-transition-field-fig}
\end{figure}

We find a nonlinear critical transition associated with the parameter $R_0$ of the linear drive.  Because the growth-rate spectrum is unchanged in these simulations, this transition demonstrates the importance of the relative injection of free energy to electrostatic energy (\ie the $\kappa$ factor) in determining the overall behavior of the turbulence.  In \Fref{gf-transition-fig} we plot the saturated energy values, each point corresponding to a separate run.  In \Fref{gf-transition-field-fig} we show a representative snapshot of the electrostatic field for sub-critical and super-critical cases.  

We note that this transition is absent if the FLR terms are neglected; see \Fref{gf-compare-fig}.  The behavior of the gyrofluid system at zeroth order is qualitatively very different from the behavior of the system with the appropriate FLR terms retained (even when $k^2 \ll 1$ is satisfied for all Fourier components).  Thus we conclude that $k^2 \ll 1$ is a singular limit of gyrokinetics (with the modified Boltzman response for electrons) in the sense that formally small (FLR) terms remain important as the size of these terms tends to zero.  This is due to the nonlocal interactions in $k$-space, \eg due to the tertiary instability that regulates the zonal flows (see \Sref{secondary-sec}), which has very fine scale contributions to its eigenfunction.  Actually, this fact casts some doubt on the quantitative validity of any fluid models of gyrokinetics that use the expansion $k^2 \ll 1$.

\section{Linearization and Secondary/Tertiary Instability}\label{secondary-sec}

The linearization of the dynamical equations about an exact solution (a monochromatic wave, which in our case is a stationary solution) gives another way to investigate the spectral evolution of fluctuations.  If an instability is present, its growth rate may be calculated analytically.  If the modes of largest growth rate are found at scales larger (smaller) than those represented in the initial condition, then it may be argued that inverse (forward) spectral transfer should be expected generally to result from such initial conditions.  This line of thinking represents an alternative way to address the problem of spectral transfer direction, based upon actual dynamics -- though the calculation is made only with a linearization of the dynamical equations and thus strictly applies only to a nearly monochromatic initial condition.  The Fj{\o}rtoft argument has the advantage that it applies to the fully nonlinear problem, with arbitrarily complex states.

An instability theory has been investigated previously for gyrokinetic linear eigenfunctions corresponding to local ion/electron temperature gradient (ITG/ETG) instability \citep{plunk-pop} under the name ``secondary instability theory.''  In that case, the velocity dependence of the initial condition (primary mode) is determined by the ITG/ETG linear theory, and thus the $\kappa$ of the primary mode, which we call $\kappa_p$, is also determined.  Here we will take an artificial parameterization of the initial condition that allows us to arbitrarily set $\kappa_p$.  

We follow \citep{plunk-pop} with the modification that the velocity dependence of the initial perturbation is chosen by hand to give the desired ratio $\kappa_p$ in the initial condition.  Let us define ${\bf k}_0$ to be the wavenumber of the initial condition (or `primary' mode).  We will examine the problem for $k_0 > 1$ and $k_0 < 1$, considering both the Boltzmann response and the modified Boltzmann response for the second case.

\subsection{$k_0 > 1$}\label{secondary-sec-subL}
First let us consider sub-Larmor scales $k_0 = |{\bf k}_0| \gg 1$.  We find that the instability is strongly affected by $\kappa_p$ in a way that agrees with expectations from the results of \citet{plunk-prl}.  We take the `no response' limit ($\Tau = 0$), which is isotropic ($\beta = \beta(|{\bf k}|)$) so we arbitrarily choose the wavenumber to be in the $y$-direction, ${\bf k}_0 = (0, k_0)$, without loss of generality.  The primary mode $g_p$ is written \footnote{Note that the convention used in \citep{plunk-pop} is to express the distribution function using $h$ instead of $g$.  Translating into that form we have $h_p = g_p + \varphi_p J_0(k_0 v) F_0$.}

\begin{equation}
g_p({\bf R}, v) = F_0[a_0 J_0(k_0 v) + a_1 J_0(\sigma k_0 v)][\e^{i {\bf k}_0\cdot {\bf R}} + \mbox{c.c.}].\label{primary-mode}
\end{equation}

\noindent The terms proportional to $a_0$ and $a_1$ are meant to provide the density and non-density contributions to free energy, respectively.  The constant $\sigma \neq 1$ is a factor that determines the effective velocity-space wavenumber of the non-density component.  Plugging this expression into \Eref{qn-g-k}, and using \Eref{gamma02-def}, we calculate the electrostatic potential of the primary mode $\varphi_p$ as

\begin{equation}
\varphi_p = \frac{\beta(k_0)}{2\pi}[a_0 \hat{\Gamma}_0(k_0, k_0) + a_1\hat{\Gamma}_0(k_0, \sigma k_0)][\e^{i {\bf k}_0\cdot {\bf r}} + \mbox{c.c.}].\label{primary-phi}
\end{equation}

\noindent Using \Eref{gamma2-large}, we see that for large arguments the second term is proportional to $\exp[-k_0^2(1-\sigma)^2/2]$ and so is negligible for sufficiently large $k_0$.  The free energy $W_{g0}$ (see \Eref{Wg0-def2}) of this mode is

\begin{equation}
W_{g0} = \left[a_0^2 \hat{\Gamma}_0(k_0) + 2 a_0 a_1 \hat{\Gamma}_0(k_0, \sigma k_0) + a_1^2 \hat{\Gamma}_0(\sigma k_0)\right].\label{modal-W}
\end{equation}
\noindent Using \Eref{E-def} the electrostatic energy $E$ may be computed as

\begin{equation}
E = \frac{\beta(k_0)}{2\pi}\left[a_0\hat{\Gamma}_0(k_0) + a_1\hat{\Gamma}_0(k_0, \sigma k_0) \right ]^2.\label{modal-E}
\end{equation}

\noindent The primary mode has a definite amount of $W$ and $E$, the ratio of which establishes the modal $\kappa$ factor $\kappa_p$.  Taking the ratio of the energies computed in Equations \eref{modal-W} and \eref{modal-E} and neglecting the terms proportional to $a_1$ in $E$, we have

\begin{eqnarray}
\kappa_p &\approx \frac{2\pi}{\beta(k_0)\hat{\Gamma}_0(k_0)}\left[1 + \frac{a_1^2\hat{\Gamma}_0(\sigma k_0)}{a_0^2\hat{\Gamma}_0(k_0)} \right]\\
&\approx \kappa_{\mathrm{min}} \left[1 + \frac{a_1^2}{a_0^2 \sigma}\right],
\end{eqnarray}

\noindent where $\kappa_{\mathrm{min}} = (1 + \Tau)\sqrt{2\pi}k_0$.  By setting $a_1 = 0$, we are thus able to reach the absolute theoretical minimal value of $\kappa$ as predicted by Equations \eref{q-asymptotic} and \eref{kappa-min}; this corresponds to all of free energy being focused in the density component.  With a non-zero choice of $a_1$, we can make $\kappa_p$ arbitrarily large.

\begin{figure}
\includegraphics[width=0.95\columnwidth]{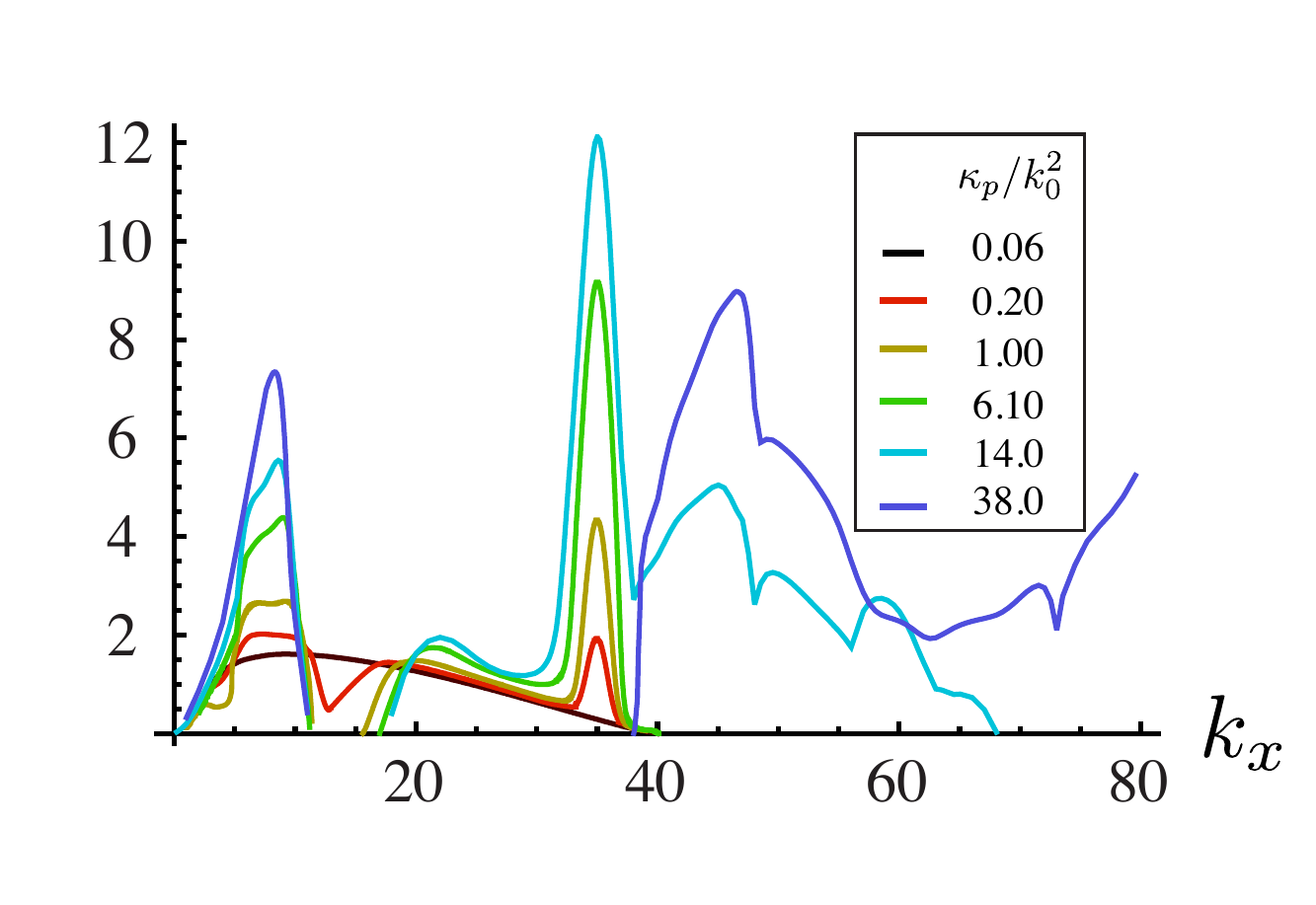}
\caption{Linear instability:  We solve for ``secondary'' modes that are perturbations to a large-amplitude monochromatic ``primary'' mode with a specified energy ratio $\kappa_p$.  Plotted is the growth rate spectrum of such modes for 6 values of $\kappa_p$ ranging from the theoretical minimum case (black) to the supercritical case (blue).  The black curve corresponds to a primary mode with no density-moment free energy, \ie a mode of minimal $\kappa_p$.  For all cases, we take $\Tau = 0$, $k_0 = 40$ and $\sigma = 0.75$.}\label{secondary-fig}
\end{figure}

Returning the calculation of the instability, note that $g_p$ is a stationary solution of \Eref{gyro-g}, being composed of a single Fourier mode.  The stability of perturbations to this solution is what we are concerned with.  Taking $g = g_p + g_s$, where $g_s \ll g_p$ is the ``secondary'' mode, the gyrokinetic equation reduces to a linear equation that represents the driving of $g_s$ by $g_p$.  We assume a normal mode solution, that is $g_s$ proportional to $\e^{i \omega_s t + i k_x X}$ and $\varphi_s$ proportional to $\e^{i \omega_s t + i k_x x}$ with $\omega_s$ the complex frequency (recall that $x$ is the coordinate in position space and $X$ is the coordinate in gyrocenter space).  The $y$-dependence of the mode is captured by an eigenmode, which is composed of harmonics of the primary wavenumber $k_x$; see \Aref{zo-GF-secondary-app} for a simple illustration of this type of problem.  An integral equation can be derived that can be solved for the eigenmode and the corresponding eigenvalue $\omega_s$; see Eqn. 12 of \citep{plunk-pop}.  We solve this integral equation numerically and plot the solutions in \Fref{secondary-fig}.

Generally the instability depends on the parameters $a_0$, $a_1$, $\sigma$ and $k_0$.  However, we focus on the single measure $\kappa_p/k_0^2$, which seems to be a good predictor of the general features of the growth rate curve.  The case of minimal $\kappa_p$ (that is, $a_1 = 0$) exhibits familiar features \citep{plunk-pop}.  The growth rate curve varies smoothly from $k_x = 0$, reaches a single clearly recognizable maximum after which it falls again to zero at a cutoff wavenumber equal to the primary wavenumber $k_0$.  It can be concluded from this evidence that the tendency of turbulence driven by modes like this $g_p$ will be to produce large-scale features in the electrostatic potential, \ie cascade the electrostatic energy inversely.

As the free energy of the primary mode is increased, we expect the instability to transform; this is indeed what happens.  From the arguments of \citet{plunk-prl} (see also \Sref{interp-sec}) our expectation is that the behavior will be most strongly affected by the control parameter $\kappa/k_0^2$.  This quantity measures the relative distribution of free energy between accessible density and non-density modes.   

\begin{figure}
\subfloat[$\kappa_p/k_0^2 = 0.06$]{\label{secondary-sideband-fig-a}\includegraphics[width=0.5\columnwidth]{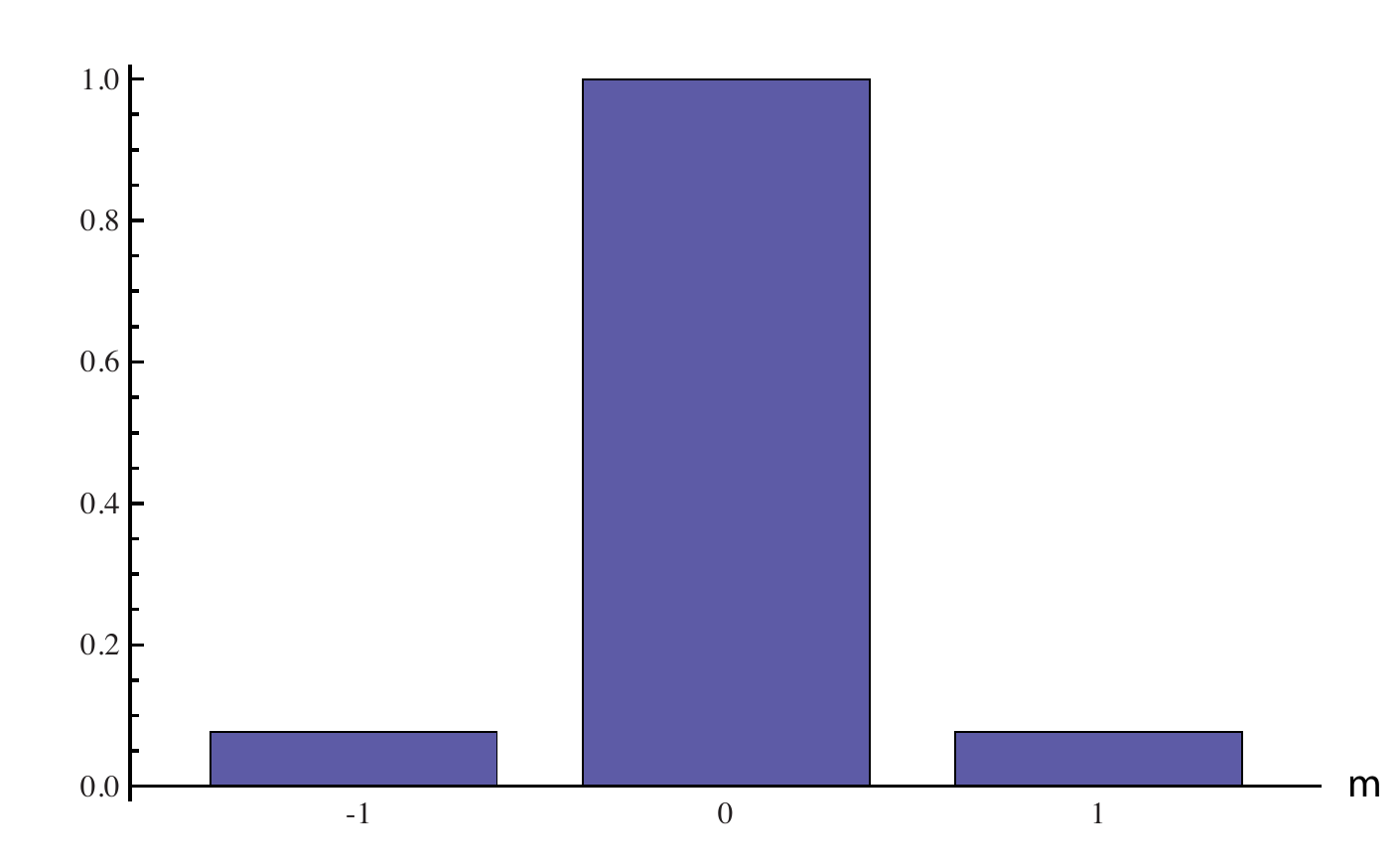}}
\subfloat[$\kappa_p/k_0^2 = 38$]{\label{secondary-sideband-fig-b}\includegraphics[width=0.5\columnwidth]{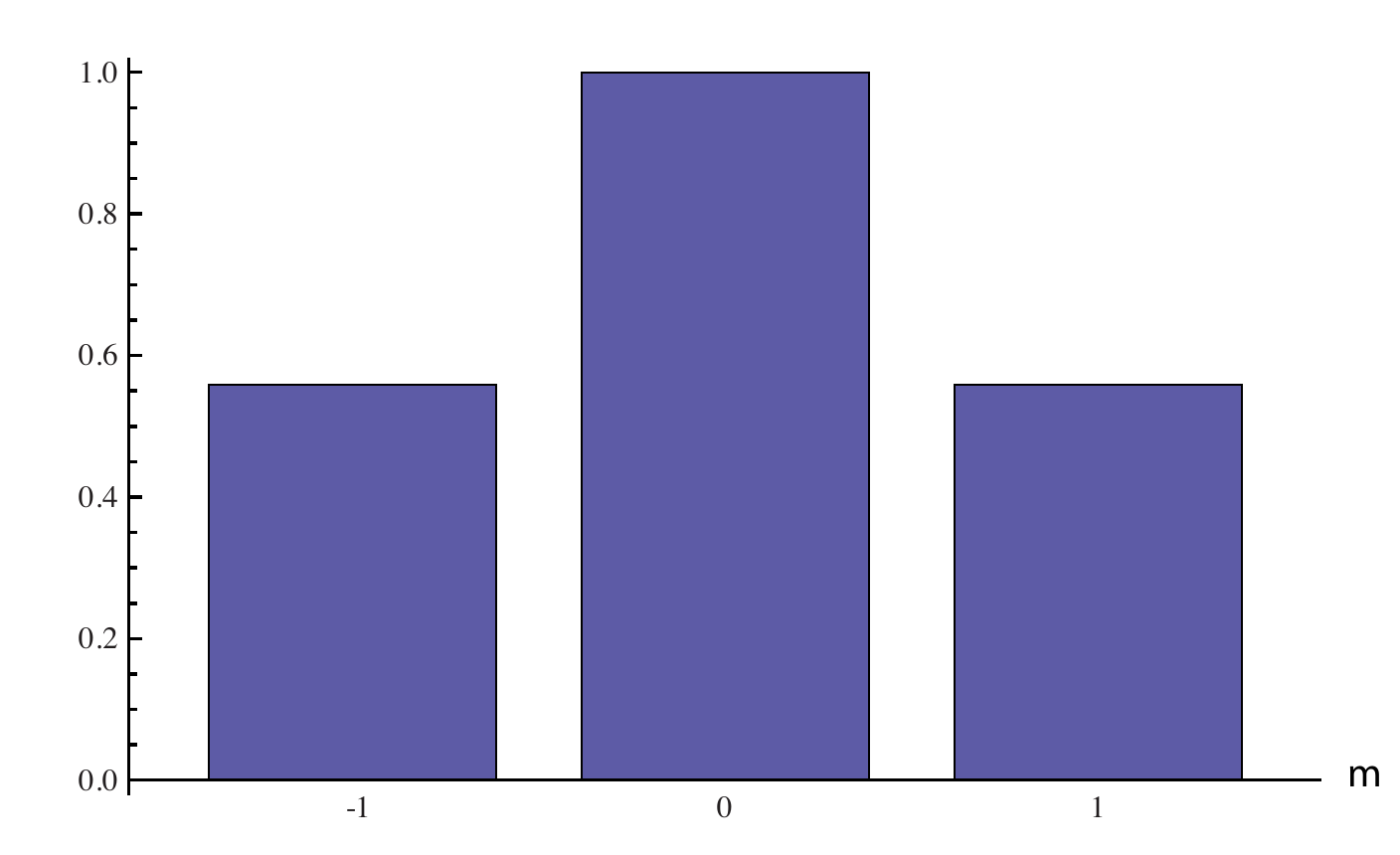}}
\caption{The relative amplitudes of sidebands (and higher harmonics not shown) of the secondary mode are higher for large $\kappa_p$.  These figures correspond to the growth rates computed for \Fref{secondary-fig} at $k_x = 9$.  This effect is even more pronounced in the quasi-singular $k_0 \ll 1$ limit and the inclusion of higher harmonics becomes necessary to accurately compute the secondary growth rate.}\label{secondary-sideband-fig}
\end{figure}

In \Fref{secondary-fig}, we show how the growth rate curve of the instability varies with $\kappa_p/k_0^2$.  Although the instability does vary somewhat with other parameters like $\sigma$, the case plotted in \Fref{secondary-fig} ($\Tau = 0$, $k_0 = 40.0$, $\sigma = 0.75$) seems representative of the overall shape (\eg peak, magnitude, cutoff) of the growth rate curve for the cases that we calculated.  We observe that (1) as $\kappa_p/k_0^2$ is increased, the growth rate is strengthened at the large-wavenumber part of the spectrum; (2) at a critical value of $\kappa_p/k_0^2 \sim {\cal O}(1)$, the mode is destabilized above the cutoff $k_0$; (3) as $\kappa_p/k_0^2$ is increased further, the global peak of the spectrum eventually appears above the primary wavenumber $k_0$, signaling a shift from inverse spectral transfer to forward spectral transfer (transfer from low-$k$ to high-$k$).  These observations echo the transition observed in the fully nonlinear numerical experiments observed by \citet{plunk-prl} and thus serve as further confirmation of the theory.  Note that the appearance of a larger growth rate at lower $k_x$ may appear to be in contradiction with our general conclusions.  This is not so.  As they become more unstable, the sidebands ($k_y = \pm k_0$) and higher harmonics of these low-$k_x$ modes grow to larger amplitudes so that over half the energy in the mode actually resides at $k = \sqrt{k_0^2 + n^2 k_x^2} > k_0$, where $n$ is the harmonic number of the secondary eigenfunction; see \citep{plunk-pop}.  This is in contrast to the small-$\kappa_p$ case where the sidebands have small relative amplitudes and most of the energy is contained in the $k_y = 0$ component; See \Fref{secondary-sideband-fig}.

Other curious features are observed, such as multiple branches of the instability, ranges in $k_x$ of stability, and oscillations in the growth rate curve.  We did not give much attention to these features; one might argue that such details are less important in fully developed turbulence than the overall shape of the growth rate spectrum.  Our focus is on the qualitative trend of the growth rate on what appears to be the strongest control parameter.  We conclude that the parameter $\kappa_p/k_0^2$, although not alone sufficient to completely characterize the instability, does constitute a reasonable predictor for the direction of spectral transfer.

\subsection{$k_0 < 1$ (Boltzmann electron response)}

Now let us turn to scales larger than the Larmor radius.  We first consider the conventional Boltzmann response for the non-kinetic species.  This is most relevant for scales a bit larger than the electron Larmor radius (but much smaller than the ion Larmor radius), \eg the range of energy injection for the electron temperature gradient (ETG) driven turbulence in a tokamak.  We represent the non-density free energy using $P_1^{(k)}$, defined in \Sref{orth-poly-sec}, instead of the Bessel function used in the previous section.  We again consider an initial condition with the mode aligned arbitrarily in the $y$-direction, ${\bf k}_0 = (0, k_0))$:

\begin{equation}
g_p = F_0[P^{(k_0)}_0 + \chi P^{(k_0)}_1][\e^{i{\bf k}_0\cdot {\bf R}} + \mbox{c.c.}] \label{gp_p1-eqn}
\end{equation}

\noindent and recall that $P_0^{(k)} = J_0(k v)\hat{\Gamma}_0^{-1/2}$ we have

\begin{equation}
\varphi_p = \frac{\beta}{2\pi}\hat{\Gamma}_0^{1/2}(k_0)[\e^{i{\bf k}_0\cdot {\bf r}} + \mbox{c.c.}].\label{phip_p1-eqn}
\end{equation}

\noindent The function $P_1^{(k)}$ can be easily computed explicitly to give

\begin{equation}
P_1^{(k)} = \frac{(2 - k^2 - v^2)}{\sqrt{4 + k^4}},
\end{equation}

\noindent By construction, the term proportional to $\chi$ in \Eref{gp_p1-eqn} gives no contribution to $\varphi_p$.  The electrostatic energy of the primary mode is simply

\begin{equation}
E = \frac{\beta}{2\pi}\hat{\Gamma}_0(k_0),
\end{equation}

\noindent and the free energy is

\begin{equation}
W_{g0} = 1 + \chi^2.
\end{equation}

\begin{figure}
\includegraphics[width=0.95\columnwidth]{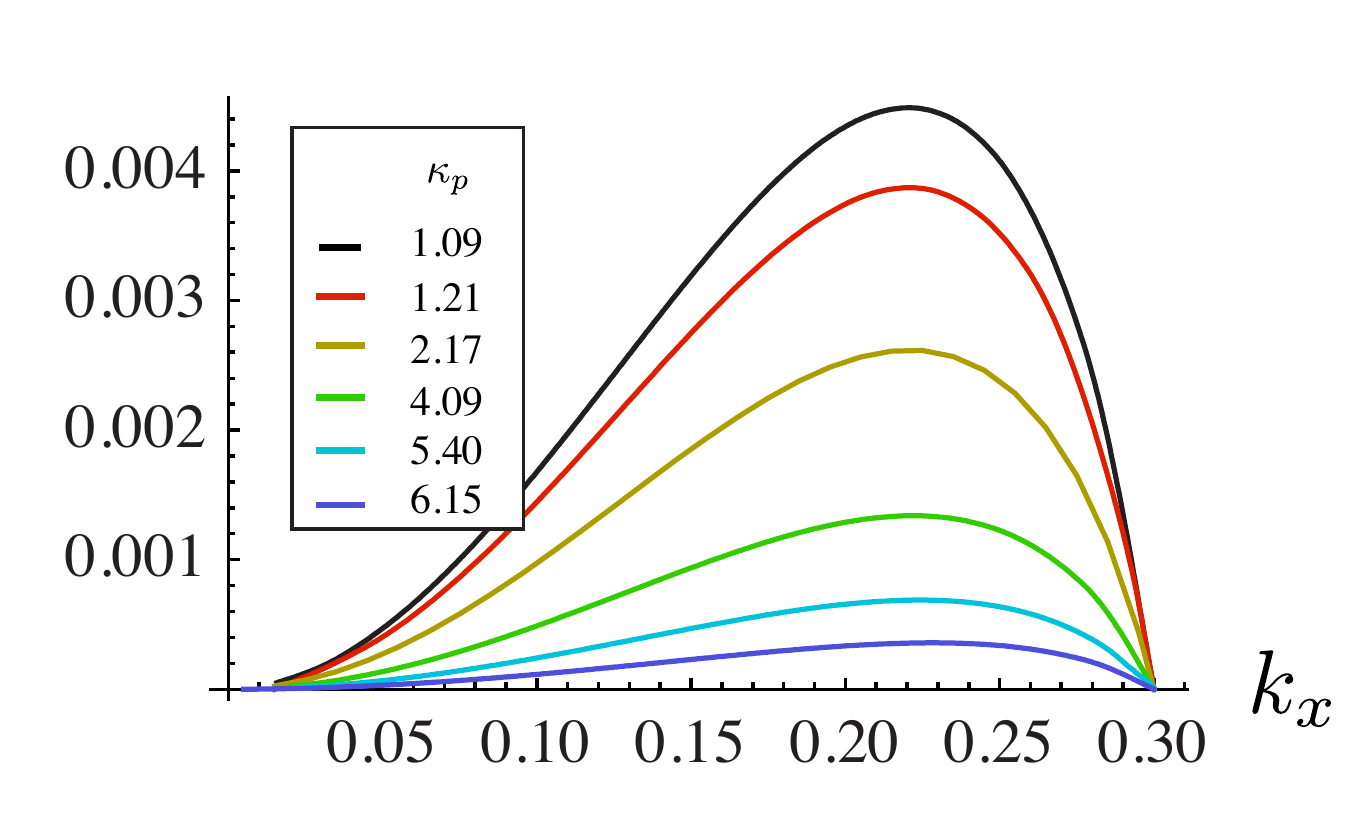}
\includegraphics[width=0.95\columnwidth]{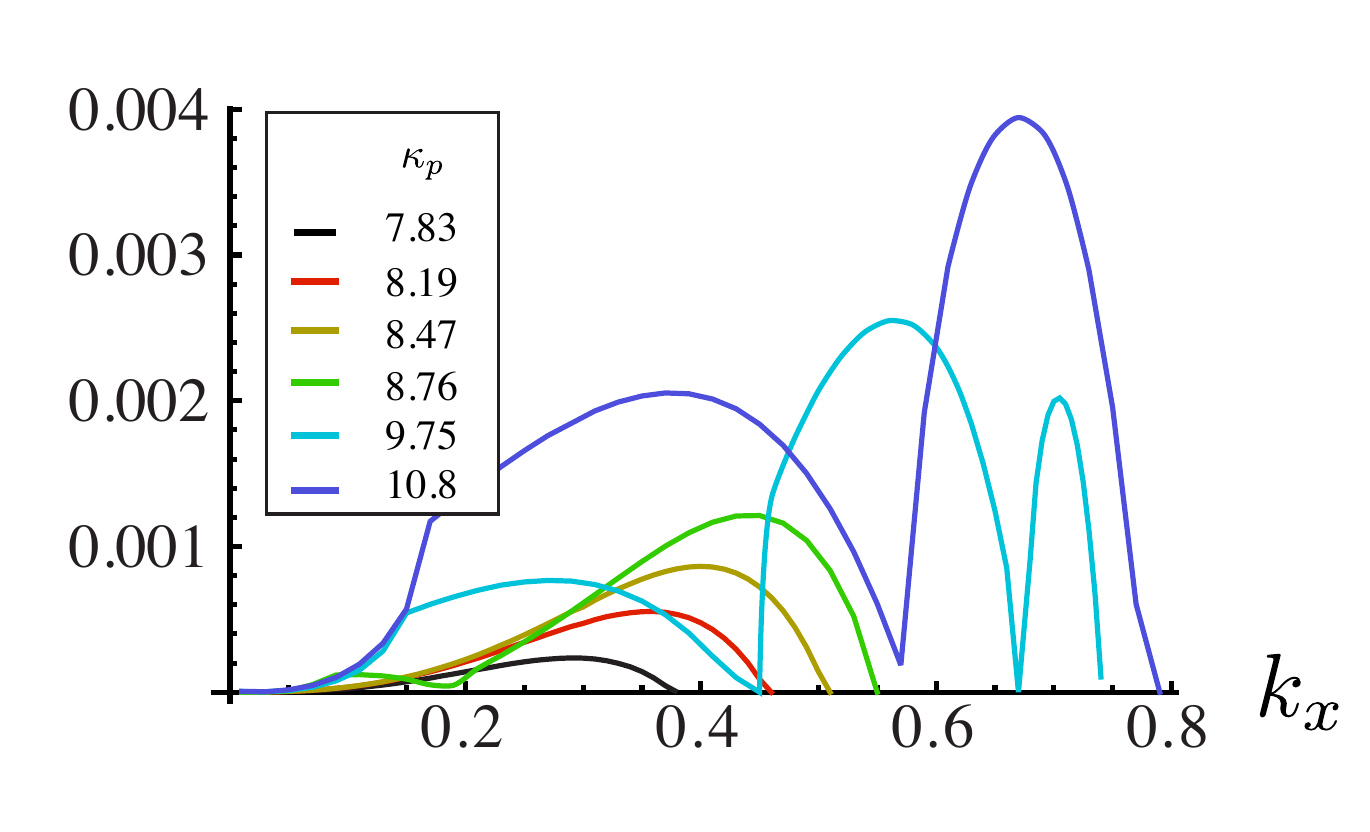}
\caption{Secondary/tertiary instability curve for $k_0 = 0.3$ with conventional Boltzmann electrons ($\Tau = 1$).}\label{secondary-fig-small-k}
\end{figure}

\noindent So the $\kappa$ factor for the initial mode is $\kappa_p = 2\pi(1 + \chi^2)/(\beta \hat{\Gamma}_0)$, which for $k^2 \ll 1$ takes the form $\kappa_p \approx (1 + \chi^2)\Tau$.  See \Fref{secondary-fig-small-k} for a plot of the growth rate curve as a function of $\kappa(\chi)$.  We take $\Tau = 1$ and $k_0 = 0.3$ and vary $\chi$.  For small $\chi$, instability is observed only for $k_x < k_0$ and gradually weakens as $\chi$ is increased from zero.  We observe absolute stability at about $\chi = 0.7$.  As $\kappa_p$ is increased further, the instability returns and presents at $k_x > k_0$ for sufficiently large $\kappa_p$.  For large $\kappa_p$, the  growth rate peaks at $k_x > k_0$ (and also develops very fine-scale structure in the $y$-dependence of its eigenmode) signaling a reversal in spectral transfer direction like the one observed for $k_0 \gg 1$ in \citet{plunk-prl}.

\subsection{$k_0 < 1$ (modified Boltzmann response)}\label{secondary-mod-response-sec}

Taking $\Tau = \tilde{\tau} =  \tau(1 - \delta(k_y))$ as in \Sref{zonal-sec}, we once more examine the case of a modified Boltzmann response for electrons.  This is the case corresponding to ion Larmor scale turbulence set with an equilibrium magnetic field lying in closed flux surfaces.  The modified response introduces anisotropy to the problem, so the direction of ${\bf k}_0$ is now important.

\begin{figure}
\includegraphics[width=0.5\textwidth]{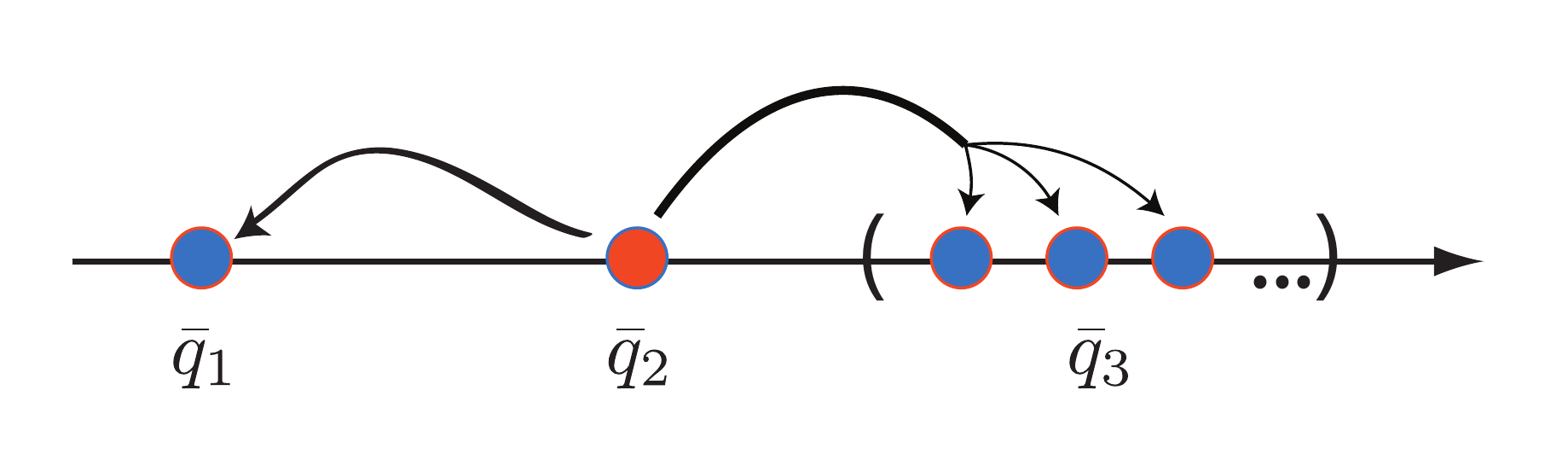}
\caption{The energy flow associated with the growth of a secondary instability must satisfy Fjortoft's constraint.  The primary mode (red dot) behaves as an energy source for the harmonics of secondary mode (blue dots).  This leads to the instability criterion $\bar{q}_1 < \bar{q}_2$ or equivalently $k_x^2 < \tau + k_0^2$.}
\label{GHM-sec-fjortoft-fig}
\end{figure}

It is informative to briefly revisit the simple case of the GHM equation.  Let us consider \Eref{GHM-eqn}, but neglect the linear terms (\ie $v_{*} = L_{*} = 0$).  If we take an initial condition with wavenumber in the y-direction, $\varphi_p = \e^{i k_0 y} + \mbox{c.c.}$, and consider perturbations about that, we find unstable modes of the form $\varphi_s = \e^{i (k_x x - \omega_s t)} \sum_n \hat{\varphi}_n \e^{i n k_0 y }$, which are unstable for $k_x^2 < \tau + k_0^2$.  There are a number of ways of deriving this instability criterion but we find an elegant method in terms of the Fj{\o}rtoft analysis of \Sref{spectral-redist-sec}.  See \Fref{GHM-sec-fjortoft-fig}.  They key idea is that the three-scale result of Equations \eref{fjortoft-type-eqn-1}-\eref{fjortoft-type-eqn-2} is actually applicable to any three sets of contiguous scales, given that the energy changes at scales within a set are all the same sign.  One simply replaces $q_1$, $q_2$ and $q_3$ with the energy-weighted averages for each of the three sets, $\bar{q}_1$, $\bar{q}_2$ and $\bar{q}_3$.  Recalling the definition of $\qGHM$ given in \Eref{GHM-spectral-constraint} we note that the harmonics of the perturbation $\varphi_s$ have $\qGHM(k_x \hat{\bf x} + n k_0 \hat{\bf y}) = \tau(1-\delta(n)) + k_x^2 + n^2 k_0^2$.  Let us take the first set to be the zeroth harmonic of the secondary mode, $\bar{q}_1 = \qGHM(k_x \hat{\bf x}) = k_x^2$, the second set to be the primary mode, $\bar{q}_2 = \qGHM({\bf k}_0) = \tau + k_0^2$, and the third to be all the remaining harmonics of the secondary mode, \ie $\bar{q}_3$ is the energy-weighted average of $\qGHM$ for remaining harmonics -- we do not need to evaluate this explicitly but just note that this quantity must be at least as big as the first harmonic, \ie $\bar{q}_3 \geq \qGHM(k_x \hat{\bf x} + k_0 \hat{\bf y}) = \tau + k_x^2 + k_0^2$.   By Fj{\o}rtofts argument, it is a simple consequence of energy and enstrophy conservation that only an intermediate scale can be a source to other scales involved in energy exchange.  Thus instability can only occur if the zeroth harmonic of $\varphi_s$ is at larger effective scale than $\varphi_p$ and so the instability criterion is $\bar{q}_1 < \bar{q}_2$ or $k_x^2 < \tau + k_0^2$.

Now if we take the initial condition to be zonal, \ie $\varphi_p = \e^{i k_0 x} + \mbox{c.c.}$, we may apply the same logic to show that perturbations are absolutely stable for $k_0 < \tau$.  This is because the mode $\varphi_s = \e^{i (k_y y - \omega_s t)} \sum_n \hat{\varphi}_n \e^{i n k_0 x }$ is composed of harmonics that all have $\qGHM({\bf k}) > \qGHM({\bf k}_0)$ since the zonal components represent the largest effective scale of the system.  We will find below that the strict stability of zonal modes is not enforced for gyrokinetics: instability can indeed occur if the zonal mode has sufficiently large $\kappa$ factor.

Let us take another detour and examine the instability driven by a poloidal ${\bf k}_0 = (0, k_0)$ mode (secondary instability) using our gyrofluid model; the fully gyrokinetic version of this instability has been studied and was not found to be especially sensitive to kinetic effects \citep{plunk-pop}.  Nonlinear transfer function plots from the numerical simulation of \citet{banon-prl} suggests that part of the electrostatic energy transfer is made via local inverse cascade.  We believe this part is composed of non-zonal fluctuations interacting to build up the zonal component; this is indeed what was observed for the gyrofluid simulations presented in \Sref{gf-sec}.  Thus the ``zeroth-order'' model (Equations \eref{nzo-phi-eqn-0}, \eref{zon-phi-eqn-0} and \eref{T-0-eqn}) derived in \Aref{lw-app} should be valid for the calculation of this instability as it is derived in the $k^2 \ll 1$ limit under the assumption of local interactions.  We derive the growth rate of the instability exactly in \Aref{zo-GF-secondary-app}.  We find that the growth rate of the mode is proportional to the factor $\sqrt{1 - r - r^{*}}$ where $r = T_{\perp 0}/\varphi_0$ is the ratio of the complex amplitudes of the temperature and potential of the primary mode.  This factor does show that the  relative amplitude and phasing of the primary mode can have a significant effect on the secondary mode.  Note that the mode is most unstable when the real part of $r$ is large and negative
  
Now we return to the fully gyrokinetic calculation.  We focus on the instability driven by a zonal ${\bf k}_0 = (k_0, 0)$ mode (tertiary instability).  As reported by \citet{rogers-prl} the tertiary instability is very sensitive to the presence of non-density moments in the zonal mode (they study temperature fluctuations).  Let us consider an initial condition composed of the zonal mode given by Equations \eref{gp_p1-eqn} and \eref{phip_p1-eqn} with ${\bf k}_0 = (k_0, 0)$.

\begin{figure}
\includegraphics[width=0.95\columnwidth]{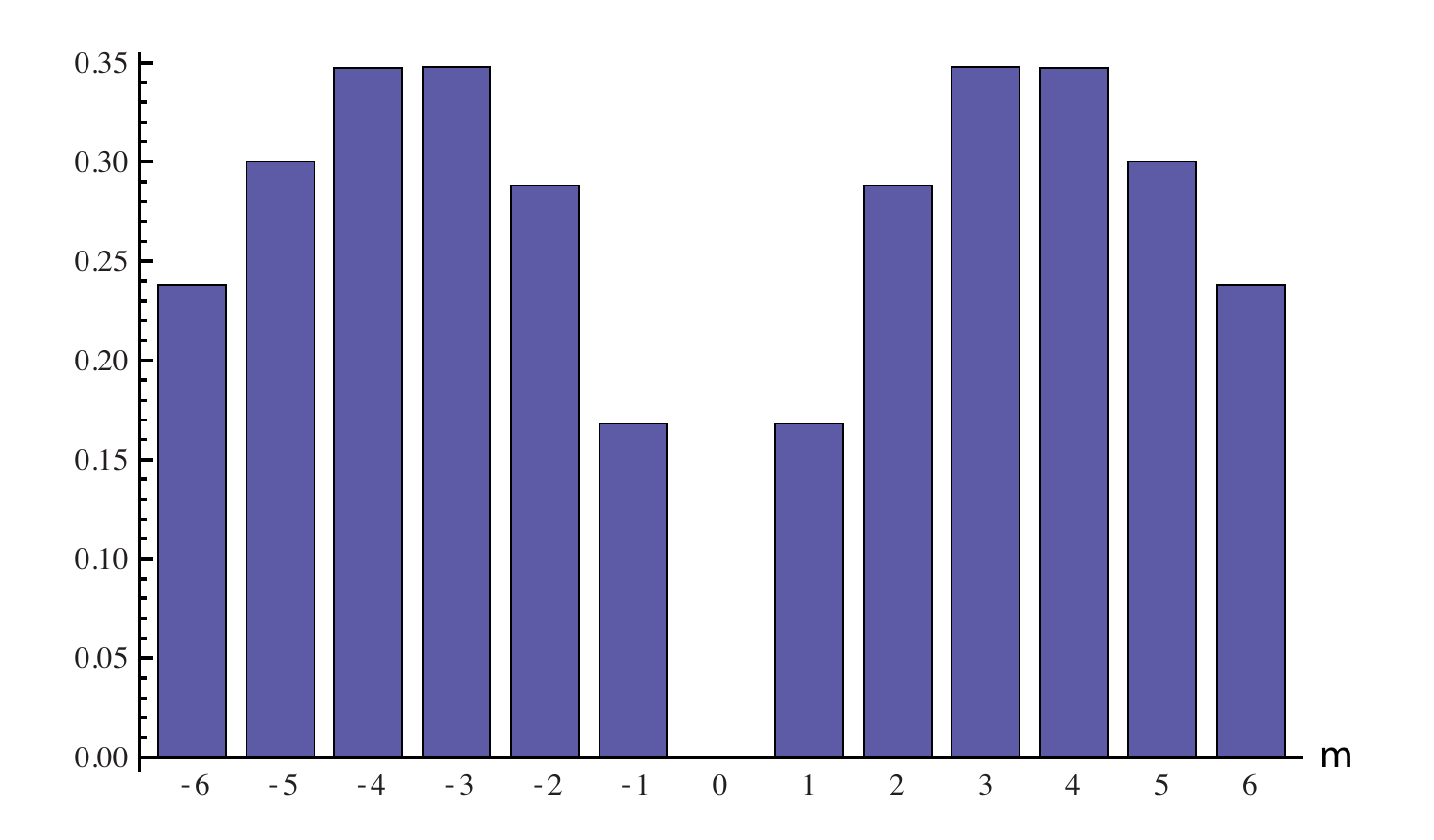}
\caption{The relative amplitude of harmonics for the gyrokinetic tertiary instability.  For this example we have $k_0 = 0.3$, $k_y = 0.43$ and $\chi = 1$.}\label{tertiary-harmonics-fig}
\end{figure}

The tertiary mode has the following features.  It is uniformly stable at sufficiently small $\kappa_p$.  Above a critical value of $\kappa_p$ (This value is numerically challenging to determine precisely due to the fact that a large number of harmonics are needed to resolve the mode near the critical $\kappa_p$; see \Fref{tertiary-harmonics-fig}.) an instability presents that generally peaks at $k_y > k_0$.  The gyrofluid model of \citet{rogers-prl} was found to peak at $k_y \sim \sqrt{k_0} \gg k_0$, which we find to be consistent with our observations but we note that the constant of proportionality is quite sensitive to $\kappa_p$.  The radial eigenfunction of the mode has no contribution from the $k_x = 0$ component.  So, the instability is composed exclusively of harmonics satisfying $|{\bf k}| > k_0$ and thus we expect long-wavelength zonal modes to release their energy in a forward cascade.  

\begin{figure}
\includegraphics[width=0.5\textwidth]{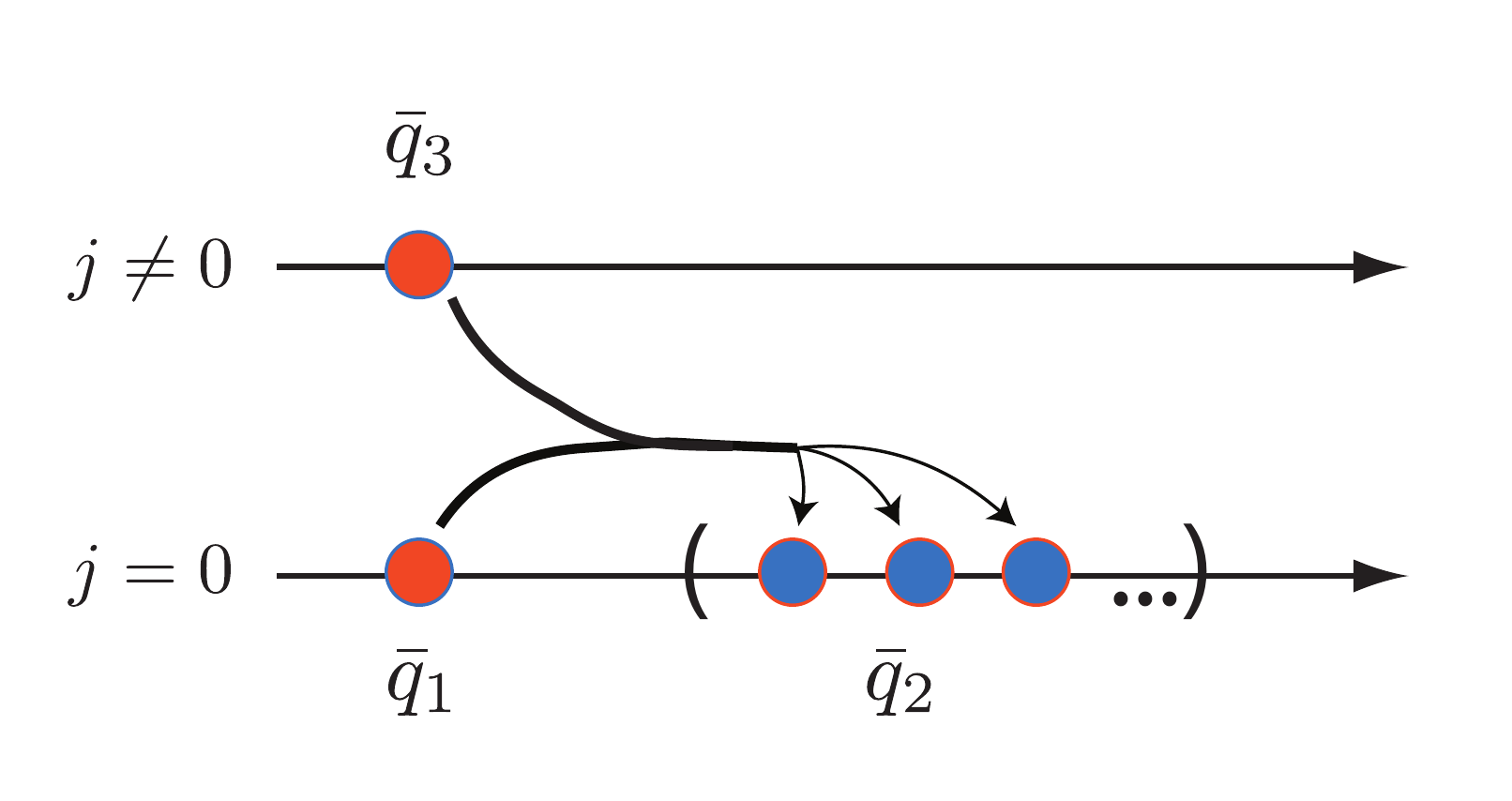}
\caption{The energy flow associated with the growth of a secondary instability must satisfy Fjortoft's constraint.  This means that non-density free energy must be present in the zonal mode (red dots) in order for it to function purely as a source of electrostatic energy for the tertiary mode (blue dots).  The zonal mode resides at $\bar{q}_1$ and the tertiary mode is at $\bar{q}_2$ (zeroth harmonic) and $\bar{q}_3$ (remaining harmonics).}
\label{GK-tertiary-fjortoft-fig}
\end{figure}

We can explain this behavior as follows.  As described in \Sref{zonal-sec}, zonal wavenumbers at $k < 1$ represent the largest effective scales of the system.  Thus, energy flow from the zonal component at $k < 1$ to the non-zonal component constitutes spectral transfer in the ``forward'' direction and generally needs a reservoir of non-density free energy to draw from.  This follows from the same kind of three-scale argument as made above for the GHM system; see \Fref{GK-tertiary-fjortoft-fig}.  If the zonal mode has only free energy in the density component, then the energy flow must occur by Fj{\o}rtoft-type transitions (see \Fref{transitions-fig-b}) and thus the zonal mode cannot serve purely as a source for the tertiary mode.  With sufficient free energy in the zonal mode, this restriction is lifted by the inclusion of Kinetic-type transitions (operating in the reverse sense of \Fref{transitions-fig-b}).  Furthermore, the instability must have $k \gg k_0$ because the FLR terms are required in the $k^2 \ll 1$ limit in order for electrostatic energy to flow from zonal to non-zonal components; this is explained in detail in \Sref{lw-app}.

\section{Direct Numerical Simulations}\label{numeric-sec}
\subsection{Method}

Our simulations of decaying gyrokinetic turbulence were done using the MPI-parallelized f95 gyrokinetic code AstroGK \citep{numata}.  AstroGK solves initial value problems of the gyrokinetic equation, \Eref{gyro-g} (with a collisional term added to the right-hand side), in the straight slab plasma with periodic boundary conditions perpendicular to a uniform background magnetic field in the $z$-direction.  To keep the problem two-dimensional, uniformity is enforced along $z$, \ie $\partial/\partial z = 0$.  The system size is $L_x = L_y = 2\pi$ in order to focus on sub-Larmor scales.  The highest resolutions employed in our runs are $256^2$ in position space and $192^2$ in velocity space, respectively, which required about 20 hours using 9,216 processor cores.

Initial conditions are constructed in Fourier space using the following functions.  A density component may be constructed by using

\begin{equation}
\hat{g}_0({\bf k}) = a_0({\bf k}) \frac{k^2}{k_0^2} \exp \left[ - \left(\frac{k-k_0}{k_{\rm w}} \right)^2 \right] J_0(kv) \, e^{-v^2}.
\end{equation}

\noindent This corresponds to free energy focused narrowly about the diagonal component $p = k_0$.  Non-density components are introduced using the initial condition

\begin{equation}
\hat{g}_1({\bf k}) = \frac{k^2}{k_1^2} \exp \left[ - \left(\frac{k-k_1}{k_{{\rm w}1}} \right)^2 \right] \,\frac{e^{-v^2}}{N_{\rm rand}}\sum_{j=1}^{N_{\rm rand}} a_j({\bf k}) \sqrt{|p_j|} J_0(|p_j|v).
\end{equation}

\noindent For each ${\bf k}$, the complex coefficient $a_0$ is chosen with a random phase and random amplitude from $(0, A_0)$.  The coefficients $a_j$, $j > 0$ are also randomly phased with amplitudes chosen from $(0, A_1)$.  The $p_j$'s are chosen randomly from the normal distribution of average $\bar{p}$ and dispersion $p_{\rm w}$.  We combine the two functions for a general initial condition

\begin{equation}
\hat{g}({\bf k}) = \hat{g}_0 + \hat{g}_1
\end{equation}

\noindent We focus on three main cases: the small $\kappa/k_0^2$ case (see \Fref{etrans-fig-a}) uses $k_0=40$, $k_{\rm w}=1$, $A_0 = 1$ and $A_1 = 0$.  The medium $\kappa/k_0^2$ case (see \Fref{etrans-fig-b}) has $k_1=20$, $k_{{\rm w}1}=8$, $N_{\rm rand} = 50$, $A_0 = 0$, $A_1 = 1$, $\bar{p}=20$ and $p_{\rm w} = 4$.  The large $\kappa/k_0^2$ case (see \Fref{etrans-fig-c}) has $k_0=5$, $k_{\rm w}=1$, $k_1=20$, $k_{{\rm w}1}=5$, $N_{\rm rand} = 50$, $A_0 = 0.04$, $A_1 = 1$, $\bar{p}=0$ and $p_{\rm w} = 3$.  

\subsection{Spectral Evolution}

\begin{figure}
\begin{center}
\subfloat{\label{ex-spectra-fig-a1}}\subfloat{\label{ex-spectra-fig-a2}}\subfloat{\label{ex-spectra-fig-b1}}\subfloat{\label{ex-spectra-fig-b2}}\subfloat{\label{ex-spectra-fig-c1}}\subfloat{\label{ex-spectra-fig-c2}}
\includegraphics[width=\columnwidth]{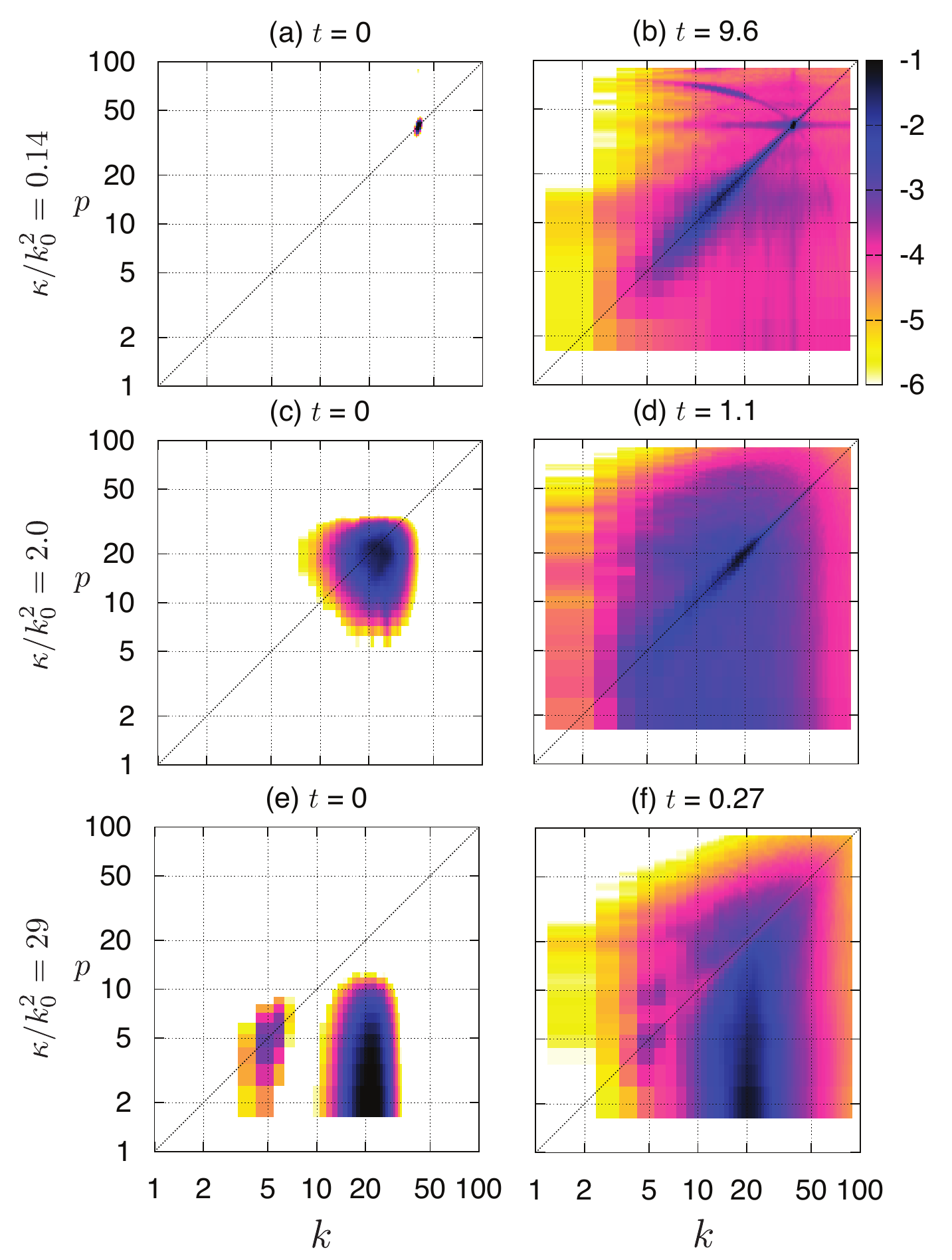}
\caption{Example spectral distributions $\log_{10} [\W(k,p)/\W]$ and initial evolution for several values of $\kappa/k_0^2$, with $\kappa = \W/E$.  Diagonals marked by dotted lines.  Figure reproduced from \citet{plunk-prl}.}
\label{ex-spectra-fig}
\end{center}
\end{figure}

\begin{figure}
\begin{center}
\subfloat[$\kappa/k_0^2 = 0.14$]{\label{espec-t-fig-a}\includegraphics[width=0.504\columnwidth]{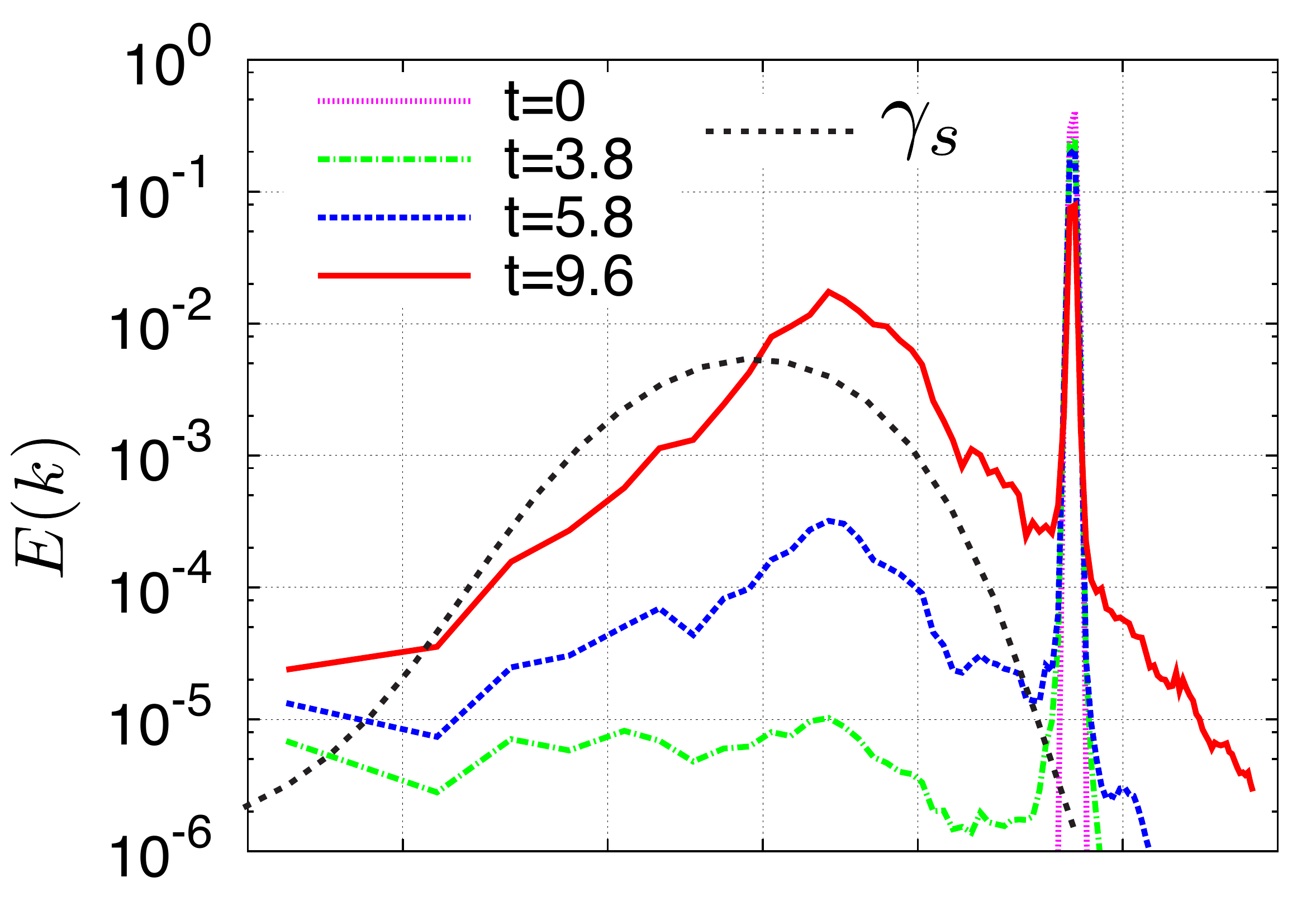}}
\subfloat[$\kappa/k_0^2 = 2.0$]{\label{espec-t-fig-b}\includegraphics[width=0.496\columnwidth]{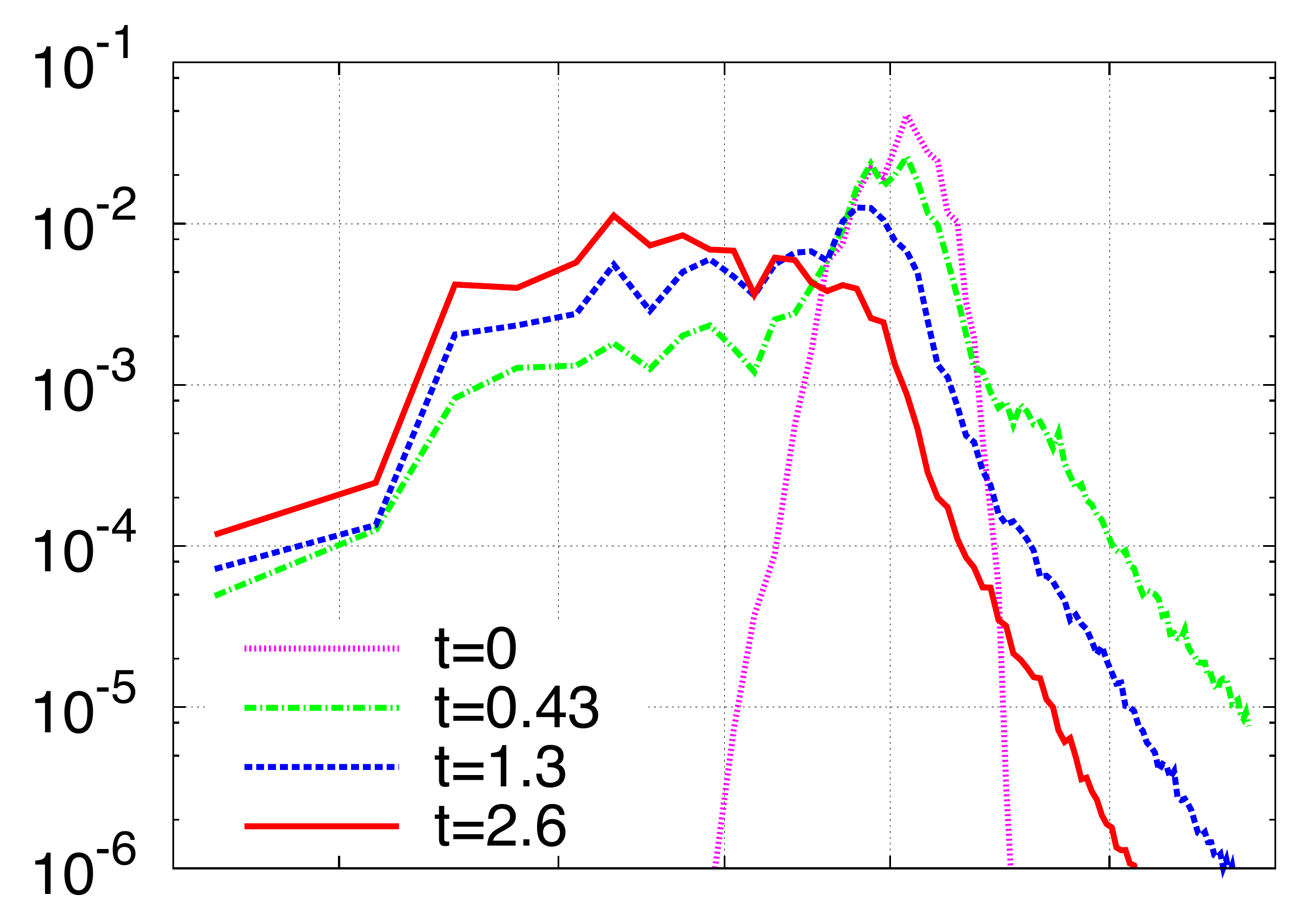}}\\
\subfloat[$\kappa/k_0^2 = 29$, $k_d = 5$, $k_0 = 20$]{\label{espec-t-fig-c}\includegraphics[width=0.505\columnwidth]{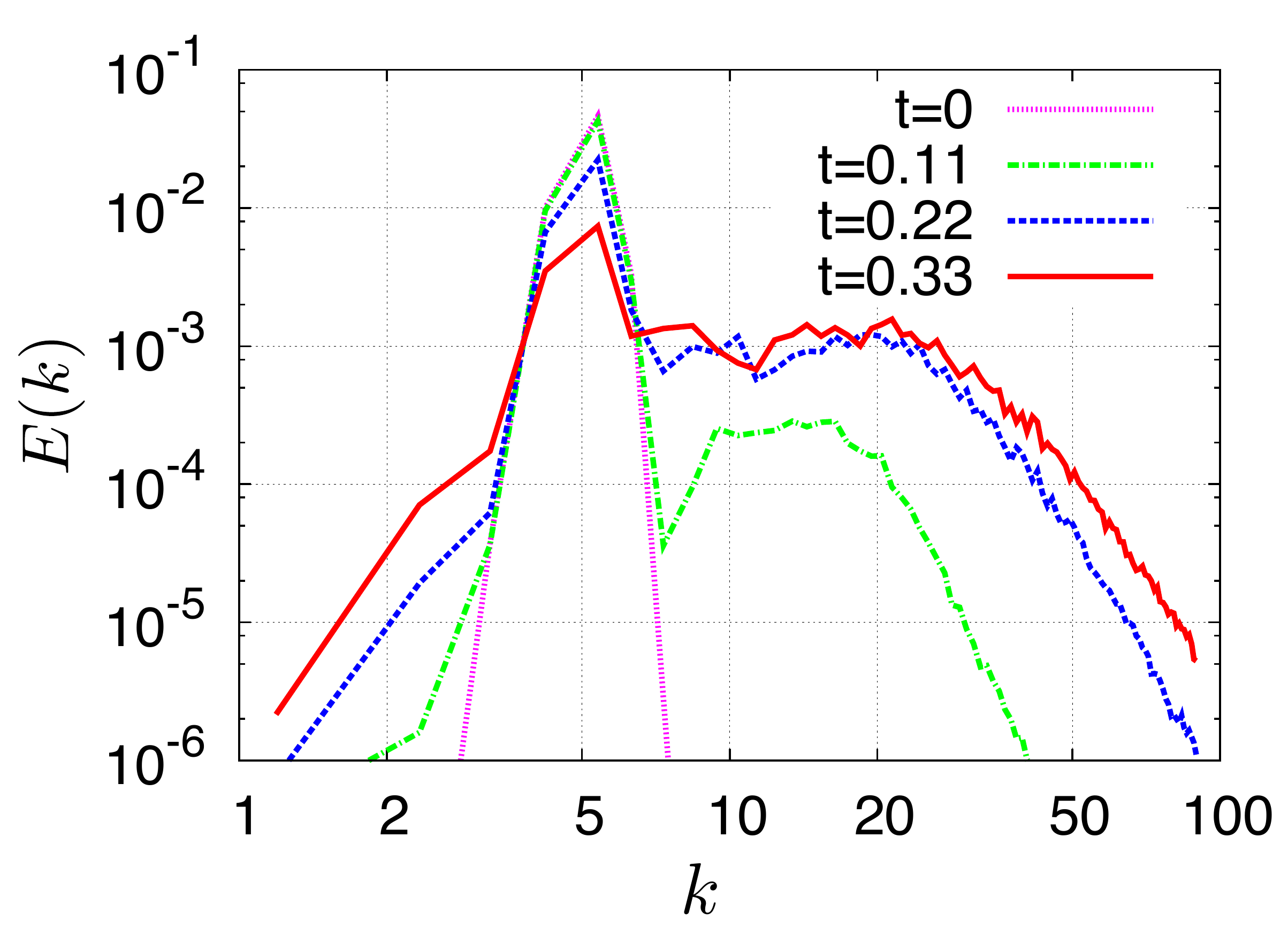}}
\subfloat[$\kappa/k_0^2 = 24$, $k_d = k_0 = 5$]{\label{espec-t-fig-d}\includegraphics[width=0.495\columnwidth]{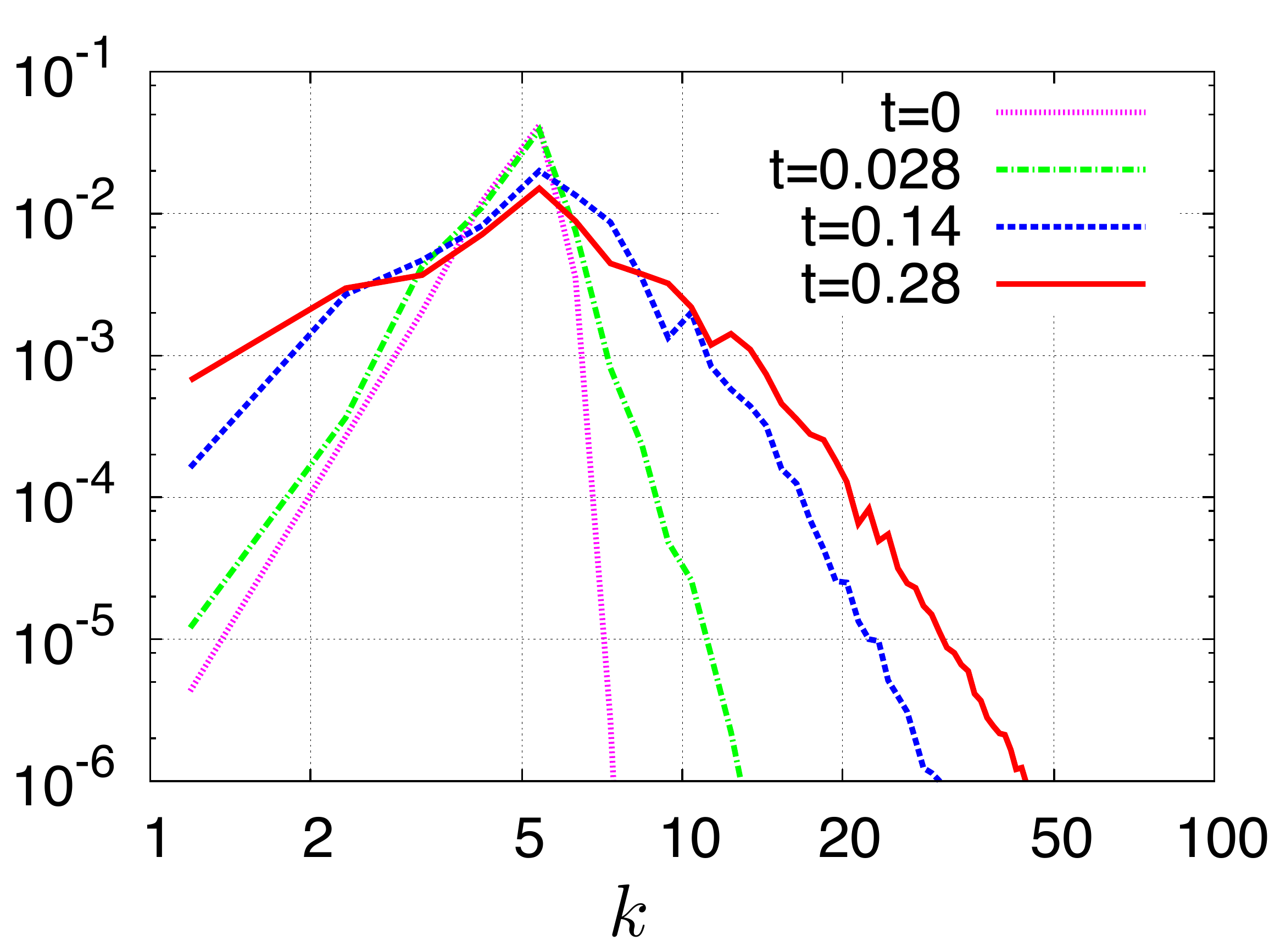}}
\caption{Evolution of the electrostatic energy spectrum.  Figure reproduced from \citet{plunk-prl}.}
\label{espec-t-fig}
\end{center}
\end{figure}

We focus on the sub-Larmor limit of \citet{plunk-prl}.  We adopt the conventions of that paper in this section: the free energy quantity is $\W = \frac{2\vcut^{(0)}}{(1 + \Tau)} W_{g1}$, and the spectral representation is the simplified Bessel series of \Eref{approx-bessel}.  We reproduce the main figures of \citet{plunk-prl} in \Fref{ex-spectra-fig} and \Fref{espec-t-fig}.

The initial ($t=0$) spectral density is concentrated on the diagonal for the small $\kappa / k_0$ case (see \Fref{ex-spectra-fig-a1}), broadened for the medium $\kappa / k_0$ case (see \Fref{ex-spectra-fig-b1}) and then further extended to the off-diagonal components for the large $\kappa/k_0$ case (see \Fref{ex-spectra-fig-c1}).

As time proceeds, the first two cases ($\kappa / k_0 \lesssim 2$) show inverse cascade along the diagonal; however, the largest $\kappa$ run ($\kappa / k_0^2=29$) shows the opposite behavior.  In this case, the initial spectra around $k \sim 20$ does not contribute a significant density component.  At a later time shot, $t = 0.27$ (see \Fref{ex-spectra-fig-c2}), the free energy becomes distributed in $k$-$p$ plane, but it has a much lower density on the strip along the diagonal as required for the conservation of $E$.
The initial diagonal component around $k \sim p \sim 5$ is the one that constitutes initial $E$, but not at a sufficient magnitude to support the conventional inverse cascade behavior.  Thus the peculiar behavior of cascade reversal may be attributed to electrostatic energy ``starvation.''  That is, the system is forced into a new mode of operation for lack of a basic resource.

\Fref{espec-t-fig} shows the time evolution of the spectral density of $E(k)$, from which the locality and direction of the nonlinear interaction may be inferred.  As shown in \Fref{espec-t-fig-a}, there is a secondary peak around $k \sim 15$ at $t=5.8$, which is separated from the initial one at $k =40$.  This shows nonlocal interaction, which is expected from the linear instability whose growth rate is indicated with the dotted curve; see \Sref{secondary-sec-subL}.  On the other hand, \Fref{espec-t-fig-b} does not show any secondary peak, and the gradual change of the spectrum indicates local inverse cascade.  \Fref{espec-t-fig-c} and \Fref{espec-t-fig-d} demonstrate nonlocal and local forward cascades of $E$.  The initial $E$ spectrum for the case of \Fref{espec-t-fig-c} is peaked at $k=5$, but a secondary peak appears around $k \sim 12$ at $t = 0.11$, which is separated from the initial one.  In \Fref{espec-t-fig-d} a mixture of forward transfer and inverse transfer of $E$ is observed.

We can analyze spectral transfer more quantitatively in terms of a transfer function for the electrostatic energy.  This construction is a particular sum of triad interaction terms that produces a function having two arguments, a ``from'' wavenumber and a ``to'' wavenumber.  There is no unique way to construct this two-scale transfer function (some authors prefer to examine the three wavenumber triad interaction terms directly) but the definition we employ is standard and also a ``natural'' definition as we explain below; see also \citet{banon-prl} and \citet{nakata} who use this type of transfer function to study simulations of tokamak turbulence.

We first introduce a shell filter in Fourier space.  Following \citet{tatsuno-jpf}, we define the shell-filtered electrostatic field as

\begin{equation}
\varphi_{K} = \sum_{{\bf k} \in \mathcal{K}} \e^{i {\bf k}\cdot {\bf r}} \hat{\varphi},
\end{equation}

\noindent where $\mathcal{K} = \{{\bf k} : K - 1/2 \leq |{\bf k}| < K + 1/2\}$.  Likewise, we may define the shell-filtered distribution function

\begin{equation}
g_{K} = \sum_{{\bf k} \in \mathcal{K}} \e^{i {\bf k}\cdot {\bf R}} \hat{g}.
\end{equation}

\noindent The electrostatic energy at shell $K$ is defined

\begin{equation}
E_K = \frac{1}{2}\sum_{{\bf k} \in \mathcal{K}}\frac{2\pi}{\beta}|\hat{\varphi}|^2.
\end{equation}

\noindent Multiplying the collisionless gyrokinetic equation, \Eref{gyro-g} with the collision operator neglected, by $\gyroavg{\varphi_K}$ and integrating over ${\bf R}$ and $v$, one obtains an evolution equation for $E_K$

\begin{equation}
\frac{d E_K}{dt} = \sum_{Q, P}T^{(E)}(P,Q; K),
\end{equation}

\noindent where we have expanded $\varphi = \sum_Q \varphi_Q$, $g = \sum_P g_P$ and introduced the three-argument electrostatic energy transfer rate $T^{(E)}(Q,P; K)$, defined

\begin{equation}
T^{(E)}(P, Q; K) = -\int vdv \frac{d^2{\bf R}}{V} \gyroavg{\varphi_K}\poiss{\gyroavg{\varphi_Q}}{g_P}.\label{tri-scale-transfer-fcn}
\end{equation}

\noindent This function represents the rate of change of energy in shell $K$ due to three-wave interactions involving wavenumbers contained in shells $P$ and $Q$.  The transfer function $T^{(E)}(P, Q; K)$ is quite general (we need not even partition the spectral domain using shells) but we note that it does not satisfy symmetry under exchange of $P$ and $Q$ like the triad interaction function of \cite{nakata}, though such a symmetrized transfer function can be constructed from it.  For our purposes, the deficiency in $T^{(E)}(P, Q; K)$ is that there is no clear identification of a source and sink for the energy, just three interacting shells.  It turns out that by requiring three simple properties a two-scale transfer function $T^{(E)}(Q, K)$ can be uniquely determined.  These properties are

\begin{itemize}
\item Antisymmetric: $T^{(E)}(Q, K) = -T^{(E)}(K, Q)$
\item Complete: $\frac{dE_K}{dt} = \sum_Q T^{(E)}(Q, K)$
\item Literal: $T^{(E)}(Q, K) = \displaystyle{\sum_{P, S}} \eta^{Q}_{P, S, K}T^{(E)}(P, S; K)$
\end{itemize}

\noindent The first condition is a basic requirement of energy conservation under shell-to-shell transfer: energy accumulates in shell $K$ due to loss at shell $Q$.  The second condition is just the requirement that when the transfer function is summed over all shells $Q$, nothing that contributes to the energy change at shell $K$ is missing.  The third condition is that the transfer function should be constructed only out of the three-shell transfer terms explicitly present in equation for the evolution of $E_K$, \ie those terms defined by \Eref{tri-scale-transfer-fcn}.  Note that if this last condition is relaxed, a more general class of transfer functions $T^{(E)}(Q, K)$ can be constructed as a sum of terms $T^{(E)}(P, S; R)$ (that is, the final index is not fixed to be $K$) \citep{carati}.  These three conditions are satisfied by simply summing $T^{(E)}(P,Q; K)$ over the index $P$, \ie by taking $\eta^{Q}_{P, S, K} = \delta_{Q, S}$.  Thus we arrive at the definition for the electrostatic energy transfer rate from shell $Q$ to shell $K$,  $T^{(E)}(Q, K)$:

\begin{equation}
T^{(E)}(Q, K) = -\int vdv \frac{d^2{\bf R}}{V} \gyroavg{\varphi_K}\poiss{\gyroavg{\varphi_Q}}{g}.
\end{equation}

\begin{figure}
\begin{center}
\subfloat[$\kappa/k_0^2 = 0.14$]{\label{etrans-fig-a}\includegraphics[width=0.504\columnwidth]{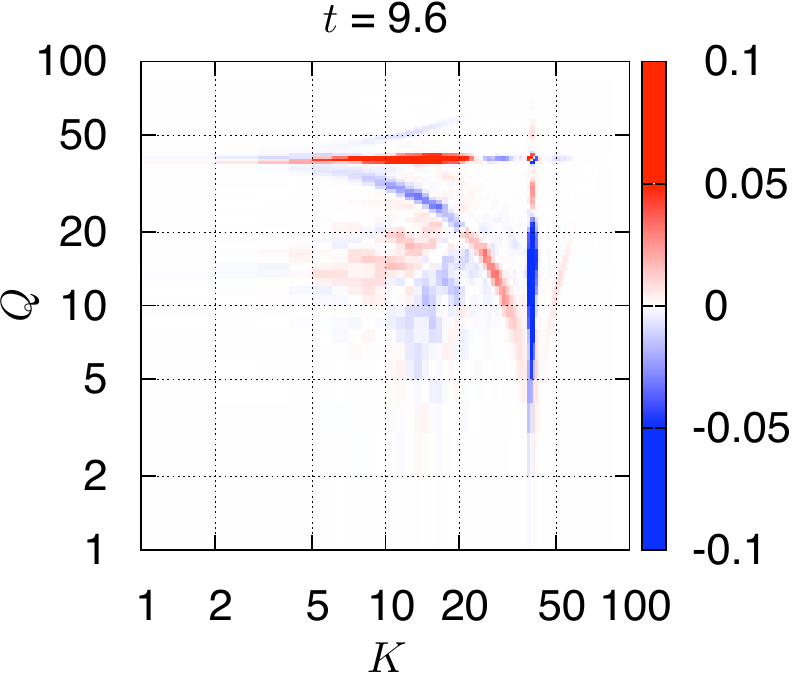}}
\subfloat[$\kappa/k_0^2 = 2.0$]{\label{etrans-fig-b}\includegraphics[width=0.496\columnwidth]{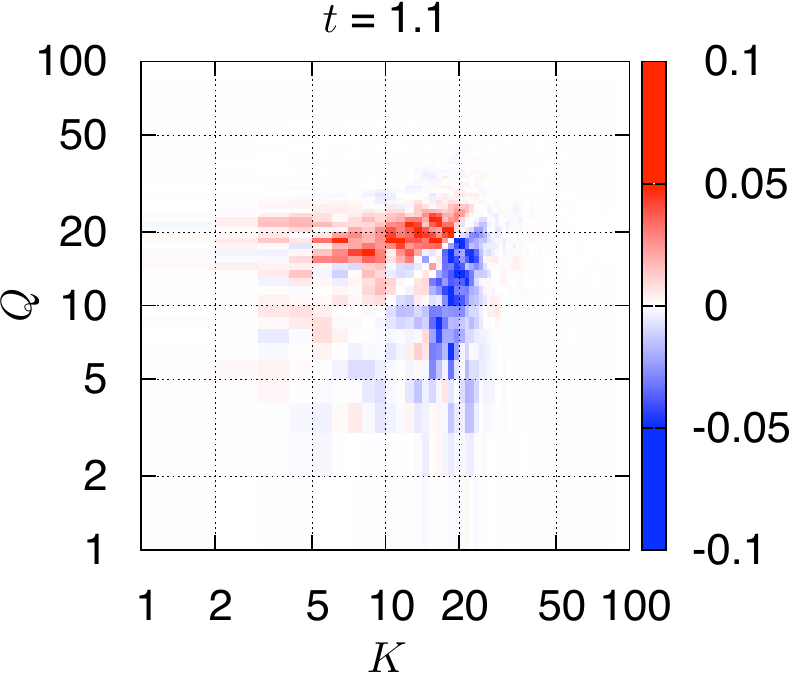}}\\
\subfloat[$\kappa/k_0^2 = 29$]{\label{etrans-fig-c}\includegraphics[width=0.505\columnwidth]{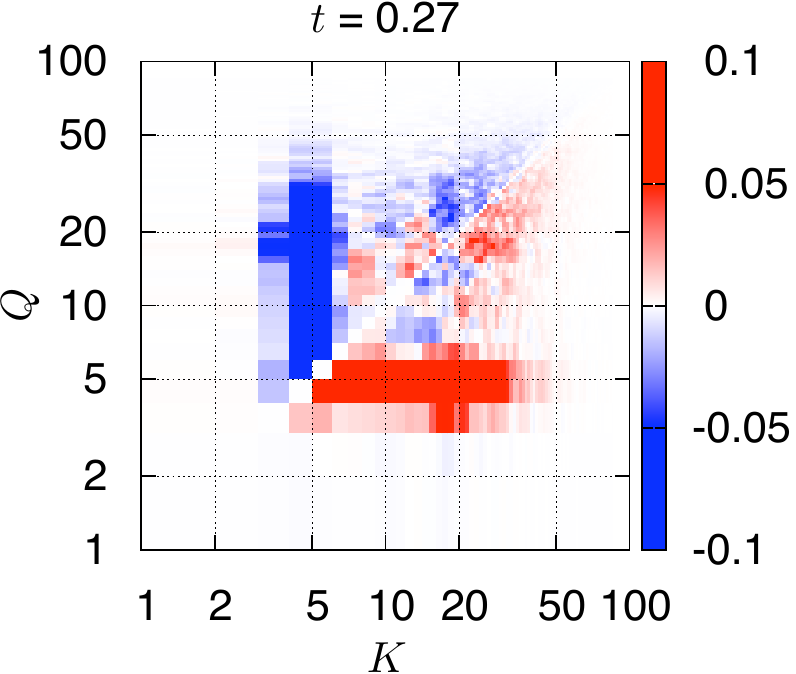}}
\caption{Nonlinear transfer function $T^{(E)}(Q, K)$ measures the instantaneous rate of nonlinear transfer of electrostatic energy from shell $Q$ to shell $K$.}
\label{etrans-fig}
\end{center}
\end{figure}

\noindent As shown in \Fref{etrans-fig}, we have computed the spectral transfer function for the cases studied in \citet{plunk-prl}.  These correspond to three regimes: (1) nonlocal inverse transfer (2) local inverse transfer and (3) forward transfer.  These transfer plots offer a direct glimpse of the energy flows between different scales and confirm that the spectral evolution observed by \citet{plunk-prl} was indeed due to nonlinear interaction.

\section{Discussion}\label{discussion-sec}

Let us review the key questions posed by this work and discuss the results we find.  As argued in \Sref{gk-energy-sec}, the free energy and electrostatic energy are measures that constrain gyrokinetic turbulence, with the former admitting some freedom in the exact definition.  The formulation of a spectral representation (see \Sref{spectral-sec}) leads to a precise statement of this constraint via \Eref{fjortoft-constraint}.  We then are able to systematically extend the arguments of Fj{\o}rtoft to establish the basic mechanism that governs the direction of spectral energy transfer.  Building on the work of \citet{plunk-prl}, we show that our constraint supports inverse cascade of electrostatic energy (spontaneous transfer to large scales), but we also find that there is additional freedom that allows for reversal of the cascade direction.  As evidence of these conclusions, we present numerical evaluation of the spectral transfer function, corresponding to the simulations of \citet{plunk-prl}; see \Sref{numeric-sec}.  By linearizing the nonlinear gyrokinetic equation about a monochromatic initial condition, we also calculate an instability that exhibits the same transition predicted by the Fj{\o}rtoft arguments; see \Sref{secondary-sec}.

We have also extended our analysis to scales larger than the Larmor radius (see \Sref{super-larmor-sec}) and have included the modified Boltzmann response, an essential effect for application of gyrokinetics to the closed flux surface geometries of magnetic confinement experiments like tokamaks and stellarators.  

By attempting to derive a simple fluid reduction of the gyrokinetic system (see \Aref{lw-app}) we arrive at an important conclusion, namely that the long-wavelength limit is singular in the nonlinear interactions between fluctuations and zonal flows.  Thus the regulation of zonal flows by turbulence appears to be an inextricably kinetic phenomenon.  That is, nonlinear phase mixing persists in the long-wavelength regime, acting as a channel for the transfer of free energy, and the turbulent viscosity on zonal flows can in principle be changed from positive to negative by the same basic mechanism as that which caused the ``cascade reversal'' observed by \citet{plunk-prl}; we demonstrate this reversal in \Sref{gf-sec} with simulations using a simple gyrofluid model.  These findings casts doubt on the validity of any gyrofluid model that treats FLR effects asymptotically.  However, it leaves open the possibility of a more sophisticated gyrofluid model with special care taken to model kinetic nonlinear interactions between zonal and non-zonal fluctuations.

In focusing on nonlinear spectral transfer, the Fj{\o}rtoft theory avoids questions involving sources and sinks (in fully gyrokinetic turbulence).  Does the dual cascade persist under these conditions?  Or, more generally, is the constraint on spectral transfer a useful tool for understanding the behavior of driven and dissipated gyrokinetic turbulence?  These questions are generally open, but there is some evidence suggesting that the answers are both yes.

There is numerical evidence that the sub-Larmor inertial-range theory \citep{schek-ppcf, plunk-jfm} gives the correct result in ion temperature gradient (ITG) tokamak turbulence \citep{banon-prl}.  There is also evidence that gyrokinetic turbulence \citep{barnes-parra} obeys an inertial-range scaling in the super-Larmor range, where most of the fluctuation energy and transport physics resides.  The scaling theory of \citet{barnes-parra} may be obtained by assuming a two-dimensional ($k_x$-$k_y$) cascade and adding to this the addendum that structure in the third dimension ($k_{\parallel}$) will develop so as to always keep the parallel streaming term large enough to compete with the nonlinear term.  However, despite the fact that this term remains large enough to be dynamically relevant, the effective damping of the electrostatic field by parallel phase mixing (linear Landau damping) seems to be sub-dominant to the nonlinear cascade rate in the inertial range.  If this were not the case, the energy spectrum would decay exponentially in a local cascade; see \ie \citet{howes2008}.  Indeed, numerical diagnostics by \citet{banon-pop} show the parallel damping to be localized to the outer scale, with the effective parallel damping rate decreasing with $k$.  Thus, at least in ITG tokamak turbulence, the electrostatic energy appears to retain the status of invariant in the inertial range (along with the free energy), and the inverse cascade of $E$ is indeed observed in simulations by \citet{banon-prl}.

In the present work, we have identified a control parameter for gyrokinetic turbulence, namely $\kappa$.  The role of $\kappa$ must be tested in realistic gyrokinetic systems, where characteristics of the magnetic field geometry and background plasma come into play.  We have focused on the nonlinear term in the gyrokinetic equation.  Other terms, associated with the background temperature and density gradients, and magnetic field geometry, may be considered generally as energy sources for the turbulence.  It is natural to then ask what the relative input of free energy to electrostatic energy is for these sources; in other words, at what ``level of $\kappa$'' is the turbulence driven.  One way to approach this question is to calculate the $\kappa$ factor for linearly unstable eigenmodes.  An evaluation of the local gyrokinetic dispersion relation for the ITG mode seems to indicate that $\kappa$ generally increases with the strength of the background ion temperature gradient, $1/L_T$.  Thus, in this case, the effect of increasing $\kappa$ may not be observed independently from the effect of increasing the growth rate of the instability.  But such a local calculation does not take into account the influence of plasma shape and the question of how $\kappa$ depends such system parameters is open.  

The control parameter $\kappa$ may help explain how the behavior of small-scale turbulence depends on large-scale features of plasma; it could also have predictive capability, especially for classes of plasma configurations that have a large parameter space, \eg stellarators, where optimizing the shape of the magnetic field for turbulence could lead to enhanced performance.  While it is natural to seek optimized magnetic configurations that minimize instability by reducing growth rates or increasing the domain of stability, the present study suggests that it is also desirable that instabilities present with optimal $\kappa$.  In the case of ion-scale turbulence (see \Sref{zonal-sec}), small $\kappa$ seems desirable to facilitate the generation of zonal flows and thereby reduce the amplitude of the turbulence in steady state.  On the other hand, in cases where the formation of large scale structures are seen to increase turbulent transport (\ie formation of streamers in ETG turbulence), one might seek to maximize $\kappa$.  A third possible case was identified by \citep{plunk-prl} where high-$\kappa$ fluctuations at sub-Larmor scales served as an energy sink for turbulence driven at large scales.  Could small-scale instabilities (like trapped particle modes) be optimized to damp the more deleterious large-scale turbulence?  Of course, further work is necessary to explore these ideas.

Other open areas include the generalization of the results of this work to systems with electromagnetic fluctuations and multiple kinetic species.  In principle, both of these extensions can be made in a straightforward fashion by considering the generalization of the electrostatic energy and free energy; the multi-species version of $E$ is given already in the appendix of \citet{schekochihin}.  Electromagnetic fluctuations will bring qualitatively new physics but it is still worthwhile investigating constrained spectral energy transfer in this context.

Perhaps the biggest open challenge is to make quantitative predictions about the spectrum of steady-state turbulence, driven by physical instabilities.  In particular we would like to know what sets the characteristic amplitude and dominant scale of the turbulence.  We would also like to characterize the anisotropy of the turbulence -- at what scale do zonal flows (or streamers, \etc) form and what is the relative amplitude of these structures compared with the non-zonal fluctuations?  These issues are for the most part beyond the scope of the present work.  Still, we can make some speculative remarks:  It seems that sufficiently small scales, to a certain extent, behave as if governed by inertial range physics.  As such, the spectrum of these fluctuations should fall off in $k$-space as a universal power law, set by the constancy of the nonlinear flux of free energy.  We are left with the challenge of describing the turbulence at the scale of energy input.  For ion-scale turbulence, it seems plausible, given the present study, that the amplitude of fluctuations relative to zonal flows depends on $\kappa$, as measured by the linear instability.  In other words, if the zonal flow amplitude is set by the usual saturation rule of $\gamma_E \sim \gamma_L$ (where $\gamma_E$ is the $E\times B$ shearing rate and $\gamma_L$ is the linear growth rate), then the amplitude of non-zonal fluctuations may be found to be proportional the zonal flow amplitude, with the constant of proportionality being an increasing function of $\kappa$.  This seems to be anecdotally supported by the gyrofluid simulations of \Sref{gf-sec}; note the trend of increasing non-zonal energy at constant zonal energy for super-critical $R_0$ shown in \Fref{gf-transition-fig}.  Perhaps similar conclusions could be drawn for other types of gyrokinetic turbulence, such as ETG or trapped-particle-driven turbulence.  It is our hope that future work will uncover such basic effects of constrained spectral energy transfer on turbulent steady states in gyrokinetics.


\section{Acknowledgements}

The authors gratefully acknowledge conversations with Alejandro Ba\~{n}on-Navarro, Greg Hammett, John Krommes, Alex Schekochihin and Jian-Zhou Zhu.  Our collaboration was generously supported by the Wolfgang Pauli Institute, the Leverhulme Trust Network, and the Isaac Newton Institute for Mathematical Sciences.  This work was supported by U.S. DOE Grant No. DESC0005106.

\appendix

\section{Completeness of $\mathcal{G}$}\label{G-completeness-sec}

To prove that $\mathcal{G}$ is complete in the weighted Hilbert space $\mathcal{L}^2((0,\infty),\e^{-u})$, we first note that a simple set of polynomials, being defined as a set $\{p_0, p_1, p_2, ... \}$ with $p_n$ a polynomial of order $n$, is a complete basis for this space; see \ie page 31 of \citep{higgins}.  Our set $\mathcal{G}$ differs from a simple set only in the function $P_0$, which is of course not a zeroth-order polynomial.  Thus, to show that $\mathcal{G}$ is complete, we must simply prove that a nonzero constant is within the span of $\mathcal{G}$.  We drop the superscript $(k)$ in what follows.

First let us define a set of polynomials $\{Q_1, Q_2, Q_3, ...\}$ such that $Q_n(u) = q_n + u^n$ where $q_n$ is a constant.  We construct this set in terms of $P_1$, $P_2$, \etc, as follows.  We take $Q_1 = P_1$, then it is easy to see that $Q_2$ can be constructed in terms of $Q_1$ and $P_2$.  Then $Q_3$ can be written in terms of $Q_1$, $Q_2$ and $P_3$ and so forth.  Now note the power series expansion for $J_0(x)$:

\begin{equation}
J_0(x) = \displaystyle{\sum_{m=0}^{\infty}}(-1)^m\frac{(x/2)^{2m}}{(m!)^2}.
\end{equation}

\noindent Using this and recalling the definition $P_0(u) = \hat{\Gamma}_0^{-1/2}(k) J_0(k\sqrt{2u})$, we can see that the quantity

\begin{equation}
c = \lambda_0 P_0(u) + \displaystyle{\sum_{m = 1}^{\infty}} \lambda_{m} Q_{m}(u),\label{c-eqn}
\end{equation}

\noindent is constant (independent of $u$) if we choose (for $m \geq 1$)

\begin{equation}
\lambda_{m} = -\lambda_0\hat{\Gamma}^{-1/2}_0(k)\frac{ (-k^2/2)^{m}}{(m!)^{2}}.
\end{equation}

\noindent To show that $c$ is nonzero we simply multiply \Eref{c-eqn} by $\e^{-u}P_0$ and integrate over $u$ (the $Q_m$ are orthogonal with $P_0$ by construction) to obtain $c = \lambda_0\hat{\Gamma}_0^{1/2}(k)/\hat{\Gamma}_0(k, 0)$, where the generalized function $\hat{\Gamma}_0(k_1, k_2)$ is defined

\begin{equation}
\hat{\Gamma}_0(k_1, k_2) \equiv \int_0^{\infty} v dv \;\e^{-v^2/2}J_0(k_1)J_0(k_2) = I_0(k_1 k_2)e^{-(k_1^2+k_2^2)/2},\label{gamma02-def}
\end{equation}

\noindent Note that $c/\lambda_0$ will become quite large at $k \gg 1$ because $\hat{\Gamma}_0(0,k)$ is exponentially small (see \Eref{gamma2-large}).  Thus, the set $\mathcal{G}$ may be a more practical representation for $k \lesssim 1$.

\section{Asymptotic Forms of Bessel Functions}\label{asymp-app}

The $n$th-order Bessel function of the first kind $J_n(x)$ has the following asymptotic form for $x \gg |n^2 - 1/4|$

\begin{equation}
J_n(x) \approx \sqrt{\frac{2}{\pi x}} \cos(x - n \pi/2 - \pi/4).\label{bessel-asymp}
\end{equation}

\noindent Using the identity $I_0(x) = J_0(i x)$, we can infer the large argument form of $I_0$ to be $I_0(x) \approx \e^x/\sqrt{2\pi x}$.  Applying this to the \Eref{gamma02-def}, we can obtain:

\begin{equation}
\hat{\Gamma}_0(k_1, k_2) \approx \frac{1}{\sqrt{2\pi k_1k_2}}\e^{-(k_1 - k_2)^2/2},\label{gamma2-large}
\end{equation}

\noindent and, since $\hat{\Gamma}_0(k) = \hat{\Gamma}_0(k, k)$, we also have

\begin{equation}
\hat{\Gamma}_0(k) \approx \frac{1}{k\sqrt{2\pi}}\label{gamma-large}
\end{equation}

For small argument, the asymptotic expressions for $J_0$ and $\Gamma_0$ are

\begin{equation}
J_0(x) \approx 1- \frac{1}{4}x^2 + \frac{1}{64}x^4 + ...\label{J0-lw-eqn}
\end{equation}

\noindent and

\begin{equation}
\Gamma_0(k) \approx 1 - k^2 + \frac{3}{4} k^4 + ...
\end{equation}

\section{Generalized Hasegawa-Mima Equation}\label{ghm-theory-app}

To illustrate a simple and familiar example of zonal flow generation by anisotropic inverse cascade, ket us consider a long-wavelength ($k^2 \ll 1$) limit of gyrokinetics that yields the HM equation, but with the modified electron response described in \Sref{zonal-sec}.  This is called the generalized Hasegawa Mima (GHM) equation \citep{smolyakov-prl, manfredi}.  We define the zonal component of $\varphi$ as the average over the system in the $y$-direction and the non-zonal part as the rest of $\varphi$.

\begin{eqnarray}\label{zon-op-def-eqn}
\zon{\varphi} &= \frac{1}{L}\int dy \varphi(y)\\
\nzo{\varphi} &= \varphi - \zon{\varphi}
\end{eqnarray}

\noindent The GHM equation can be derived in the limit $\tau \sim k^{2} \ll 1$.  It is written

\begin{equation}\label{GHM-eqn}
\partial_t(\tau \nzo{\varphi} - \nabla^2\varphi) + \poiss{\varphi}{\tau \nzo{\varphi} - \nabla^2\varphi}  + v_{*} \partial_y \nzo{\varphi}=  L_* \nzo{\varphi}
\end{equation}
  
\noindent where we have included a source term on the right-hand side that gives linear instability.  In the absence of this term, this system conserves two quantities, $\EGHM$ and $\ZGHM$, whose spectral densities are given

\begin{eqnarray}
\EGHM({\bf k}) &= \frac{1}{2} \left(\tilde{\tau} + k^2 \right)|\varphi({\bf k})|^2\\
\ZGHM({\bf k}) &= \frac{1}{2} \left( \tilde{\tau} + k^2 \right)^2|\varphi({\bf k})|^2
\end{eqnarray}

\noindent These spectral densities satisfy the constraint

\begin{equation}
\qGHM = \frac{\ZGHM({\bf k})}{\EGHM({\bf k})} =  \tilde{\tau} + k^2.\label{GHM-spectral-constraint}
\end{equation}

\noindent This ratio sets an effective wavenumber that corresponds to the square root of the quantity $\qGHM$.  Thus, the development of anisotropic flows in this system is not only due to the anisotropy of the drift waves (introduced by the term $v_{*} \partial_y \nzo{\varphi}$) but also by anisotropy in the nonlinearity (and nonlinear invariants).  This effect is very strong.  In fact, the inclusion of unstable linear modes in this model leads a system with pathological behavior, which can be seen directly from energy enstrophy balance equations as follows.  For the remainder of this section we drop the subscript on $\qGHM$.

Consider the following thought experiment.  Stationary turbulence of the driven GHM system must satisfies energy and enstrophy balances

\begin{equation}
\frac{dE}{dt} = \sum_{\bf k} \gamma({\bf k}) E({\bf k}) = 0,
\end{equation}

\noindent and

\begin{equation}
\frac{dZ}{dt} = \sum_{\bf k} q \gamma({\bf k}) E({\bf k}) = 0
\end{equation}

\noindent Now let us partition the spectrum using $q({\bf k})$:  Large-scale zonal flows reside at $q \leq \tau$.  We assume there is instability ($\gamma({\bf k}) \geq 0$) from $\tau$ (because non-zonal modes have $q > \tau$) to $q^{\prime}$ and damping ($\gamma({\bf k}) < 0$) at larger $q$.  The rate of change of energy at these scale ranges due to the linear source term can be expressed
 
\begin{eqnarray}
\varepsilon_Z &= \displaystyle{\sum_{{\bf k}:\; q \leq \tau}} \gamma({\bf k}) E({\bf k})\\
\varepsilon_p &= \displaystyle{\sum_{{\bf k}:\; \tau < q \leq q^{\prime}}} \gamma({\bf k}) E({\bf k}),\;\;\;\; \gamma \geq 0\\
\varepsilon_D &= \displaystyle{\sum_{{\bf k}:\; q^{\prime} < q}} \gamma({\bf k}) E({\bf k}),\;\;\;\; \gamma < 0.
\end{eqnarray}

\noindent Then by energy and enstrophy balance we have

\begin{equation}
\varepsilon_p + \varepsilon_Z + \varepsilon_D = 0,\label{ghm-E-injection-balance}
\end{equation}

\noindent and

\begin{equation}
q_p\varepsilon_p + q_Z\varepsilon_Z + q_D\varepsilon_D = 0,\label{ghm-Z-injection-balance}
\end{equation}

\noindent  where it can easily be shown that 

\begin{equation}
q_Z < q_p < q_D.
\end{equation}

\noindent A trivial solution of Equations \eref{ghm-E-injection-balance} and \eref{ghm-Z-injection-balance} is $\varepsilon_p = \varepsilon_Z = \varepsilon_D = 0$; this can be realized if neutrally stable zonal flows grow, by nonlinear interaction with non-zonal fluctuations, to large amplitude and quench non-zonal fluctuations, yielding a state of stationary zonal flows and no fluctuations.  Non-trivial solutions to Equations \eref{ghm-E-injection-balance} and \eref{ghm-Z-injection-balance} (\ie those with non-zero fluctuations) require $\varepsilon_p, \varepsilon_Z, \varepsilon_D \neq 0$.  This should all be quite familiar, following the discussion around Equations \eref{fjortoft-type-eqn-1}-\eref{fjortoft-type-eqn-2}.  By analogy with those results, we can conclude that if injection at the intermediate scales is non-zero, \ie $\varepsilon_p > 0$ (this must be true of any components in the drive range are at non-zero amplitude), then we must have damping at {\it both} the zonal scales $\varepsilon_Z < 0$ and the dissipation scales $\varepsilon_D < 0$.  This result should not be surprising.  Indeed, if energy injection remains positive, then it is expected to continually cascade to larger and larger scales unless there is some sink to absorb the flux.

This mechanism of zonal flow generation by inverse cascade makes the GHM equation a questionable candidate for modeling tokamak turbulence because linear processes other than collisional dissipation do not damp all zonal flows \citep{rosenbluth-hinton-residual}.  In gyrokinetic turbulence, finite Larmor radius (FLR) effects couple the zonal flow dynamics to fluctuations in the ion temperature, which break the enstrophy conservation satisfied by the GHM nonlinearity.  This opens up a channel for energy flow from the zonal flows back into the non-zonal fluctuations, which provides a nonlinear mechanism to limit the zonal flow amplitude and allows for a (non-trivial) saturated state to exist in absence of linear damping on the zonal flows.

\section{Two-dimensional gyrokinetics in the long-wavelength limit, and including zonal flows}
\label{lw-app}

In this Appendix, we examine the fluid moment hierarchy that arises in the long-wavelength limit of the two-dimensional gyrokinetic equation.  We find it useful to examine this in detail because an asymptotic limit can impose new features on the problem and these features may not be anticipated from a general understanding of the gyrokinetic system.  

An interesting feature of this limit is that it is quasi-singular in the sense that the electrostatic energy of non-zonal fluctuations is perfectly conserved at zeroth order (and the zonal flows give no contribution to energy).  One must include high order terms in order to to capture nonlinear phase mixing and the energetics of the zonal flows.  This is reminiscent of energy conservation in the gyrokinetic equation (for fluctuations), where a higher order term, the parallel nonlinear (PNL), must be included to give non-trivial energy conservation; see the discussion in \Sref{gk-energy-sec}.  In that case the PNL term is not dynamically relevant (but its appearance in the gyrokinetic equation may be justified by other considerations such as the value of conservation laws for certain numerical schemes).  Here, however, we argue that it is asymptotically consistent to include the additional FLR terms and that the energy balance that this captures is an important part of the dynamics.  This is due to the fact that the limit $k^2 \ll 1$ is singular in that ostensibly small terms are made relevant by nonlocal interactions.  

We define the small parameter $\epsilon_{lw} = k^2$ (which is distinct from and subsidiary to the gyrokinetic expansion parameter $\epsilon = \rho/L \ll \epsilon_{lw}$).  Let us expand the gyrokinetic equation (neglecting collisions), where we use the small-argument forms of $J_0$ and $\Gamma_0$ of \Aref{asymp-app}.  Including terms up to ${\cal O}(\epsilon_{lw}^2)$ we obtain \footnote{It may be worth noting that in this equation $\varphi$ and $g$ are evaluated at gyrocenter position ${\bf R}$ and $\bnabla$ operates in ${\bf R}$-space.  On the other hand, quantities in \Eref{qn-lw-eqn} are evaluated at particle position ${\bf r}$ and $\bnabla$ is there taken to operate in this space.}

\begin{equation}
\frac{\partial g}{\partial t} + \poiss{(1+ \frac{v^2}{4} \nabla^2 + \frac{v^4}{64}\nabla^4)\varphi}{g} = 0,\label{gk-lw-eqn}
\end{equation}

\noindent and \Eref{qn-g} becomes

\begin{equation}
\tau\nzo{\varphi} - (\nabla^2+\frac{3}{4}\nabla^4)\varphi = 2\pi \int_0^{\infty} vdv (1+ \frac{v^2}{4} \nabla^2 + \frac{v^4}{64}\nabla^4) g,\label{qn-lw-eqn}
\end{equation}

\noindent where the zonal and non-zonal components have been defined in \Eref{zon-op-def-eqn} and we have taken the modified Boltzmann response for electrons by setting $\Tau = \tilde{\tau}$ in the definition of $\beta$:

\begin{equation}
\beta({\bf k}) = \frac{2\pi}{1 + \tilde{\tau} - \Gamma_0(k)},
\end{equation}

\noindent where $\tilde{\tau} = \tau(1 - \delta(k_y))$ and $\delta$ is the discrete delta function having value $1$ when its argument is zero.  An important feature of this asymptotic limit is that there is a singular relationship between the zonal parts of $\varphi$ and $g$:

\begin{equation}
\zon{\varphi} \sim k^{-2} \int vdv \; \zon{g},
\end{equation}

\noindent which implies that the zonal component of $\varphi$ is due to small corrections in $g$:

\begin{equation}
\zon{\varphi}^{(0)} \sim \zon{g}^{(1)},\;\;\; \zon{\varphi}^{(1)} \sim \zon{g}^{(2)},\mbox{ \etc},\label{qn-singular-eqn}
\end{equation}

\noindent where we have expanded $\varphi = \varphi^{(0)} + \varphi^{(1)} + ...$ and $g = g^{(0)} + g^{(1)} + ...$ with superscripts indicating ordering in $\epsilon_{lw}$.  Because of the singular relationship between $\zon{\varphi}$ and $\zon{g}$, we will find the dynamics of $\zon{\varphi}$ at one order higher in the expansion of \Eref{gk-lw-eqn}.  

\subsection{Zeroth-order system}

The fluid moment hierarchy begins with equations for $\varphi$ found by integrating \Eref{gk-lw-eqn} over velocity and using \Eref{qn-lw-eqn}.  We find first the equation for $\nzo{\varphi}$ at dominant order

\begin{equation}
\frac{\partial \tau \nzo{\varphi}^{(0)}}{\partial t} + \poiss{\zon{\varphi}^{(0)}}{\tau \nzo{\varphi}^{(0)}} = 0.\label{nzo-phi-eqn-0}
\end{equation}

\noindent Then at next order,

\begin{equation}
\eqalign{
\frac{\partial}{\partial t}\tau \nzo{\varphi}^{(1)} &+ \frac{\partial}{\partial t}(-\nabla^2 \varphi^{(0)}) + \poiss{\varphi^{(1)}}{\tau \nzo{\varphi}^{(0)}}  \nonumber\\
&+ \poiss{\varphi^{(0)}}{\tau \nzo{\varphi}^{(1)} - \nabla^2\varphi^{(0)} - \nabla^2T_{\perp}^{(0)}} + \poiss{\nabla^2 \varphi^{(0)}}{T_{\perp}^{(0)}} \nonumber\\
&+ \nabla^2\poiss{\varphi^{(0)}}{T_{\perp}^{(0)}} = 0,}\label{phi-eqn-1}
\end{equation}

\noindent where $T_{\perp}$ is half the (ion gyrocenter) perpendicular temperature,

\begin{equation}
T_{\perp} = 2\pi \int vdv \frac{v^2}{4} g.
\end{equation}

\noindent Now we may obtain a dynamical equation for $\zon{\varphi}^{(0)}$ by averaging \Eref{phi-eqn-1}

\begin{equation}
\eqalign{\frac{\partial}{\partial t}(-\nabla^2 \zon{\varphi}^{(0)}) + \zon{\poiss{\varphi^{(0)}}{-\nabla^2\varphi^{(0)} - \nabla^2T_{\perp}^{(0)}}} + \zon{\poiss{\nabla^2 \varphi^{(0)}}{T_{\perp}^{(0)}}} \nonumber \\
+ \zon{\nabla^2\poiss{\varphi^{(0)}}{T_{\perp}^{(0)}}} = 0.}\label{zon-phi-eqn-0}
\end{equation}

\noindent The equation for $T_{\perp}^{(0)}$ is found directly taking from the zeroth order part of \Eref{gk-lw-eqn},

\begin{equation}
\frac{\partial T_{\perp}^{(0)}}{\partial t} + \poiss{\varphi^{(0)}}{T_{\perp}^{(0)}} = 0.\label{T-0-eqn}
\end{equation}

\noindent One is tempted to stop here.  Indeed, Equations \eref{nzo-phi-eqn-0}, \eref{zon-phi-eqn-0} and \eref{T-0-eqn} comprise a closed zeroth-order system (ZOS), \ie a system having dependence only on zeroth order quantities $\varphi^{(0)}$, $T_{\perp}^{(0)}$ and not coupling to any other moments in the fluid moment hierarchy.  However, such a set of equations has no mechanism for energy transfer {\it from the zonal flows to the non-zonal fluctuations}.  This can be demonstrated (though we do not do it here) by solving the tertiary instability problem \citep{rogers-prl} (which asks how non-zonal fluctuations can spontaneously grow in the presence of a large amplitude zonal flow).  Another way to see this is to note that the energy of $\nzo{\varphi}^{(0)}$ is conserved under interactions with $\zon{\varphi}^{(0)}$ as is evident from \Eref{nzo-phi-eqn-0}.  Indeed, the electrostatic energy under this expansion is at dominant order just the energy of the zeroth-order non-zonal field,

\begin{equation}
E \approx \frac{1}{2V}\int d^2{\bf r} \; \left[ \tau (\nzo{\varphi}^{(0)})^2 + 2 \tau \nzo{\varphi}^{(1)}\nzo{\varphi}^{(0)} + |\bnabla \varphi^{(0)}|^2 + ... \right ].\label{E-lw}
\end{equation}

\noindent In other words the ZOS conserves just the non-zonal energy (the first term in \Eref{E-lw}), so although the zonal flows (being energetically irrelevant) can grow (or be damped) under the influence of the non-zonal fluctuations, they cannot feed back in a way that causes significant growth or decay of those fluctuations.  (Indeed, \citet{diamond-zonal} refer to zonal flows as ``modes of minimal inertia.'')  In the ZOS, the effect of zonal flows on non-zonal fluctuations is only conservative shearing, as captured by Equations \eref{nzo-phi-eqn-0} and \eref{T-0-eqn}.  Note that whatever order the fluctuations are calculated (\ie $\varphi^{(0)}$, $\varphi^{(1)}$, $\varphi^{(2)}$, \etc), there will always be an energetically irrelevant part of $\zon{\varphi}$ that is, however, dynamically relevant.

However, the tertiary instability, by which energy flows from zonal to non-zonal fluctuations, has been shown to be an important mechanism for regulation of zonal flows \citep{rogers-prl}.  As the role of zonal flows is so central to the turbulent state (see for instance \citep{waltz-holland}), processes that regulate zonal flows will regulate the full turbulent state.  \footnote{Note that the ZOS may be adequate for regimes where the tertiary instability is not important, such as below the nonlinear critical gradient for ITG where zonal flows can completely quench the turbulence on timescales smaller than collisional damping.}  We will later give a simple two-field system that includes FLR terms that capture this tertiary ``energy channel'', but let us complete the computation of the gyrofluid system under the naive $k^2 \ll 1$ expansion to include the dynamics of first order fields.

\subsection{Higher-order systems}

\noindent Now we introduce some more fields in the moment hierarchy and then give equations for these quantities up to first order.  We define the next two perpendicular velocity moments of $g$ as

\begin{equation}
\chi = 2\pi \int vdv \frac{v^4}{8} g,
\end{equation}

\noindent and

\begin{equation}
\varpi = 2\pi \int vdv \frac{v^6}{16} g.
\end{equation}

\noindent Then we have

\begin{eqnarray}
&\frac{\partial T_{\perp}^{(0)}}{\partial t} + \poiss{\varphi^{(0)}}{T_{\perp}^{(0)}} = 0\label{psi-0-eqn}\\
&\frac{\partial \chi^{(0)}}{\partial t} + \poiss{\varphi^{(0)}}{\chi^{(0)}} = 0\label{chi-0-eqn}
\end{eqnarray}

\noindent So, at dominant order the moments $T_{\perp}^{(0)}$, $\chi^{(0)}$, \etc, cascade independently to larger $k$, though all moments higher than $T_{\perp}^{(0)}$ do so passively (without affecting $\varphi$).  Thus, at zeroth order, each moment has an independent quadratic invariant $V^{-1}\int d^2{\bf r} \left[(T_{\perp}^{(0)})^2, (\chi^{(0)})^2, ...\right]$, that contributes to the total free energy.  The mixing of these energy channels is obtained by including FLR terms.  Continuing the expansion we have

\begin{eqnarray}
&\frac{\partial T_{\perp}^{(1)}}{\partial t} + \poiss{\varphi^{(0)}}{T_{\perp}^{(1)}} + \poiss{\varphi^{(1)}}{T_{\perp}^{(0)}}  + \poiss{\nabla^2\varphi^{(0)}}{\chi^{(0)}/2}= 0,\label{psi-1-eqn}\\
&\frac{\partial \chi^{(1)}}{\partial t} + \poiss{\varphi^{(0)}}{\chi^{(1)}} + \poiss{\varphi^{(1)}}{\chi^{(0)}}  + \poiss{\nabla^2\varphi^{(0)}}{\varpi^{(0)}/2}= 0\label{chi-1-eqn},
\end{eqnarray}

\noindent and so forth.  Here the nonlinear phase mixing appears, opening a channel for the flow of free energy.  From this moment hierarchy we now wish to obtain a simple two-field system that conserves $W_{g0}$.  To do this, we employ the Laguerre polynomials as a basis for $g$ (see \Aref{Laguerre-appendix}) and truncate at second order in the Laguerre hierarchy.  Note that our solution for $\varphi^{(0)}$, $\nzo{\varphi}^{(1)}$ and $T_{\perp} = T_{\perp}^{(0)} + T_{\perp}^{(1)}$ is equivalent to solution of $g_0$ and $g_1$ at zeroth and first order; see Equations \eref{laguerre-qn-1}-\eref{laguerre-T}.  We may then truncate the hierarchy by setting $g_2 = 0$, $g_3 = 0$, \etc, and so by \Eref{laguerre-chi} we can set $\chi^{(0)}  = 2 T_{\perp}^{(0)} + \tau \nzo{\varphi}^{(0)}/2$ in \Eref{psi-1-eqn}.  

\subsection{Simple gyrofluid model}

There is a fundamental problem in solving the system of equations, up to first order fluctuations, separately as presented in the previous section.  Taking \Eref{phi-eqn-1} at face value, we see that the crucial FLR terms affect the evolution of small corrections $\nzo{\varphi}^{(1)}$.  If $\nzo{\varphi}^{(1)}$ is really dynamically relevant then it must feed back on $\varphi^{(0)}$, which it does not do in this system.  In fact, the system comprised of Equations \eref{nzo-phi-eqn-0}, \eref{phi-eqn-1} and \eref{T-0-eqn} is actually not closed since $\zon{\varphi}_1$, which appears in the third term of \Eref{phi-eqn-1}, is undetermined (the dynamical equation for $\zon{\varphi}_1$ can be obtained at one order higher but we will not give it here).

We can justify the inclusion of the FLR terms in the equation for $\varphi_0$, however, by allowing for ``nonlocal interactions.''  We argue that they contribute at dominant order and so they are not really higher order.  Implicit in derivation so far is an assumption of locality of interactions, \ie the ordering in terms of $k^2$ is assumed to be true ``scale-by-scale,'' and nonlinear interactions are assumed to occur among fluctuations on comparable scales.  This is actually not correct for the tertiary instability, which couples large-scale zonal flows to fine-scale fluctuations.  It was shown by \citet{rogers-prl} that the tertiary growth rate peaks for $k_t \sim \sqrt{\bar{k}}$, where $k_t$ is the characteristic wavenumber of the tertiary instability and $\bar{k}$ is the wavenumber of the zonal flow.  Thus we include the FLR terms in the equation for $\varphi^{(0)}$ and, dropping superscripts, write

\begin{equation}
\eqalign{\partial_t(\tau \nzo{\varphi} - \nabla^2\varphi) + \poiss{\varphi}{\tau \nzo{\varphi} - \nabla^2\varphi}  +  \poiss{\varphi}{- \nabla^2T_{\perp}} + \poiss{\nabla^2 \varphi}{T_{\perp}} \nonumber\\
+ \nabla^2\poiss{\varphi}{T_{\perp}}  = 0,}\label{gf-eqn-unforced-1}
\end{equation}

\noindent Note that \Eref{gf-eqn-unforced-1} is the same as \Eref{gf-eqn-1} but without the additional source terms.  \Eref{gf-eqn-unforced-1} conserves the full (gyrofluid) electrostatic energy $\Egf$

\begin{equation}
\Egf = \frac{1}{2V}\int d^2{\bf r} \; \left[ \tau \nzo{\varphi}^2 + |\bnabla \varphi|^2\right].\label{E-gf-def}
\end{equation}

Combining the zeroth order and first order equations for $T_{\perp}$ and employing the truncation as described above, we obtain

\begin{equation}
\partial_t T_{\perp} + \poiss{\varphi}{T_{\perp}} + \poiss{\nabla^2\varphi}{2T_{\perp}- \tau \nzo{\varphi}/2} = 0,\label{gf-eqn-unforced-2}
\end{equation}

\noindent which is \Eref{gf-eqn-2} with the forcing terms on the right-hand side omitted.  In summary, Equations \eref{gf-eqn-unforced-1} and \eref{gf-eqn-unforced-2} comprise a model for nonlinear interactions in 2D gyrokinetics that includes energy flow from zonal to non-zonal components, allowing for saturation of zonal flows by nonlinear interaction.  Note that although the invariant $\Egf$ is exactly conserved, the inclusion of FLR terms in these equations means that the truncated version of $W_{g0}$ is now only approximately conserved.  This is due to the appearance of higher order terms in the equation under the change of variables between fluid moments $\varphi$, $T_{\perp}$, \etc and the Laguerre components $g_0$, $g_1$, \etc.

\begin{figure*}
\includegraphics[width=\columnwidth]{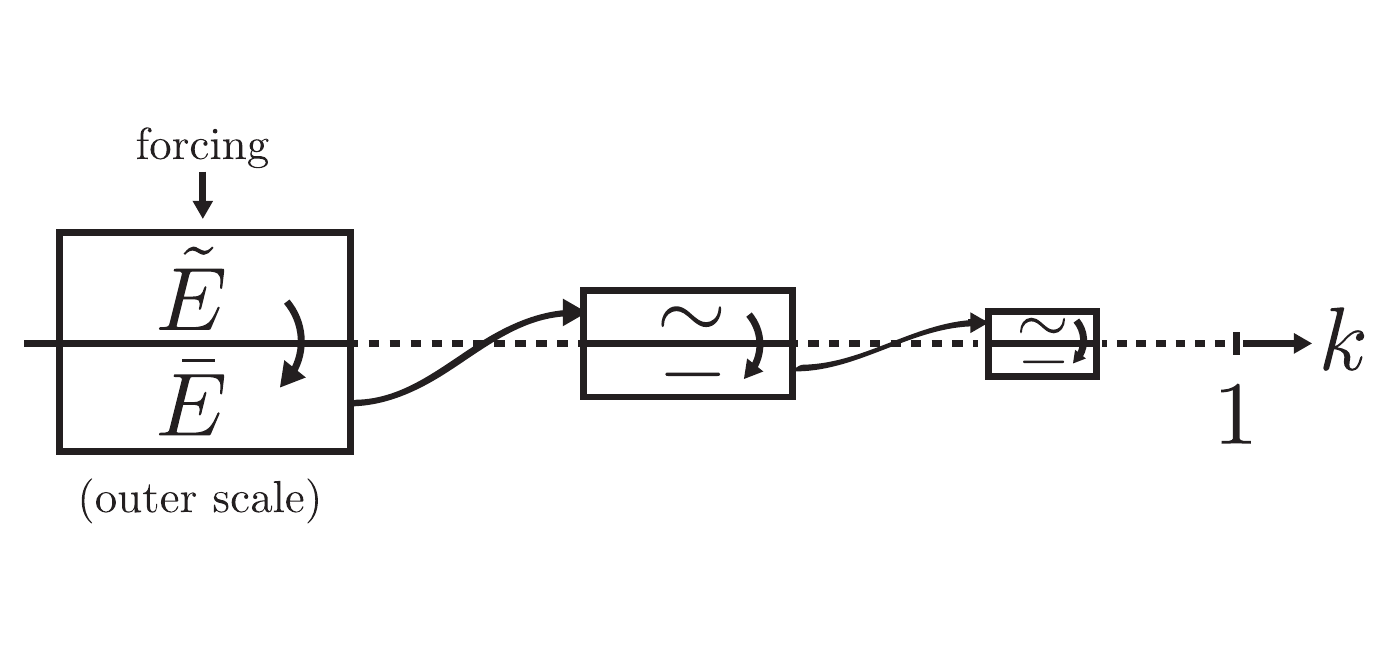}
\caption{Electrostatic energy transfer at long wavelengths ($k^2 \ll 1$) includes a crucial nonlocal transfer from zonal component to fine-scale non-zonal fluctuations.  This makes asymptotically consistent the inclusion of FLR terms in the equation for $\nzo{\varphi}$.}
\label{energy-flow-scales-fig}
\end{figure*}

Schematically, the picture of electrostatic energy flows consistent with this ordering may be summarized by \Fref{energy-flow-scales-fig}, where we consider local forcing at the largest scale of the system.  This picture is also compatible with the spectral transfer of $E$ observed in gyrokinetic ITG simulations \citep{banon-prl}, which shows the coexistence of nonlocal forward transfer and local inverse transfer (though not separated into zonal and non-zonal components).

\begin{figure}
\begin{center}
\subfloat[Zeroth order gyrofluid model.]{\label{gf-compare-fig-a}\includegraphics[width=0.504\columnwidth]{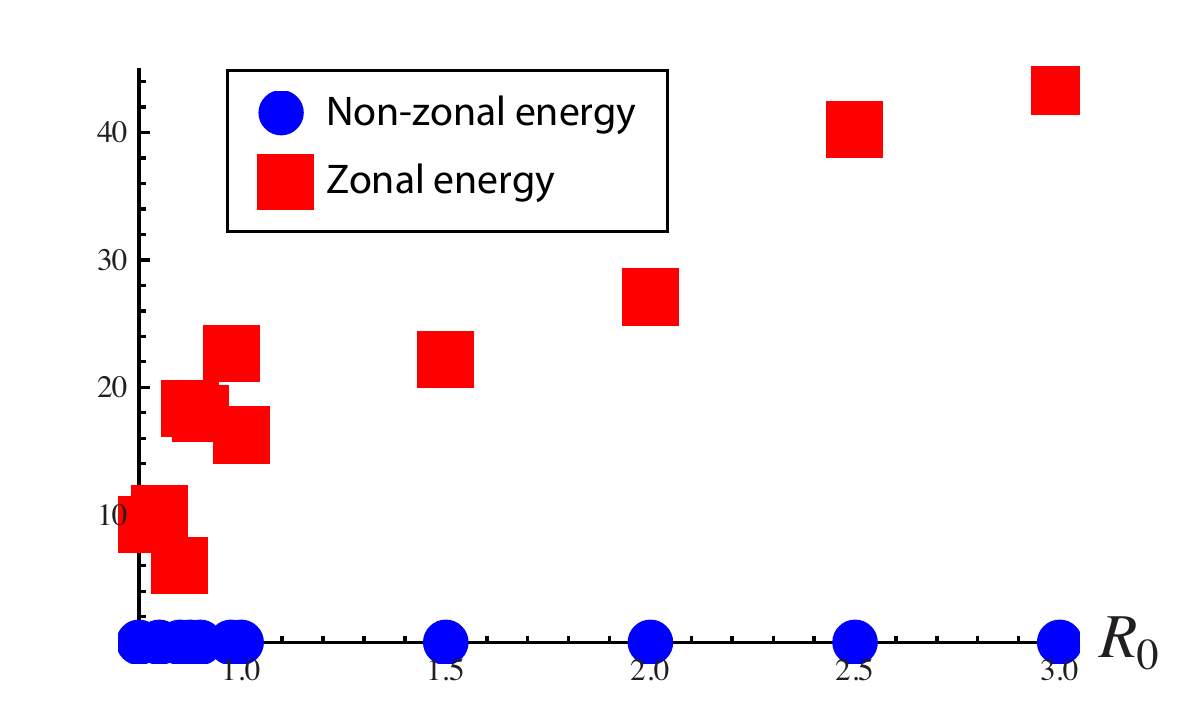}}
\subfloat[Gyrofluid model including FLR terms.]{\label{gf-compare-fig-b}\includegraphics[width=0.504\columnwidth]{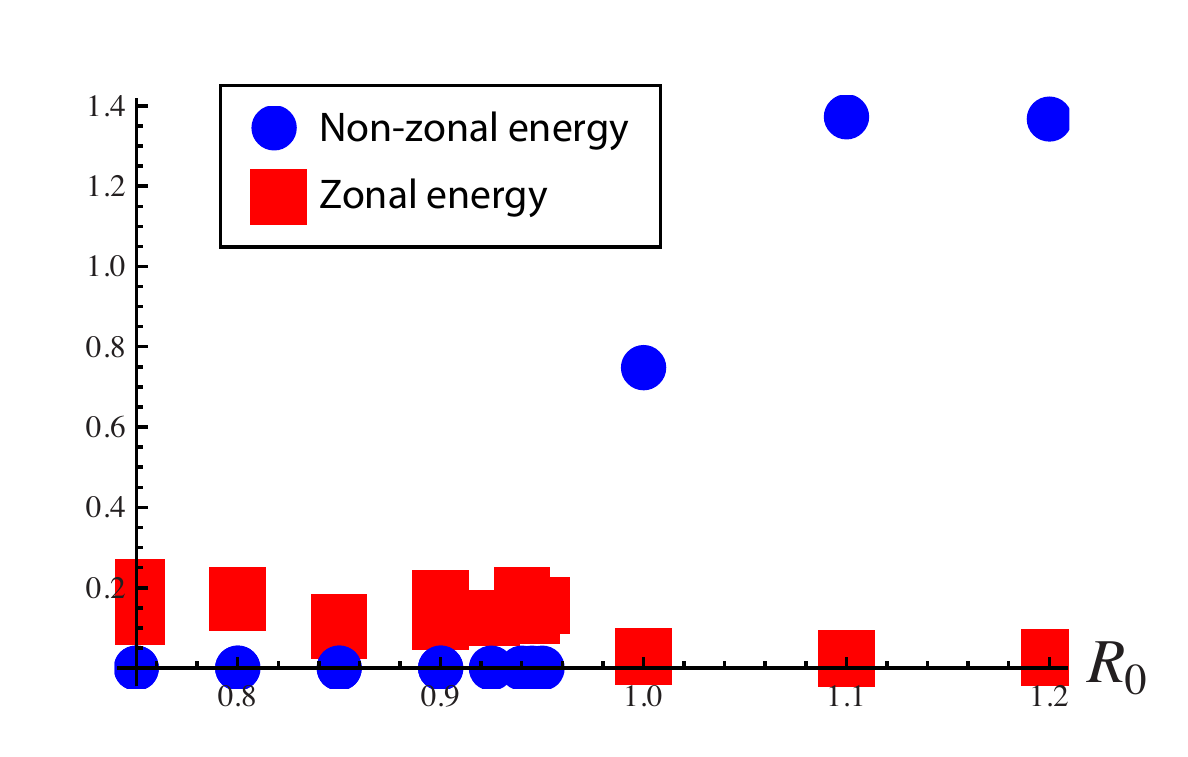}}
\end{center}
\caption{Nonlinear critical transition absent with zeroth-order model.}\label{gf-compare-fig}
\end{figure}

By direct numerical simulation we have compared the gyrofluid model that includes FLR effects with the ZOS (Equations \eref{nzo-phi-eqn-0}, \eref{zon-phi-eqn-0} and \eref{T-0-eqn}).  These runs were made with just 60 fourier modes but the linear drive terms and corresponding parameters are otherwise the same as those reported in \Sref{gf-sec}.  \Fref{gf-compare-fig} shows the striking difference in the behavior of these systems.  (Note that although for these runs the spectrum peaks around the instability drive $k \lesssim 0.25$, we have found that the difference does not diminish even if the magnitudes of the wavenumbers of the system are decreased further to represent the asymptotic limit $k^2 \rightarrow 0$.)  The nonlinear critical transition is absent for the ZOS and the amplitudes of the fluctuations at saturation are different by well over an order of magnitude.

\subsection{Fluid moment hierarchy in Laguerre polynomials}\label{Laguerre-appendix}

A convenience of the long-wavelength limit is that the Bessel function may be expanded as a low-order polynomial and one can easily use a standard set of orthogonal polynomials to represent the fluid moment hierarchy and decompose the velocity space dependence of $g$.  We choose the Laguerre polynomials, which have the desired weighting function so that the free energy $W_{g0}$ can be written as a simple sum of squared Laguerre coefficients.

The Laguerre polynomials, denoted $L_n(u)$ where $u = v^2/2$, are orthogonal with

\begin{equation}
\int du \e^{-u} L_n(u) L_m(u) = \delta(n - m)
\end{equation}

\noindent It follows that if we define

\begin{equation}
g = \frac{1}{2\pi}\sum_n L_n \e^{-u} g_n({\bf R}, u),
\end{equation}

\noindent then the quantity $W_{g0}$ is simply

\begin{equation}
W_{g0} = \frac{1}{2V}\int d^2{\bf R} \;\sum_n g_n^2.\label{laguerre-Wg0}
\end{equation}

\noindent Then, writing \Eref{qn-lw-eqn} up to first order in $\epsilon_{lw} = k^2$ we have

\begin{equation}
\tau\nzo{\varphi} - \nabla^2\varphi = (1 + \frac{1}{2}\nabla^2)g_0 - \frac{1}{2}\nabla^2 g_1,
\end{equation}
 
\noindent which implies

\begin{eqnarray}
&\tau\nzo{\varphi}^{(0)} = \nzo{g}_0^{(0)},\label{laguerre-qn-1}\\
&\zon{g}_0^{(0)} = 0,\label{laguerre-qn-2}\\
&\tau\nzo{\varphi}^{(1)} - \nabla^2 \varphi^{(0)} = g_0^{(1)} +\frac{1}{2}\nabla^2 g_0^{(0)} - \frac{1}{2}\nabla^2 g_1^{(0)}.\label{laguerre-qn-3}
\end{eqnarray}

\noindent We may also express the moments of the previous section as

\begin{eqnarray}
T_{\perp} &= \frac{1}{2}(g_0 - g_1) \label{laguerre-T}\\
\chi &= g_2 - 2 g_1 + g_0,\label{laguerre-chi}
\end{eqnarray}

\noindent and so forth.  The moment hierarchy may be written simply

\begin{eqnarray}
\partial_t g_0^{(0)} + \poiss{\varphi^{(0)}}{g_0^{(0)}} = 0,\\
\partial_t g_1^{(0)} + \poiss{\varphi^{(0)}}{g_1^{(0)}} = 0,
\end{eqnarray}

\noindent and

\begin{eqnarray}
\partial_t g_0^{(1)} + \poiss{\varphi^{(1)}}{g_0^{(0)}} + \poiss{\varphi^{(0)}}{g_0^{(1)}} + \poiss{\nabla^2\varphi^{(0)}}{\frac{1}{2}(g_0^{(0)} - g_1^{(0)})} = 0,\\
\partial_t g_1^{(1)} + \poiss{\varphi^{(1)}}{g_1^{(0)}} + \poiss{\varphi^{(0)}}{g_1^{(1)}} \nonumber\\
+ \poiss{\nabla^2\varphi^{(0)}}{-\frac{1}{2}g_0^{(0)} + \frac{3}{2}g_1^{(0)} - g_2^{(0)}} = 0,
\end{eqnarray}

\noindent which upon truncating the hierarchy, $g_2 \rightarrow 0$, is equivalent to the gyrofluid equations of the previous section.

\subsection{Linearization and instability for GHM and zeroth order gyrofluid system}\label{zo-GF-secondary-app}

We derived a criterion for the secondary instability of the GHM equation in \Sref{secondary-mod-response-sec}; see \Fref{GHM-sec-fjortoft-fig}.  Now we repeat the proof by a different method involving continued fractions.  Let the primary mode be 

\begin{equation}
\varphi_p = \varphi_0\exp{i k_0 y} + \mbox{c.c.},
\end{equation}

\noindent where $\varphi_0$ is a positive real amplitude, `c.c.' denotes the complex conjugate and $k_0$ is the primary wavenumber.  The secondary mode, $\varphi_s \ll \varphi_p$, satisfies the equation

\begin{equation}
\partial_t(\tau\nzo{\varphi}_s - \nabla^2\varphi_s) + \poiss{\varphi_p}{\tau\nzo{\varphi}_s - \nabla^2\varphi_s} + \poiss{\varphi_s}{\tau\varphi_p - \nabla^2\varphi_p} = 0.\label{mhm-secondary-eqn}
\end{equation}

\noindent By Floquet theory, the normal mode solution is of the form

\begin{equation}
\varphi_s = \exp{(i\mu y - i\omega t + i k_x x)}\displaystyle{\sum_n}\;c_n\exp{i n k_0 y},\label{mhm-sec-mode-form}
\end{equation}

\noindent where the complex frequency of the mode is $\omega$, the $x$-wavenumber is $k_x$ and the free parameter $\mu$ is the Floquet exponent, which we can assume satisfies $-k_0/2 < \mu < k_0/2$. Substituting this into \Eref{mhm-secondary-eqn}, separating the equation into Fourier components results in a system of equations for the coefficients $\{c_n\}$.  Following \citep{meshalkin-sinai, frenkel}, a solution to this system may be found by continued fractions.  This solution is purely growing (zero real frequency) and the dispersion relation for this mode is written:

\begin{equation}
-z_0 = 
\cfrac{1}{z_{1} + 
\cfrac{1}{z_{2} + 
\cfrac{1}{z_{3} +...
}}} 
+
\cfrac{1}{z_{-1} + 
\cfrac{1}{z_{-2} + 
\cfrac{1}{z_{-3} +...
}}} \label{mhm-secondary-disp-reln}
\end{equation}

\noindent where $z_n = \frac{\gamma_s}{k_xk_0\varphi_0}\frac{s_n^2}{s_n^2 - k_0^2- \tau}$, $s_n^2 = \tau[1 - \delta(nk_0 + \mu)] + k_x^2 + (n k_0 + \mu)^2$, $\gamma_s = -i \omega_s$ is the secondary growth rate and $\delta$ is the discrete delta-function.  Now, noting that $z_n > 0$ for $|n| \geq 1$, the right-hand side of \Eref{mhm-secondary-disp-reln} is positive.  Thus, we must have $z_0 < 0$, which leads to the instability criterion $k_x^2 < k_0^2 + \tau$.

Now we turn to the ``secondary'' instability of the ZOS, \Eref{nzo-phi-eqn-0}, \eref{zon-phi-eqn-0} and \eref{T-0-eqn}.  Borrowing some of the definitions already given above for the GHM secondary, we consider a monochromatic primary mode, with wavenumber $k_0$, but now include a corresponding temperature wave 

\begin{equation}
T_{\perp p} = T_{\perp 0} \exp{i k_0 y} + \mbox{c.c.}
\end{equation}

\noindent where $T_{\perp 0}$ is a complex amplitude.  This introduces two new parameters: $\delta\theta = \mbox{Arg}[T_{\perp 0}]$, the phase difference between the perturbed temperature and electrostatic potential, and $|T_{\perp 0}|/\varphi_0$, the relative amplitude of the perturbed temperature.  The secondary mode now consists of two parts, $\varphi_s \ll \varphi_p$ and a temperature perturbation $T_{\perp s} \ll T_{\perp p}$.  The secondary instability equations are

\begin{eqnarray}
\partial_t\nzo{\varphi}_s + \poiss{\zon{\varphi}_s}{\varphi_p} = 0,\label{sec-warm-eqn-1}\\
\partial_t (-\nabla^2\zon{\varphi_s}) + \zon{\poiss{\varphi_p}{-\nabla^2 \nzo{\varphi}_s}} + \zon{\poiss{\nzo{\varphi}_s}{-\nabla^2\varphi_p}} + \nabla^2\zon{\poiss{\varphi_p}{\nzo{T}_{\perp s}}} \nonumber\\
+ \nabla^2\zon{\poiss{\nzo{\varphi}_s}{T_{\perp p}}} -\zon{\poiss{\varphi_p}{\nabla^2\nzo{T}_{\perp s}}} - \zon{\poiss{\nzo{\varphi}_s}{\nabla^2T_{\perp p}}} + \zon{\poiss{\nabla^2\varphi_p}{\nzo{T}_{\perp s}}} \nonumber\\
+ \zon{\poiss{\nabla^2\nzo{\varphi}_s}{T_{\perp p}}} = 0,\label{sec-warm-eqn-2}\\
\partial_tT_{\perp s} + \poiss{\varphi_p}{T_{\perp s}} + \poiss{\varphi_s}{T_{\perp p}} = 0.\label{sec-warm-eqn-3}
\end{eqnarray}

\noindent The form of $\varphi_s$ is given by \Eref{mhm-sec-mode-form}, and form of $T_{\perp s}$ is

\begin{equation}
T_{\perp s} = \exp{(i\mu y - i\omega t + i k_x x)}\displaystyle{\sum_n}\;d_n\exp{i n k_0 y}.\label{sec-mode-form-temp}
\end{equation}

\noindent We plug Equations \eref{mhm-sec-mode-form} and \eref{sec-mode-form-temp} into Equations \eref{sec-warm-eqn-1}-\eref{sec-warm-eqn-3} and, after some algebra, obtain a system of equations for the coefficients ${c_n}$ and ${d_n}$.  From this system we derive a dispersion relation for the secondary complex frequency $\omega_s$.  For $\mu \neq 0$, the zonal component of the secondary mode is zero and we are left with \Eref{sec-warm-eqn-3}; for this case, the solution is purely oscillatory.  

For $\mu = 0$, we find

\begin{equation}
z^2 = 2(1 -r -r^*),
\end{equation}

\noindent where $z = \frac{\gamma_s}{k_xk_0\varphi_0}$ and $r = \frac{T_{\perp 0}}{\varphi_0}$.

\bibliographystyle{unsrtnat}
\bibliography{big-fjortoft}

\end{document}